\documentclass[preprint,aps,prc,endfloats*,tightenlines,showpacs,superscriptaddress,nofootinbib]{revtex4-1}
\usepackage{amsmath,amssymb,amsfonts}
\usepackage[dvips]{graphicx,color}
\usepackage{dcolumn}
\usepackage{epsfig}
\usepackage{bm}

\usepackage{ulem}
\usepackage{color}

\newcommand{\ket}[1]{|{#1}\rangle}
\newcommand{\bra}[1]{\langle{#1}|}

\newcommand{\slas}[1]{\not\!{#1}}

\def\ohalf{\textstyle{\frac{1}{2}}}
\def\thalf{\textstyle{\frac{3}{2}}}
\def\fhalf{\textstyle{\frac{5}{2}}}
\def\shalf{\textstyle{\frac{7}{2}}}

\begin{document}

\title{Dynamical coupled-channels model of $K^- p$ reactions (I): Determination of partial-wave amplitudes} 
\author{H. Kamano}
\affiliation{Research Center for Nuclear Physics, Osaka University, Ibaraki, Osaka 567-0047, Japan}
\author{S. X. Nakamura}
\affiliation{Department of Physics, Osaka University, Toyonaka, Osaka 560-0043, Japan}
\author{T.-S. H. Lee}
\affiliation{Physics Division, Argonne National Laboratory, Argonne, Illinois 60439, USA}
\author{T. Sato}
\affiliation{Department of Physics, Osaka University, Toyonaka, Osaka 560-0043, Japan}

\begin{abstract}
We develop a dynamical coupled-channels model of $K^- p$ reactions,
aiming at extracting the parameters associated with hyperon resonances and providing 
the elementary antikaon-nucleon scattering amplitudes that can be used for
investigating various phenomena in the strangeness sector such as 
the production of hypernuclei from kaon-nucleus  reactions.
The model consists of 
(a)~meson-baryon ($MB$) potentials $v_{M'B',MB}$ 
derived from the phenomenological SU(3) Lagrangian, and 
(b)~vertex interactions $\Gamma_{MB,Y^*}$ for describing the decays of the bare excited 
hyperon states ($Y^*$) into $MB$ states.
The model is defined in a channel space spanned by the
two-body $\bar K N$, $\pi \Sigma$, $\pi \Lambda$, $\eta \Lambda$, and $K\Xi$ states and also 
the three-body $\pi\pi\Lambda$ and $\pi\bar K N$ states that 
have the resonant $\pi \Sigma^*$ and $\bar{K}^* N$ components, respectively.
The resulting coupled-channels scattering equations satisfy 
the multichannel unitarity conditions
and account for the dynamical effects arising from the off-shell rescattering processes.
The model parameters are determined by fitting the available data of 
the unpolarized and polarized observables of the 
$K^- p \to \bar K N, \pi \Sigma, \pi \Lambda, \eta \Lambda, K\Xi$ 
reactions in the energy region from  the threshold  to  invariant mass $W=2.1$ GeV. 
Two models with equally good $\chi^2$ fits to the data have been constructed.
The partial-wave amplitudes obtained from the constructed models are compared with the results 
from a recent partial-wave analysis by the Kent State University group.
We discuss the differences between these three analysis results.
Our results at energies near the threshold suggest that 
the higher partial waves should be treated on the same footing as
the $S$ wave if one wants to understand the nature of $\Lambda(1405)1/2^-$ 
using the data below the $\bar K N$ threshold, as will be provided by
the J-PARC E31 experiment.
\end{abstract}


\maketitle

\section{Introduction}

The spectrum and structure of baryons with nonvanishing strangeness ($S$) quantum number,
the hyperons ($Y^*$), are currently much less understood than 
the  $N^*$ and $\Delta^*$ excited states of the nucleon.
As pointed out in Ref.~\cite{zhang2013a}, 
the past partial-wave analyses~\cite{pwa-1,pwa-2,pwa-3,pwa-4,pwa-5,pwa-6,pwa-7}
for investigating $Y^*$ were mostly performed using the Breit-Wigner parametrization
and did not extract the resonance parameters defined by 
the poles and residues of the scattering amplitudes.
In fact, the values of poles and residues for the $Y^*$ resonances are not given 
by the Particle Data Group (PDG)~\cite{pdg2012}, unlike the $N^*$ and $\Delta^*$ resonances.
To establish the $Y^*$ mass spectrum, more extensive investigations are needed
theoretically and experimentally.

In this work, we develop a dynamical coupled-channels (DCC) model for antikaon-nucleon ($\bar KN$) reactions 
within a Hamiltonian  formulation developed
in Refs.~\cite{msl07,jlms07,jlmss08,kjlms09-1,jklmss09,kjlms09-2,knls10,knls13,kpi2pi13}.
The $\bar K N$ reactions are particularly suitable for studying
$Y^*$ with $S=-1$, $\Lambda^*$ and $\Sigma^*$, since those appear as direct $s$-channel
processes in the reactions.
Following the formulation of Ref.~\cite{msl07}, we derive the model Hamiltonian by using the
unitary transformation method.
The interaction Hamiltonian consists of 
(a)~meson-baryon ($MB$) potentials $v_{M'B',MB}$ derived from
the phenomenological SU(3) Lagrangian, and 
(b)~vertex interactions $\Gamma_{MB,Y^*}$ for describing
the decays of  bare excited hyperon states ($Y^*=\Lambda^*, \Sigma^*$) into $MB$ states.
One can show that resulting coupled-channels scattering amplitudes satisfy 
the multichannel unitarity conditions.
Furthermore, they account for the dynamical effects arising from the off-shell
rescattering processes. 
We will apply the model to analyze all of the available data of the
unpolarized and polarized observables of
the $K^- p \to \bar K N, \pi \Sigma, \pi \Lambda, \eta\Lambda, K\Xi$ reactions
from the threshold up to $W=2.1$ GeV, where $W$ is the total scattering energy in the center-of-mass frame.

The $\bar KN$ reaction data  included in our analysis are similar to those 
used in the partial-wave analysis by the Kent State University (KSU) group~\cite{zhang2013}.
It is useful to note here that there are some connections and differences between our dynamical 
approach and  the KSU analysis. In their single-energy partial-wave analysis,  
a multichannel $K$-matrix model developed by Manley~\cite{manley2003,shre2012} 
was used to guide/constrain their determinations of partial-wave amplitudes from fitting the data.
It was pointed out in Ref.~\cite{lee2005} that this $K$-matrix model can be derived 
from a dynamical model based on a Hamiltonian, such as the one employed in this work,
by taking the on-shell approximation to evaluate the meson-baryon propagators in
the scattering equations. 
The full amplitudes in both approaches can be written in terms of  vertices $f_{MB,B^*}$ describing
the decay of the excited baryons ($B^*=$ $N^*$, $\Delta^*$ or $Y^*$) 
into $MB$ states and the ``background'' scattering 
$T$ matrix $t^{\text{bg}}$  between the considered $MB$ states. 
The parameters associated with the vertices $f_{MB,B^*}$ are treated purely phenomenologically. 
The main differences between two approaches 
are in the treatments of  $t^{\text{bg}}$. 
In our dynamical approach, the on- and off-shell matrix elements
of $t^{\text{bg}}$ in \textit{all} partial waves are determined by the \textit{same} 
parameters of the constructed meson-exchange potentials $v^{\text{mex}}$. 
In the KSU approach, the needed background scattering matrix $\omega^{(\pm)}_{\text{bg}}$ defined 
by the on-shell matrix elements of $t^{\text{bg}}$ are parametrized in terms of 
unitary matrices and their parameters in each partial wave are adjusted independently 
of other partial waves in fitting the data.
This difference makes the KSU approach more efficient in fitting the data. 
Perhaps mainly because the amount and the quality
of the data in each energy bin ($20$ MeV) could be very different,
the partial-wave amplitudes determined in the KSU's single-energy analysis could be 
not smooth in energy.
Thus they impose ``smoothness'' as an additional condition in finalizing their results. 
This is a reasonable approach since they
also verify their final results by showing that the observables calculated 
from their partial-wave amplitudes are in agreement with the data. 
In our dynamical approach, the parameters associated with the potential
$v^{\text{mex}}$ and the vertices $f_{MB,B^*}$ are adjusted to fit the data of 
observables in all considered energies.
Thus the determined partial-wave amplitudes in all partial waves depend on the \textit{same} 
set of the parameters of the constructed meson-baryon potentials. 
This makes the fits to the data of the observables of $\bar KN$ reactions more difficult 
than the KSU analysis.  
Furthermore, solving the coupled-channels equations in a dynamical approach 
is rather time consuming.

As discussed previously~\cite{knls13}, the purpose
of taking a much more complicated dynamical model to analyze the meson-baryon reaction data
is not only to determine the partial-wave amplitudes for resonance
extractions, but also to provide an understanding of the
dynamical content of the extracted baryon resonances. 
Here we further point out the following three motivations of our approach:
\begin{enumerate}
\item 
It is well recognized, as discussed in Refs.~\cite{tabakin,shkl11,rosetta}, that
the data from complete or overcomplete
measurements are needed to have a model \textit{independent} (up to a common phase)
determination of partial-wave amplitudes for extracting hadron resonances.
As will be detailed in Sec.~\ref{sec:results}, 
the data needed for our analysis are still rather incomplete.
Thus it is desirable to fit the available data within a reaction model 
that is constrained by the well-established physics.
Our DCC model and a similar coupled-channels model developed
in the J\"ulich $\pi N$ analysis~\cite{juelich,juelich12} are constrained by 
the meson-exchange mechanisms. 
Both approaches are motivated by the success of
the meson-exchange models of $NN$ interactions~\cite{machleidt}
and $\pi N$ and $\gamma N$ reactions
in the $\Delta$ (1232) region~\cite{pj-91,gross,sl96,sl01,pasc,pitt-ky,fuda,ntuanl}.
Thus there are good reasons to assume that a DCC model, 
as formulated in Refs.~\cite{msl07} and~\cite{juelich},
can minimize the uncertainties due to the lack of complete data 
in extracting nucleon resonances. 
Our motivation here is similar to the dispersion-relation approaches~\cite{hohler,cutkosky}.
To reduce the experimental uncertainties in determining partial-wave amplitudes 
for resonance extractions, it is necessary to impose various theoretical assumptions,
such as the choices of the subtraction terms, the input needed
for crossing symmetry, and the asymptotic behavior of the amplitudes, 
to solve the considered dispersion relations.

\item To extract resonances from the data, the parametrization of the scattering amplitudes
should have essential analytical properties such as 
the left-hand cuts and $2\rightarrow 2$ and $2 \rightarrow 3$ cut singularities. 
These properties can be  built in straightforwardly by defining our model Hamiltonian
in terms of the meson-exchange mechanisms. 
Consequently, the background amplitudes
of different partial waves are determined simultaneously, while the KSU model and
other similar analysis models~\cite{said,bg2012,cmb,zegrab,pitt-anl} do not have this
advantage.

\item The dynamical model constructed in this work can make predictions on 
some unmeasurable transition amplitudes with theoretical constraints, 
while the pure phenomenological models such as the KSU model cannot.
In addition, our model can  account for the off-shell effects due to the $\bar KN$ 
rescattering processes. 
Those effects are known to be important for a quantitative understanding of 
the production of hypernuclei and kaonic nuclei in kaon-induced nuclear reactions~\cite{k-nucl}. 
Our dynamical model thus has a great advantage also in the applications 
to various reaction systems in the strangeness sector that are relevant
to the recent experimental efforts at J-PARC~\cite{jparc}.

\end{enumerate}

Here we also note that most of the previous investigations of $Y^*$ based on
coupled-channels models (e.g., Refs.~\cite{landau-1,shinmura,ucm-1}) focus on studying 
the resonances extracted from the $S$-wave amplitudes of $\bar KN$ scattering at low energies. 
Higher partial waves were also considered in Ref.~\cite{landau-2}, 
but the channels and the data considered in this analysis are much
more limited than what we will present in this paper.
Similarly, a coupled-channels $\bar K N$ model developed in Ref.~\cite{juelich-kbn} 
is also limited to the threshold region.
There also exist model studies based on tree diagrams (e.g., Ref.~\cite{kon11}) of
$Y^*$, which are obviously different from the coupled-channels approaches.
In parallel with extracting the $Y^*$ resonance parameters from the experimental data
on the basis of reaction approaches,
there are activities to compute the real energy spectrum of QCD in the $S=-1$ hyperon sector
within the lattice QCD framework~\cite{lattice-spec}
by imposing the (anti)periodic boundary condition.
Furthermore, several attempts are also performed 
to extract complex resonance parameters from such real energy spectrum 
of QCD (see, e.g., Ref.~\cite{lattice-res}).

Our first task is to determine the model parameters by fitting
the available  data of $K^- p$ reactions from the threshold to $W=2.1$ GeV.
The partial-wave amplitudes of the $\bar KN$ reactions obtained from 
the constructed models are then compared 
with the results from the recent single-energy partial-wave analysis~\cite{zhang2013} 
of the KSU group. 
These two results will be presented in this paper.
The $Y^*$ resonance parameters, which are extracted from our partial-wave amplitudes
by using the analytic continuation method developed in Ref.~\cite{ssl}, 
will be presented in a separate paper~\cite{knls14a}.

In Sec.~\ref{sec:dcc}, we recall the coupled-channels formulation of Ref.~\cite{msl07} to
write down the scattering equations for investigating $\bar KN$ reactions.
The fits to the data are presented in Sec.~\ref{sec:results}. 
In Sec.~\ref{sec:discussion}, we present the partial-wave amplitudes
obtained from our models and compare them with the KSU results.
The threshold parameters (scattering lengths and effective ranges) 
and the predicted $K^- p$ reaction total cross section are also presented.
Summary and discussions on necessary future works are given in Sec.~\ref{sec:summary}.

\section{Dynamical Coupled-Channels Model}
\label{sec:dcc}

\begin{figure}
\includegraphics[clip,width=0.8\textwidth]{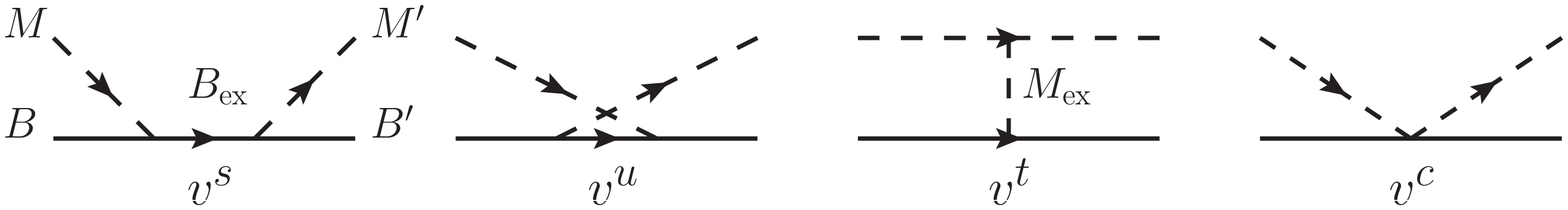}
\caption{\label{fig:mex} Exchange mechanisms in $v_{M'B',MB}$}
\end{figure}

Following the formulation of Ref.~\cite{msl07}, we assume that the Hamiltonian of the considered
systems is
\begin{equation}
H=H_0 + H_I ,
\label{eq:h}
\end{equation}
where $H_0$ is the free Hamiltonian and 
the interaction  Hamiltonian $H_I$ for $\bar{K}N$ reactions can be written as
\begin{equation}
H_{I} = \sum_{M'B',MB}v_{M'B',MB} + \sum_{Y^*,MB}\left(\Gamma_{MB,Y^*} 
+ \Gamma_{Y^*,MB} \right) 
+(h_{\pi \bar{K},\bar{K}^*} + h_{\bar{K}^*, \pi \bar{K}}) ,
\label{eq:hi}
\end{equation}
where $MB = \bar K N, \pi \Sigma, \pi \Lambda , \eta\Lambda, K\Xi,  \pi \Sigma^*, \bar{K}^* N$;
$v_{M'B',MB}$ is the meson-baryon exchange potentials derived from the phenomenological SU(3) Lagrangian;
$\Gamma_{MB,Y^*}$ are the vertex interactions describing the decays of bare
excited hyperon states ($Y^*=\Lambda^*, \Sigma^*$) to $MB$ states; 
and $h_{\pi \bar K, \bar{K}^*}$ describes the decay of $\bar{K}^*$ to $\pi \bar{K}$ state.
As shown in Fig.~\ref{fig:mex}, the meson-baryon exchange potentials $v_{M'B',MB}$
consist of the tree diagrams of $s$-channel and $u$-channel baryon exchanges, 
$t$-channel meson exchanges, and contact terms.
We consider the ground state baryons belonging to the flavor SU(3) 
octet and decuplet representations for the $u$-channel exchange baryons,
while only the ground state octet baryons are considered for the $s$-channel exchange baryons.
This is because the $s$-channel decuplet baryon exchanges
are taken into account via the $Y^*$-excitation term as described below.
For the $t$-channel processes, however,
the ground-state octet vector, scalar, and pseudoscalar mesons are considered as exchanged particles.
We list the Lagrangian used in our derivations and the explicit forms of $v_{M'B',MB}$
in Appendices~\ref{app:lag} and~\ref{app:pot}, respectively.
(Table~\ref{tab:vmbmb} in Appendix~\ref{app:pot} summarizes the exchanged hadrons 
included in $v_{M'B',MB}$.)

Here we note that there is no vertex interactions associated 
with the ground states of the considered
baryons ($N, \Lambda, \Sigma,\Xi$) and mesons ($\pi, K, \bar K,\eta$) in 
the considered Hamiltonian, Eqs.~(\ref{eq:h}) and~(\ref{eq:hi}). 
Such vertex interactions are eliminated from the starting
Lagrangians by employing a unitary transformation method, 
as detailed in Refs.~\cite{sl96,sko}, and absorbed in the two-body interactions $v_{M'B',MB}$.
As also discussed in Refs.~\cite{sl96,msl07}, this greatly simplifies the scattering 
equation since there is no complications due to the mass renormalization
in satisfying the unitarity condition.
Accordingly, we simply use the physical masses for these ground states hadrons in
our formulation. 
On the other hand, the vertex interaction $\Gamma_{MB,Y^*}$ will dress the
mass of the bare $Y^*$ state.

Following the standard collision theory~\cite{goldberger}, the $T$ matrix elements
defined by the Hamiltonian, Eqs.~(\ref{eq:h}) and~(\ref{eq:hi}), are of the following form
\begin{equation}
\bra{\beta}T(E)\ket{\alpha} = 
\bra{\beta}H_I\ket{\alpha} 
+ \sum_{\gamma}
\bra{\beta}H_I\ket{\gamma} 
\bra{\gamma} \frac{1}{E-H_0+i\epsilon}\ket{\gamma} \bra{\gamma}T(E)\ket{\alpha} ,
\label{eq:scatt-t}
\end{equation}
where the model space is spanned by 
the bare $Y^*$, (quasi-)two-body ($\bar{K}N$, $\pi\Sigma$, $\pi\Lambda$, $\eta\Lambda$, $K\Xi $,  
$\pi \Sigma^*$, $\bar K^* N$),
and three-body ($\pi\pi\Lambda, \pi \bar{K} N$) states.
As seen in Eq.~(\ref{eq:hi}), the interaction Hamiltonian $H_I$ is energy independent 
and hence it is straightforward to follow the procedures in Ref.~\cite{goldberger} to show that 
the scattering $T$ matrix defined by Eq.~(\ref{eq:scatt-t})
satisfies the two- and three-body unitary condition in the considered multichannel space.
Here we also note that the unitarity condition in our formulation has the three-body
unitary cuts that are similar to what are in the formulation of Aaron-Amado-Young~\cite{aay},
although two approaches are rather different.

To facilitate the numerical calculations, it is more convenient to cast Eq.~(\ref{eq:scatt-t})
into a form such that the amplitudes can be written as a sum of a ``nonresonant'' (background)
term, which is determined only by the exchange interactions $v_{M'B',MB}$, 
and a term describing the formation of resonant $Y^*$ states. 
This can be done by following the steps in Appendix~B of Ref.~\cite{msl07}.
Because this coupled-channels formulation  has been given in detail
in Refs.~\cite{msl07,jlms07,jlmss08,jklmss09,kjlms09-1,kjlms09-2,knls10,knls13,kpi2pi13}, here we
only present concisely the resulting  equations that are used in the
calculations.

By applying the projection operator method~\cite{feshbach} on Eq.~(\ref{eq:scatt-t}), 
we can cast the partial-wave components of the $T$ matrix elements of the meson-baryon reactions, 
$M(\vec k) + B(-\vec k) \to M'(\vec k') + B'(-\vec k')$, into the following form
\begin{equation}
T_{M'B',MB}(k',k;W) = t_{M'B',MB}(k',k;W) + t^R_{M'B',MB}(k',k;W),
\label{eq:tmbmb}
\end{equation}
where $W$ is the total energy, $k$ and $k'$ are the 
meson-baryon relative momenta in the center-of-mass frame.
[The label ``$MB$'' also specifies quantum numbers (spin, parity, isospin, etc.)
associated with the channel $MB$.]
The nonresonant amplitudes $t_{M'B',MB}(k',k;W)$ in Eq.~(\ref{eq:tmbmb})
are defined by a set of coupled-channels integral equations,
\begin{eqnarray}
t_{M'B',MB}(k',k;W) &=& V_{M'B',MB}(k',k;W) 
\nonumber \\
&& 
+ \sum_{M^{''} B^{''}} \int_{C_{M''B''}} k''^{2}dk'' V_{M'B',M''B''}(k',k'';W)
\nonumber\\
&&
\qquad\qquad\qquad
\times
G_{M''B''}(k'';W) t_{M''B'',MB}(k'',k;W).
\label{eq:cc-eq}
\end{eqnarray}
Here $C_{M''B''}$ is the integration path,
which is taken from $0$ to $\infty$ for the physical $W$;
the summation $\sum_{M''B''}$ runs over the orbital angular momentum and total spin indices
for all $M''B''$ channels allowed in a given partial wave;
$G_{M''B''}(k'';W)$ are the meson-baryon Green's functions.
Defining  $E_\alpha(k)=[m^2_\alpha + k^2]^{1/2}$ with $m_\alpha$ being
the mass of a particle $\alpha$,
the meson-baryon Green's functions in the above equations are
\begin{eqnarray}
G_{MB}(k;W)=\frac{1}{W-E_M(k)-E_B(k) + i\epsilon},
\label{eq:prop-stab}
\end{eqnarray}
for the stable $\bar K N$, $\pi \Sigma$, $\pi \Lambda$, $\eta \Lambda$, and $K\Xi$ channels, and
\begin{equation}
G_{MB}(k;W)=\frac{1}{W-E_M(k)-E_B(k) -\Sigma_{MB}(k;W)},
\label{eq:prop-unstab}
\end{equation}
for the unstable $\pi\Sigma^*$ and $\bar{K}^* N$ channels.
The details of the self-energy, $\Sigma_{MB}(k;W)$ in Eq.~(\ref{eq:prop-unstab}),
are given in Appendix~\ref{app:self}.
The meson-baryon Green's functions are responsible for the unitarity cuts in the $T$
matrix elements due to the opening of 
the two-body ($\bar K N, \pi \Sigma, \pi \Lambda, \eta \Lambda, K\Xi,\pi\Sigma^*,\bar K^* N$) 
channels as well as the three-body ($\pi\pi\Lambda, \pi \bar K N$) channels.

The driving terms of Eq.~(\ref{eq:cc-eq}) are
\begin{equation}
V_{M'B',MB}(k',k;W) = v_{M'B',MB}(k',k) + Z^{(E)}_{M'B',MB}(k',k;W). 
\label{eq:veff-mbmb}
\end{equation}
Here the potentials $v_{M'B',MB}(k',k)$ are the partial-wave components of 
$v_{M'B',MB}$ in Eq.~(\ref{eq:hi}).
Within the unitary transformation method~\cite{sko,sl96}
used in the derivation, those potentials are energy independent.
The energy-dependent $Z^{(E)}_{M'B',MB}(k',k;W)$ terms in Eq.~(\ref{eq:veff-mbmb})
are the effective one-particle-exchange potentials~\cite{msl07},
which are derived with the projection operator method~\cite{feshbach} 
and contain the singularities owing to the three-body unitarity cuts.
We note here that the similar $Z$-diagram mechanisms are also in the formulation
of Ref.~\cite{aay}.
The $Z$-diagram mechanisms $Z^{(E)}_{M'B',MB}(k',k;W)$
are the exchange processes of the self energy $\Sigma_{MB}(k;W)$ 
in the meson-baryon Green's functions. 
Both terms are necessary for maintaining
the three-body unitarity in the $T$ matrix elements.
The procedures for evaluating the partial-wave matrix elements of $Z^{(E)}_{M'B',MB}(k',k;W)$
are explained in detail in Appendix~E of Ref.~\cite{msl07}.

In this first attempt to construct a $\bar K N$ model, however, 
we neglect the $Z^{(E)}_{M'B',MB}(k',k;W)$ terms for the simplicity.
This partly violates the three-body unitarity, but we expect its influence is tiny
since we have observed that $Z^{(E)}_{M'B',MB}(k',k;W)$ have only few percent effects 
on the total cross sections in our previous calculations~\cite{knls13} for $\pi N$ reactions.
This of course needs to be improved along with other necessary tasks in the future, 
as will be discussed in Sec.~\ref{sec:summary}.

The second term on the right-hand side of Eq.~(\ref{eq:tmbmb}) 
is the $Y^*$-excitation term defined by
\begin{eqnarray} 
t^R_{M'B',MB}(k',k;W)= \sum_{Y^*_n, Y^*_m}
\bar{\Gamma}_{M'B',Y^*_n}(k';W) [D(W)]_{n,m}
\bar{\Gamma}_{Y^*_m, MB}(k;W) .
\label{eq:tmbmb-r} 
\end{eqnarray}
Here the dressed $Y^\ast \to MB$ and $MB \to Y^\ast$ vertices are, respectively, defined by
\begin{equation}
\bar\Gamma_{MB,Y^\ast}(k;W) =
\Gamma_{MB,Y^\ast}(k) +
\sum_{M'B'}\int_{C_{M'B'}} q^2 dq t_{MB,M'B'}(k,q;W) G_{M'B'}(q,W)\Gamma_{M'B',Y^\ast}(q),
\label{eq:dress-1}
\end{equation}
\begin{equation}
\bar\Gamma_{Y^\ast,MB}(k;W) =
\Gamma_{Y^\ast,MB}(k) +
\sum_{M'B'}\int_{C_{M'B'}} q^2 dq \Gamma_{Y^\ast,M'B'}(q) G_{M'B'}(q,W) t_{M'B',MB}(q,k;W),
\label{eq:dress-2}
\end{equation}
with $\Gamma_{MB,Y^\ast}(k)$ being the bare $Y^\ast\to MB$ decay vertex.
The inverse of the dressed $Y^\ast$ propagators is defined by
\begin{equation}
[D^{-1}(W)]_{n,m} = (W - M^0_{Y^*_n})\delta_{n,m} - [\Sigma_{Y^\ast}(W)]_{n,m},
\label{eq:nstar-g}
\end{equation}
where $M_{Y^*}^0$ is the mass of the bare $Y^*$ state 
and the $Y^\ast$ self-energies $\Sigma_{Y^\ast}(W)$ are given by 
\begin{eqnarray}
[\Sigma_{Y^\ast}(W)]_{n,m}
&=&\sum_{MB}\int_{C_{MB}}k^2 dk\Gamma_{Y^*_n,MB}(k) G_{MB}(k;W)
\bar{\Gamma}_{MB,Y^*_m}(k;W).
\label{eq:nstar-sigma}
\end{eqnarray}
We emphasize here that in general
the $Y^\ast$ propagators $D(W)$ becomes nondiagonal and multivalued in complex $W$
owing to the meson-baryon interactions in the coupled-channels system.
This makes the relation between bare states and physical resonances highly
nontrivial.
For instance, it has been demonstrated in Ref.~\cite{sjklms10} that within a coupled-channels
system a naive one-to-one correspondence between bare states and physical resonances
does not hold in general.

Equations~(\ref{eq:tmbmb})-(\ref{eq:nstar-sigma}) define the DCC 
model used in our analysis.
In the absence of theoretical input, the DCC model, as well as all hadron reaction models,
has parameters that can only be determined phenomenologically
from fitting the data. 
The exchange potentials $v_{M'B',MB}$ depend on the coupling constants 
and the cutoffs of form factors that qualitatively characterize the
finite sizes of hadrons.
While the values of some of the model parameters
can be estimated from the flavor SU(3) relations,
we allow most of them to vary in the fits.
The $s$-channel and $u$-channel mechanisms of $v_{M'B',MB}$ 
($v^s$ and $v^u$ in Fig.~\ref{fig:mex}) include at each meson-baryon-baryon
vertex a form factor of the form
\begin{eqnarray}
F(\vec{k},\Lambda)=\left(\frac{\Lambda^2}{\vec{k}^2+\Lambda^2}\right)^2,
\label{eq:ff}
\end{eqnarray}
with $\vec{k}$ being the meson momentum. For the meson-meson-meson
vertex of $t$-channel mechanism ($v^t$), Eq.~(\ref{eq:ff}) is
also used with $\vec{k}$ being the momentum of the exchanged meson.
For the contact term ($v^c$) we regularize it by
$F(\vec{k'},\Lambda')F(\vec{k},\Lambda)$.
The bare vertex functions in Eqs.~(\ref{eq:dress-1}) 
and~(\ref{eq:dress-2}) are parametrized as
\begin{eqnarray}
{\Gamma}_{MB(LS),Y^\ast}(k)
&=& \frac{1}{(2\pi)^{3/2}}\frac{1}{\sqrt{m_N}}C_{MB(LS),Y^\ast}
\left(\frac{\Lambda_{Y^\ast}^2}{\Lambda_{Y^\ast}^2 + k^2}\right)^{(2+L/2)}
\left(\frac{k}{m_\pi}\right)^{L} ,
\label{eq:gmb}
\end{eqnarray}
where $L$ and $S$ denote the orbital angular momentum and spin of the $MB$ state, respectively
[note that $\Gamma_{Y^\ast,MB}(k)= \Gamma_{MB,Y^\ast}^\dag(k)$].
All of the possible $(L,S)$ states in each partial wave 
included in our coupled-channels calculations are listed in Table.~\ref{tab:pw}.
The vertex function~(\ref{eq:gmb}) behaves as $k^L$ at $k\sim 0$ and $k^{-4}$ for $k\to\infty$.
The coupling constant $C_{MB(LS),Y^*}$ and the cutoff $\Lambda_{Y^*}$
are adjusted along with the bare masses $M^0_{Y^*}$  in the fits.

\begin{table}
\caption{
The orbital angular momentum $(L)$ and total spin ($S$) of each $MB$ channel allowed in a given partial wave.
In the first column, partial waves are denoted with the conventional notation $l_{I2J}$ as well as ($I$,$J^P$).
\label{tab:pw}}
\begin{ruledtabular}
\begin{tabular}{ccccccccccc}
$l_{I2J}$ $(I,J^P)$ &\multicolumn{10}{c}{$(L,S)$ of the considered partial waves}\\
\cline{2-11}
                       &$\bar K N$  &$\pi \Sigma$&$\pi \Lambda$&$\eta\Lambda$&$K\Xi$     &\multicolumn{2}{c}{$\pi \Sigma^*$}&\multicolumn{3}{c}{$\bar K^* N$}      \\
                       &            &            &             &             &           &$(\pi \Sigma^*)_1$&$(\pi \Sigma^*)_2$ &$(\bar K^* N)_1$&$(\bar K^* N)_2$&$(\bar K^* N)_3$\\
\hline
$S_{01}$ $(0,\ohalf^-)$&($0,\ohalf$)&($0,\ohalf$)&--          &($0,\ohalf$)&($0,\ohalf$)&($2,\thalf$)& --         &($0,\ohalf$)&($2,\thalf$)& --         \\
$S_{11}$ $(1,\ohalf^-)$&($0,\ohalf$)&($0,\ohalf$)&($0,\ohalf$)&--          &($0,\ohalf$)&($2,\thalf$)&--          &($0,\ohalf$)&($2,\thalf$)& --         \\
$P_{01}$ $(0,\ohalf^+)$&($1,\ohalf$)&($1,\ohalf$)&--          &($1,\ohalf$)&($1,\ohalf$)&($1,\thalf$)&--          &($1,\ohalf$)&($1,\thalf$)& --         \\
$P_{03}$ $(0,\thalf^+)$&($1,\ohalf$)&($1,\ohalf$)&--          &($1,\ohalf$)&($1,\ohalf$)&($1,\thalf$)&($3,\thalf$)&($1,\ohalf$)&($1,\thalf$)&($3,\thalf$)\\
$P_{11}$ $(1,\ohalf^+)$&($1,\ohalf$)&($1,\ohalf$)&($1,\ohalf$)&--          &($1,\ohalf$)&($1,\thalf$)&--          &($1,\ohalf$)&($1,\thalf$)& --         \\
$P_{13}$ $(1,\thalf^+)$&($1,\ohalf$)&($1,\ohalf$)&($1,\ohalf$)&--          &($1,\ohalf$)&($1,\thalf$)&($3,\thalf$)&($1,\ohalf$)&($1,\thalf$)&($3,\thalf$)\\
$D_{03}$ $(0,\thalf^-)$&($2,\ohalf$)&($2,\ohalf$)&--          &($2,\ohalf$)&($2,\ohalf$)&($0,\thalf$)&($2,\thalf$)&($2,\ohalf$)&($0,\thalf$)&($4,\thalf$)\\
$D_{05}$ $(0,\fhalf^-)$&($2,\ohalf$)&($2,\ohalf$)&--          &($2,\ohalf$)&($2,\ohalf$)&($2,\thalf$)&($4,\thalf$)&($2,\ohalf$)&($2,\thalf$)&($4,\thalf$)\\
$D_{13}$ $(1,\thalf^-)$&($2,\ohalf$)&($2,\ohalf$)&($2,\ohalf$)&--          &($2,\ohalf$)&($0,\thalf$)&($2,\thalf$)&($2,\ohalf$)&($0,\thalf$)&($2,\thalf$)\\
$D_{15}$ $(1,\fhalf^-)$&($2,\ohalf$)&($2,\ohalf$)&($2,\ohalf$)&--          &($2,\ohalf$)&($2,\thalf$)&($4,\thalf$)&($2,\ohalf$)&($2,\thalf$)&($4,\thalf$)\\
$F_{05}$ $(0,\fhalf^+)$&($3,\ohalf$)&($3,\ohalf$)&--          &($3,\ohalf$)&($3,\ohalf$)&($1,\thalf$)&($3,\thalf$)&($3,\ohalf$)&($1,\thalf$)&($3,\thalf$)\\
$F_{07}$ $(0,\shalf^+)$&($3,\ohalf$)&($3,\ohalf$)&--          &($3,\ohalf$)&($3,\ohalf$)&($3,\thalf$)&($5,\thalf$)&($3,\ohalf$)&($3,\thalf$)&($5,\thalf$)\\
$F_{15}$ $(1,\fhalf^+)$&($3,\ohalf$)&($3,\ohalf$)&($3,\ohalf$)&--          &($3,\ohalf$)&($1,\thalf$)&($3,\thalf$)&($3,\ohalf$)&($1,\thalf$)&($3,\thalf$)\\
$F_{17}$ $(1,\shalf^+)$&($3,\ohalf$)&($3,\ohalf$)&($3,\ohalf$)&--          &($3,\ohalf$)&($3,\thalf$)&($5,\thalf$)&($3,\ohalf$)&($3,\thalf$)&($5,\thalf$)\\     
\end{tabular}
\end{ruledtabular}
\end{table}

\section{Results of the fit}
\label{sec:results}

As already mentioned in the previous sections, we determine the model parameters by fitting
the available data of unpolarized and polarized observables of 
$K^- p \to \bar K N, \pi\Sigma, \pi\Lambda, \eta\Lambda, K\Xi$ 
from the threshold up to $W = 2.1$ GeV. 
The procedure and strategy for the fitting, e.g., criteria how many bare $Y^*$ states
are included in each partial wave, are essentially the same as
those employed in our coupled-channels analysis of $N^*$ resonances~\cite{knls13},
and we will not repeat it here.
The number of the data of each observable included in our fits is listed 
in Table~\ref{tab:data-chi2}.
Our database is similar to what were used in
the KSU single-energy partial-wave analysis~\cite{zhang2013}.
It is known that for the considered pseudoscalar-meson-baryon scattering,
the $complete$ data for determining partial-wave amplitudes 
need to include spin-rotation observables ($\beta$, $R$, or $A$). 
As seen in Table~\ref{tab:data-chi2}, there exist no data 
for such spin-rotation observables that can be included in our fits.
We thus have enough uncertainties of the constraints by the data
to construct two models, called Model A and Model B. 
As mentioned in the introduction, solving the coupled-channels equations
is rather time consuming compared to the on-shell approaches.
As a result, it is quite difficult to accomplish a detailed error estimation
of the partial-wave analyses within an acceptable time.
Instead, here we shall regard the discrepancies between the partial-wave amplitudes
from Models A and B as a measure of the ``error'' of the determined amplitudes,
resulting from the incompleteness of the data.
Here we also note that Models A and B have not only different sets of model parameters,
but also different forms for the vector-meson-exchange processes
in $v_{M'B',MB}$ with $MB,M'B'=\bar K N, \pi\Sigma, \pi\Lambda, \eta\Lambda, K\Xi$:
familiar vector-meson-exchange diagrams are used in Model A, while
in Model B a hybrid of the so-called Weinberg-Tomozawa (WT) terms and modified 
vector-meson-exchange diagrams is employed. 
The latter is intended to make a clear comparison with recent studies 
on the near-threshold phenomena in $S$-wave such as $\Lambda(1405)1/2^-$ (see e.g., Ref.~\cite{ucm-1}).
The details are explained in Appendix~\ref{app:pbpb-t-v-2}.

In our fits, we have also made an effort to find the well established decuplet baryon
$\Sigma^*(1385)$ with $S=-1$, $J^P=3/2^+$, and $I=1$.
However, the corresponding resonance parameters cannot be
constrained directly by 
the $K^- p$ reaction data included in our fits since $\Sigma^*(1385)3/2^+$
is below the $\bar K N$ threshold.
We therefore take the pole mass of $\Sigma^*(1385)3/2^+$, $1381-i20$ MeV~\cite{sstar-mass}, as ``data'' 
and determine the model parameters such that this resonance pole is  reproduced.

We next discuss how we perform the minimization of $\chi^2$. We follow the
most commonly used definition
\begin{equation}
\chi^2 = \sum_{O}\sum_{i,j}
\frac{[O^{\text{model}}(E_i,\theta_j)-O^{\text{exp.}}(E_i,\theta_j)]^2}{[\delta O^{\text{exp.}}(E_i,\theta_j)]^2},
\label{eq:chi2}
\end{equation}
where $O^{\text{model}}(E_i,\theta_j)$ is the observable $O$ at the energy $E_i$ and the angle $\theta_j$ calculated from the model parameters, while $O^{\text{exp.}}(E_i,\theta_j)$ and
$\delta O^{\text{exp.}}(E_i,\theta_j)$ are the central value and 
the statistical error of the experimental data.
There are more sophisticated minimization procedures accounting
for separately the systematical and statistical errors.
Thus some of the discrepancies between our final
results and the data, as will be presented in the next section,
could be partly due to our use of Eq.~(\ref{eq:chi2}) for $\chi^2$.
Such a more careful fitting procedure will be desirable when we move to our next
analysis including more complete data from future experiments.

The $\chi^2$ values from the fits are listed in Table~\ref{tab:data-chi2}. 
The $\chi^2/{\rm d.o.f.}$ value computed with the entire database is found to be 2.91 (3.03) 
for Model A (Model B). 
These values are larger than $\chi^2/{\rm d.o.f.} = 1$,
yet acceptable in determining the model parameters as presented in Appendix~\ref{app:model-para}.
There are mainly two origins for $\chi^2/{\rm d.o.f.} > 1$.
One is the existence of the data that are inconsistent and/or conflicting with each other.
As discussed later in Sec.~\ref{sec:pis}, 
an obvious example is the recoil polarization $P$ of $K^- p \to \pi^0 \Sigma^0$,
where the data from different analyses show a clear inconsistency.
Accordingly, the $\chi^2/{\rm data}$ values computed only with the data for this observable
become large, $\chi^2/{\rm data}\sim 6$.
At least, the observables providing $\chi^2/{\rm data} > 4$
are likely to contain such inconsistent and/or conflicting data.
Another is because at present only the statistical errors are taken into account in the fits,
as done in most theoretical analyses for baryon spectroscopy.
Incorporation of the systematic uncertainties in the fits 
will improve the $\chi^2$ values, and we will leave this to our future works.

\begin{figure}
\includegraphics[clip,width=\textwidth]{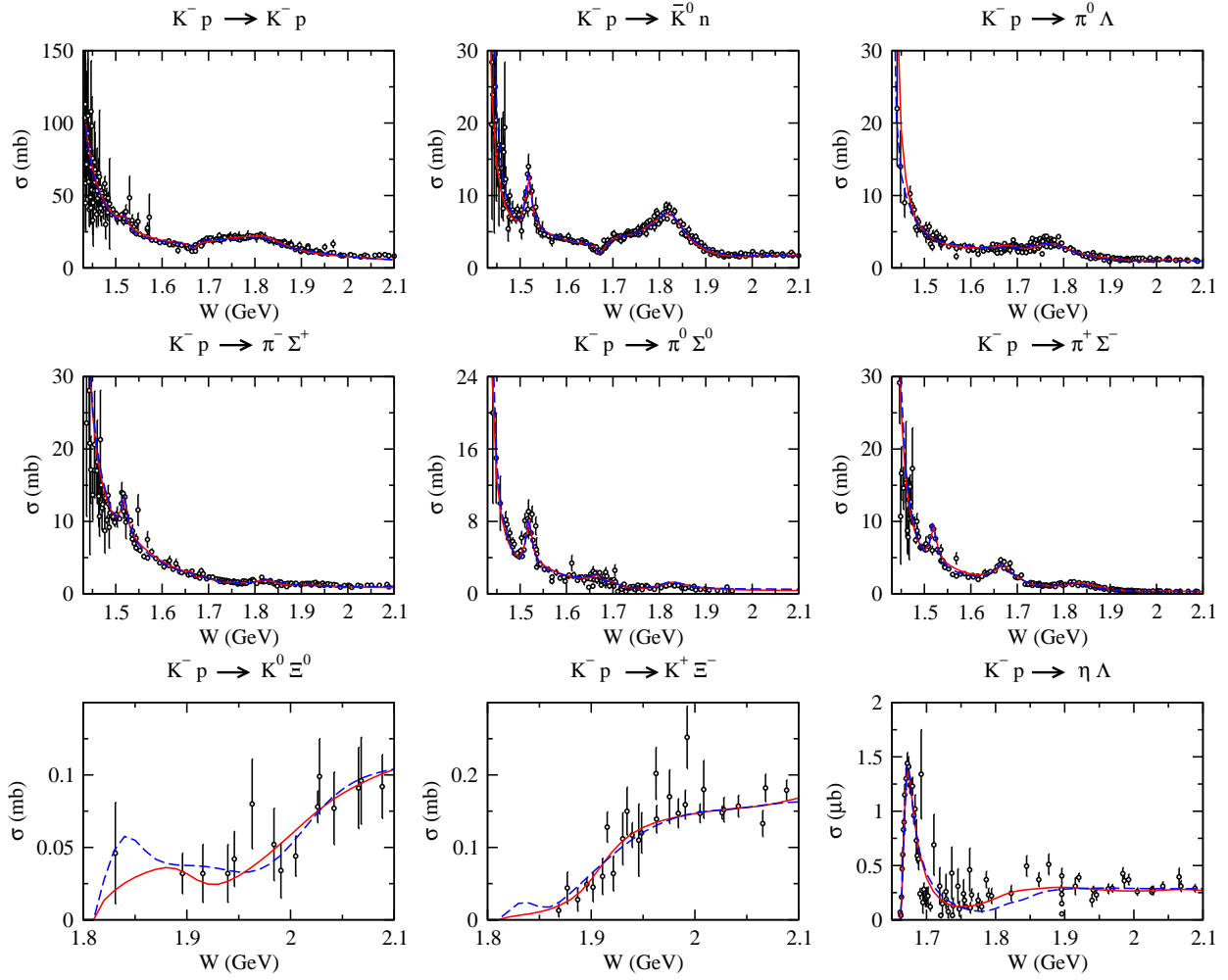}
\caption{(Color online)
Total cross sections of the $K^- p$ reactions up to $W= 2.1$ GeV.
The red solid curves (blues dashed curves) are the fitted results of Model A (Model B).
The same applies to Figs.~\ref{fig:kmp-dc-1}-\ref{fig:kpxm-dc} below.
}
\label{fig:tcs}
\end{figure}

According to the discussions in the last paragraph and the fact that
the data included in the fits are far from complete,
the $\chi^2$ values may not give accurate assessments of the constructed models.
It is therefore necessary to show that our fits are indeed very good.
We begin by showing in Fig.~\ref{fig:tcs} that we are able to give very good fits to the
total cross section data of the considered 
$K^- p \to K^- p, \bar{K}^0 n, \pi^0 \Lambda$ (upper row), 
$K^- p \to \pi^- \Sigma^+, \pi^0\Sigma^0,\pi^+\Sigma^-$ (middle row), and
$K^- p \to K^0 \Xi^0, K^+\Xi^-, \eta \Lambda$ (bottom row) reactions.
Here, it is noted that we have only included the recent BNL data~\cite{el-data} 
for $K^- p \to \eta \Lambda$ near the threshold.
The differences between Models A and B are significant only in
$K^- p \to K^0 \Xi^0, K^+\Xi^-$ near the threshold
and $K^- p \to \eta \Lambda$ at $1.78 \lesssim W\lesssim 1.85$ GeV, where the data are poor.
In Table~\ref{tab:data-chi2}, we see that the $\chi^2/{\rm data}$ values for
the $K^- p \to \eta \Lambda$ total cross section are very large: $\chi^2/{\rm data}\sim 8$.
This also comes from conflicting data. 
As can be seen in Fig.~\ref{fig:tcs}, at $W\sim 1.89$ GeV there are five data points
with small statistical errors in the $K^- p \to \eta \Lambda$ total cross section.
Theoretical curves cannot be within the errors of all of the five data points,
and this results in a large $\chi^2$ value.
In fact, if we eliminate these five data points, the resulting 
$\chi^2/$data value for the $K^- p \to \eta \Lambda$ total cross section
is reduced by more than 50\%.

In the next subsections, we will show in more detail the quality of our fits to the
differential cross sections ($d\sigma/d\Omega$), polarizations ($P$), and 
their product ($P \times d\sigma/d\Omega$) for each of the considered reactions.

\begin{table}
\caption{\label{tab:data-chi2}
Observables and number of the data considered in this coupled-channels analysis.
References for the data are listed in the fourth column.
Resulting ``$\chi^2$/data'' values for Model A (Model B) are listed in the fifth (sixth) column,
while ``$\chi^2$/d.o.f'' values are listed in bold face at the bottom-right of the table.
}
\begin{ruledtabular}
\begin{tabular}{lcrccc}
Reactions &Observables & No. of data & Data references &\multicolumn{2}{c}{$\chi^2$/data}\\
\cline{5-6}
&&&& Model A& Model B\\
\hline
$K^- p \to K^- p$ &$d\sigma/d\Omega$& 3962 &\cite{Dau68,And70,Arm70,Alb71,Con71,Abe75,Ada75,Gri75,Con76,Mas76}& 3.07& 2.98\\
& $P$ &510&\cite{Dau68,And70,Alb71}& 2.04& 2.08 \\
&$\sigma$&253&\cite{tcs-data}& 4.03& 4.02 \\ 
\\
$K^- p \to \bar K^0 n$ & $d\sigma/d\Omega$& 2950&\cite{Arm68,Arm70,Gri75,Jon75,Con76,Mas76,Als78,prakhov}&2.67 & 2.75 \\
&$\sigma$&260&\cite{tcs-data}&5.49 & 4.75 \\ 
\\
$K^- p \to \pi^- \Sigma^+$ &$d\sigma/d\Omega$&1792&\cite{Arm68,Arm70,Ber70,Jon75,Con76,Gri75}& 3.37& 3.49\\ 
&$P$&418&\cite{Arm68,Jon75,Con76}& 1.30& 1.28\\ 
&$P\times d\sigma/d\Omega$&177& \cite{Arm70}& 1.33 & 2.33 \\
&$\sigma$&173&\cite{tcs-data}& 3.27& 3.42\\ 
\\
$K^- p \to \pi^0 \Sigma^0$ &$d\sigma/d\Omega$&580&\cite{Arm70,Bax73,Lon75,manweiler,prakhov}& 3.68& 3.50\\ 
&$P$&196&\cite{manweiler,prakhov}& 6.39& 5.80\\ 
&$P\times d\sigma/d\Omega$&189&\cite{Arm70}& 1.24  & 1.24 \\ 
&$\sigma$&125&\cite{tcs-data}& 5.66& 6.40 \\ 
\\
$K^- p \to \pi^+ \Sigma^-$ &$d\sigma/d\Omega$&1786&\cite{Arm68,Arm70,Ber70,Gri75,Jon75,Con76}& 2.56 & 2.18\\ 
&$\sigma$&181&\cite{tcs-data}& 3.08 & 2.44 \\ 
\\
$K^- p \to \pi^0 \Lambda$ &$d\sigma/d\Omega$&2178&\cite{Arm68,Arm70,Ber70-2,Bax73,Gri75,Jon75,Lon75,Con76,prakhov}& 2.59& 3.71\\ 
&$P$&693&\cite{Arm68,Ber70-2,Jon75,Con76,prakhov}& 1.41 & 1.73 \\ 
&$P\times d\sigma/d\Omega$&176&\cite{Arm70}&  1.46  & 1.52   \\ 
&$\sigma$&207&\cite{tcs-data}& 3.99 & 4.20 \\ 
\\
$K^- p \to \eta \Lambda$ & $d\sigma/d\Omega$& 160&\cite{Arm70,el-data}& 2.69 & 2.03 \\ 
&$P$& 18&\cite{el-data}& 0.94& 3.83 \\ 
&$\sigma$&78&\cite{tcs-data,el-data}& 7.62& 8.53\\ 
\\
$K^- p \to K^0 \Xi^0$ &$d\sigma/d\Omega$&33&\cite{kx-data}& 1.24 & 1.61\\
&$\sigma$&15&\cite{tcs-data}& 0.46 & 0.35\\
\\
$K^- p \to K^+ \Xi^-$ &$d\sigma/d\Omega$&92&\cite{kx-data}& 2.05 & 1.74 \\
&$\sigma$&27&\cite{tcs-data}& 0.97 & 1.34 \\ 
\\
Total& &17229&& 2.87 &  2.98 \\
     & &     && \textbf{2.91} &  \textbf{3.03} 
\end{tabular}
\end{ruledtabular}
\end{table}

\subsection{$K^- p \to \bar KN$}

\begin{figure}
\includegraphics[clip,width=\textwidth]{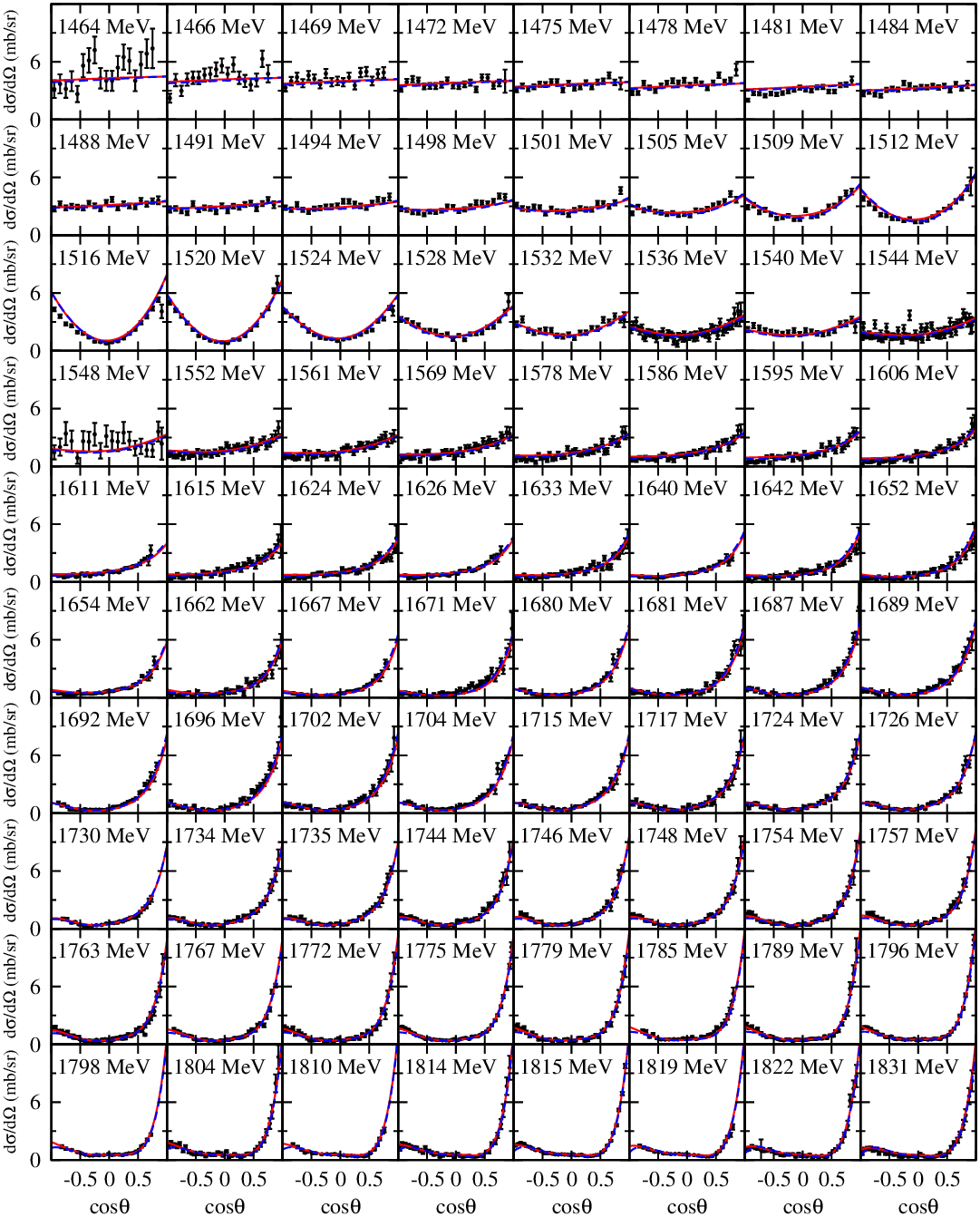}
\caption{(Color online)
$d\sigma/d\Omega$ of $K^-p \to K^-p$.
}
\label{fig:kmp-dc-1}
\end{figure}

\begin{figure}
\includegraphics[clip,width=\textwidth]{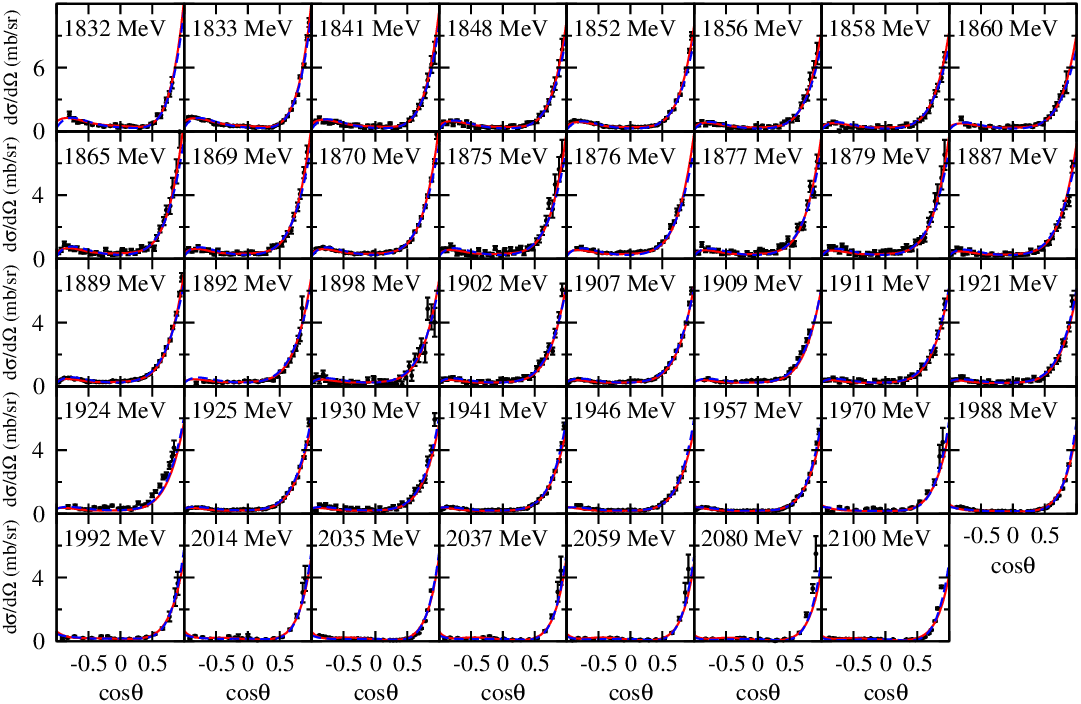}
\caption{(Color online)
$d\sigma/d\Omega$ of $K^-p \to K^-p$ (continued).
}
\label{fig:kmp-dc-2}
\end{figure}

\begin{figure}
\includegraphics[clip,width=\textwidth]{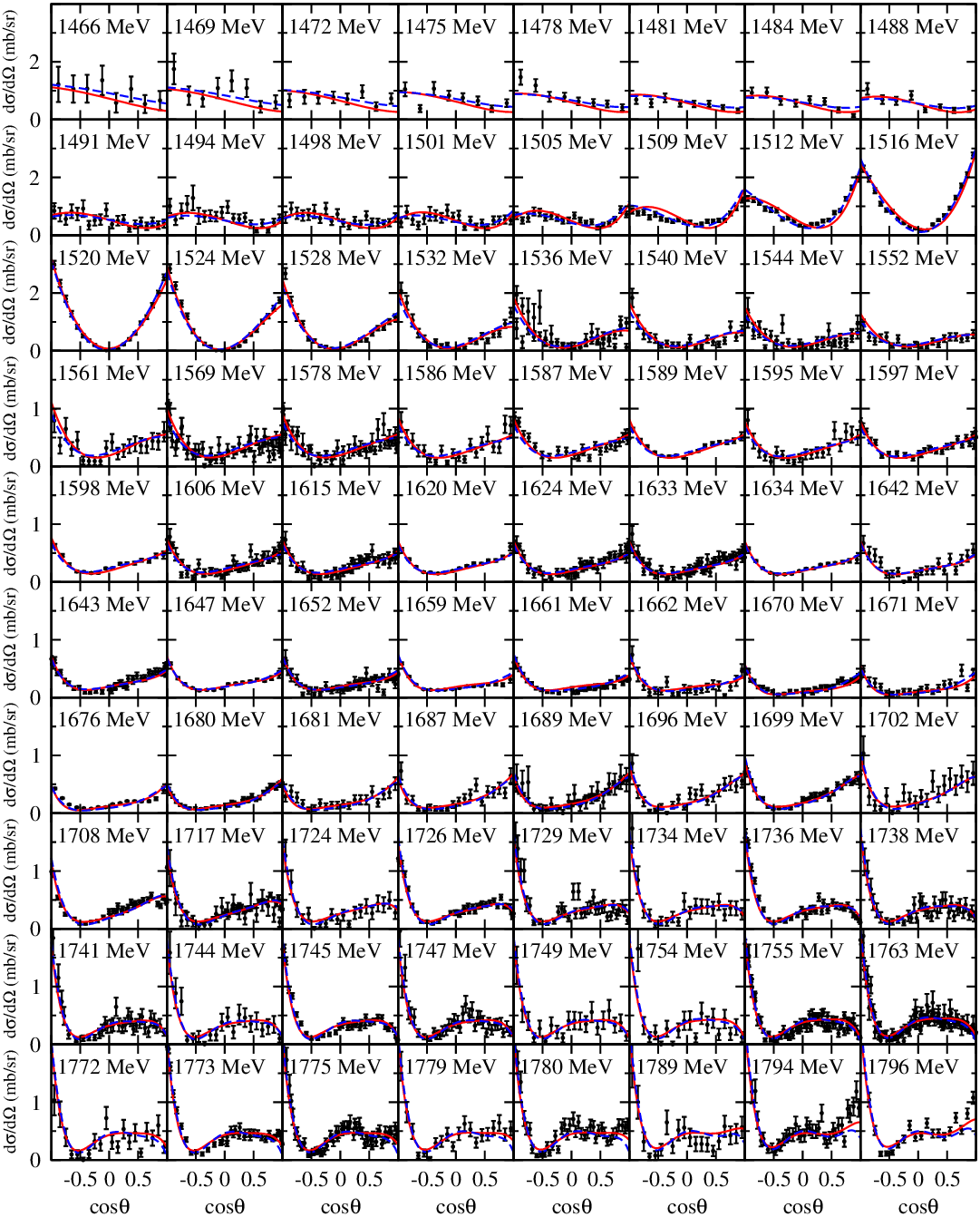}
\caption{(Color online)
$d\sigma/d\Omega$ of $K^-p \to \bar K^0 n$.
}
\label{fig:k0n-dc-1}
\end{figure}

\begin{figure}
\includegraphics[clip,width=\textwidth]{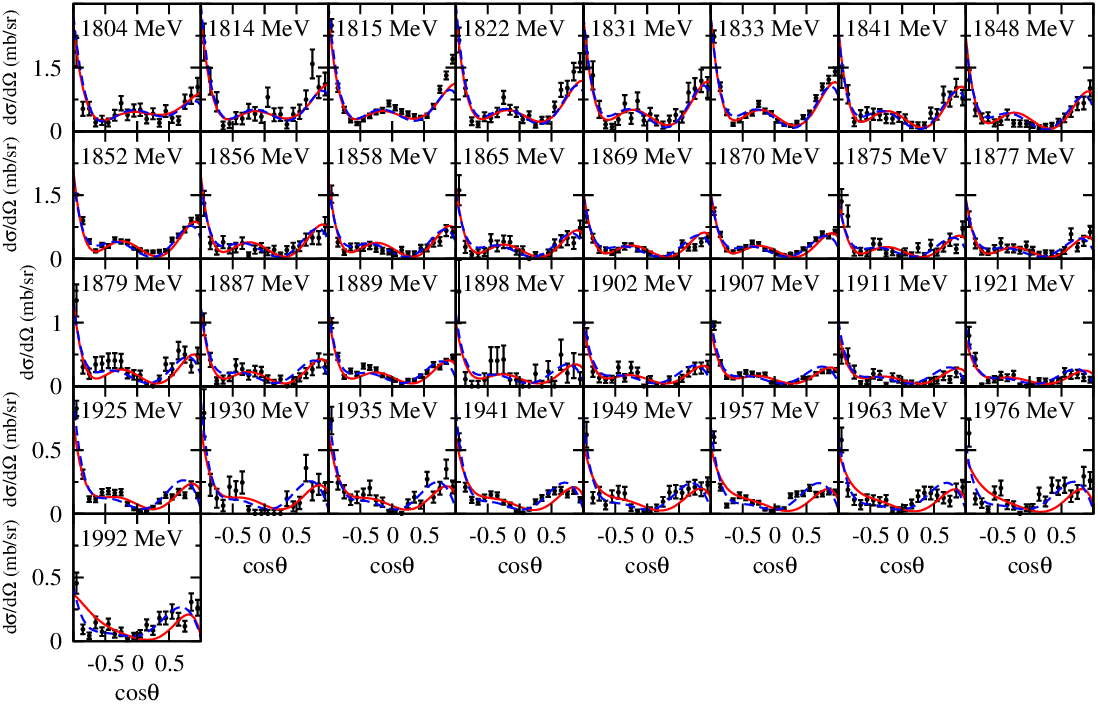}
\caption{(Color online)
$d\sigma/d\Omega$ of $K^-p \to \bar K^0 n$ (continued).
}
\label{fig:k0n-dc-2}
\end{figure}

Our fits to the data of $d\sigma/d\Omega$ are shown in
Figs.~\ref{fig:kmp-dc-1} and~\ref{fig:kmp-dc-2} for the elastic $K^- p \to K^- p$, and
in Figs.~\ref{fig:k0n-dc-1} and~\ref{fig:k0n-dc-2} for the charge-exchange $K^- p \to  \bar{K}^0 n$. 
We see that the data for the elastic $K^- p \to K^- p$ scattering are rather extensive and accurate.
The data  for $K^- p \to  \bar{K}^0 n$ are a little less accurate,
but are sufficient for playing an important role in the coupled-channels fits.
In Figs.~\ref{fig:kmp-dc-1}-\ref{fig:k0n-dc-2}, we see that both Models A and B can fit the data equally well.
We note that the data at low $W= 1464$-$1469$ MeV have rather large errors
and no data exist at $W$ closer to the $\bar K N$ threshold. 
This must be further improved for extracting accurately the physics relevant 
to the $\bar K N$ threshold region such as $\Lambda(1405)1/2^-$ and the $\bar K N$ scattering length.

\begin{figure}
\includegraphics[clip,width=\textwidth]{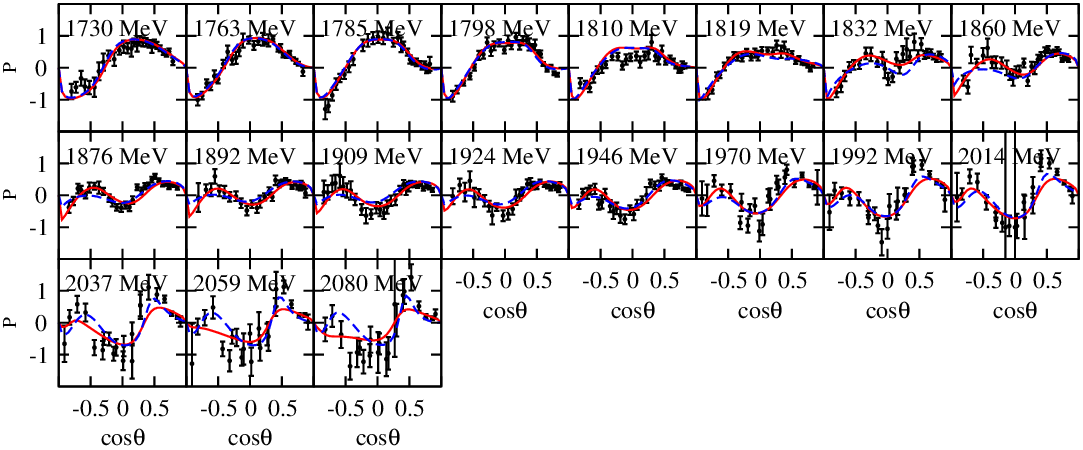}
\caption{(Color online)
$P$ of $K^-p \to K^-p$.
}
\label{fig:kmp-p}
\end{figure}

The data for the polarization $P$ are very limited for $K^- p \to K^- p$.
In fact, we could not find any data at $W \lesssim 1.7$ GeV.
Both models can describe these data well, as shown in Fig.~\ref{fig:kmp-p}.
There is no polarization data for $K^- p \to \bar{K}^0 n$.

\subsection{$K^- p \to \pi\Sigma$}
\label{sec:pis}

\begin{figure}
\includegraphics[clip,width=\textwidth]{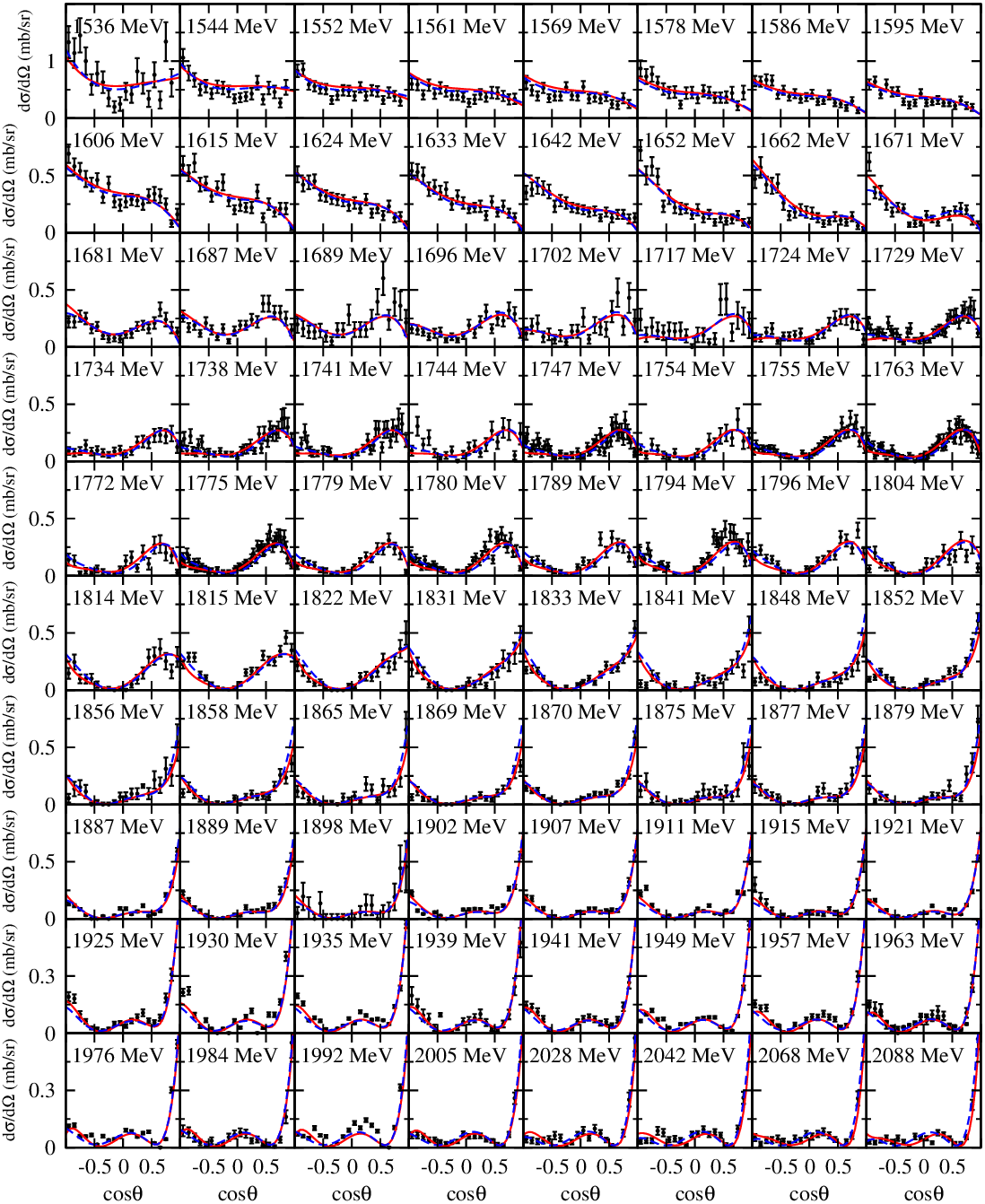}
\caption{(Color online)
$d\sigma/d\Omega$ of $K^-p \to \pi^- \Sigma^+$.
}
\label{fig:pimsp-dc}
\end{figure}

\begin{figure}
\includegraphics[clip,width=\textwidth]{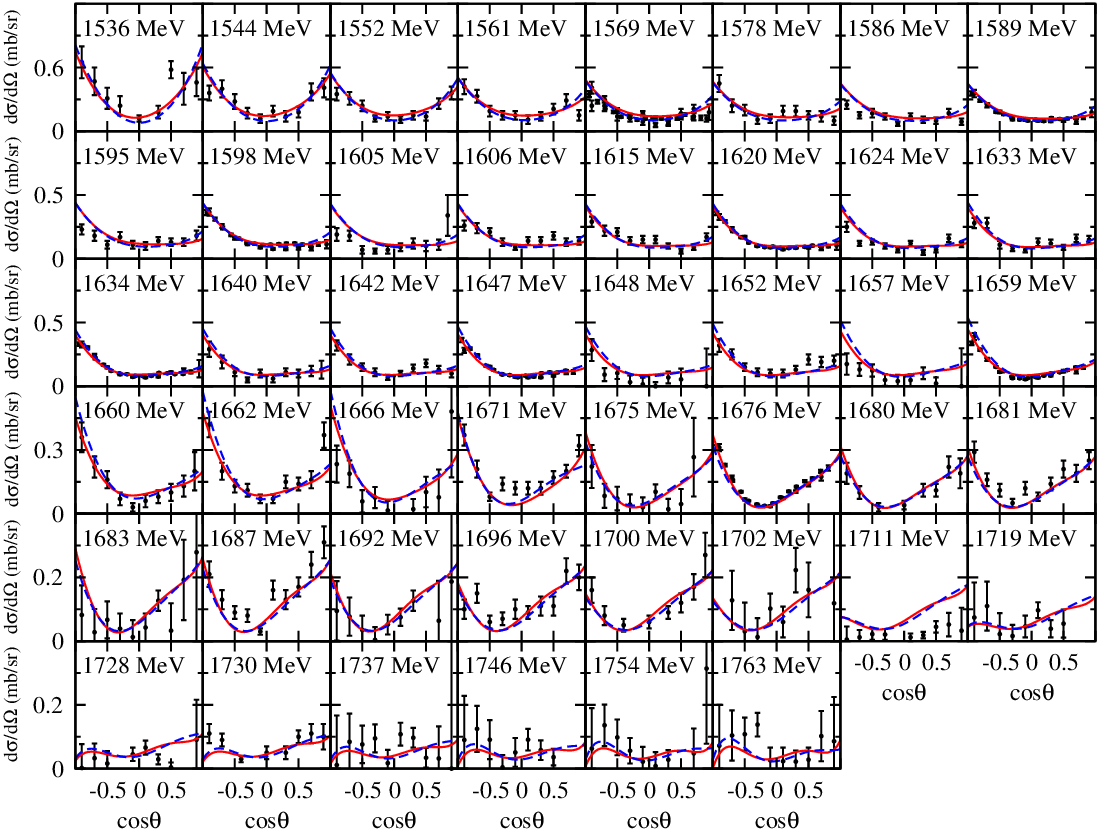}
\caption{(Color online)
$d\sigma/d\Omega$ of $K^-p \to \pi^0 \Sigma^0$.
}
\label{fig:pi0s0-dc}
\end{figure}

\begin{figure}
\includegraphics[clip,width=\textwidth]{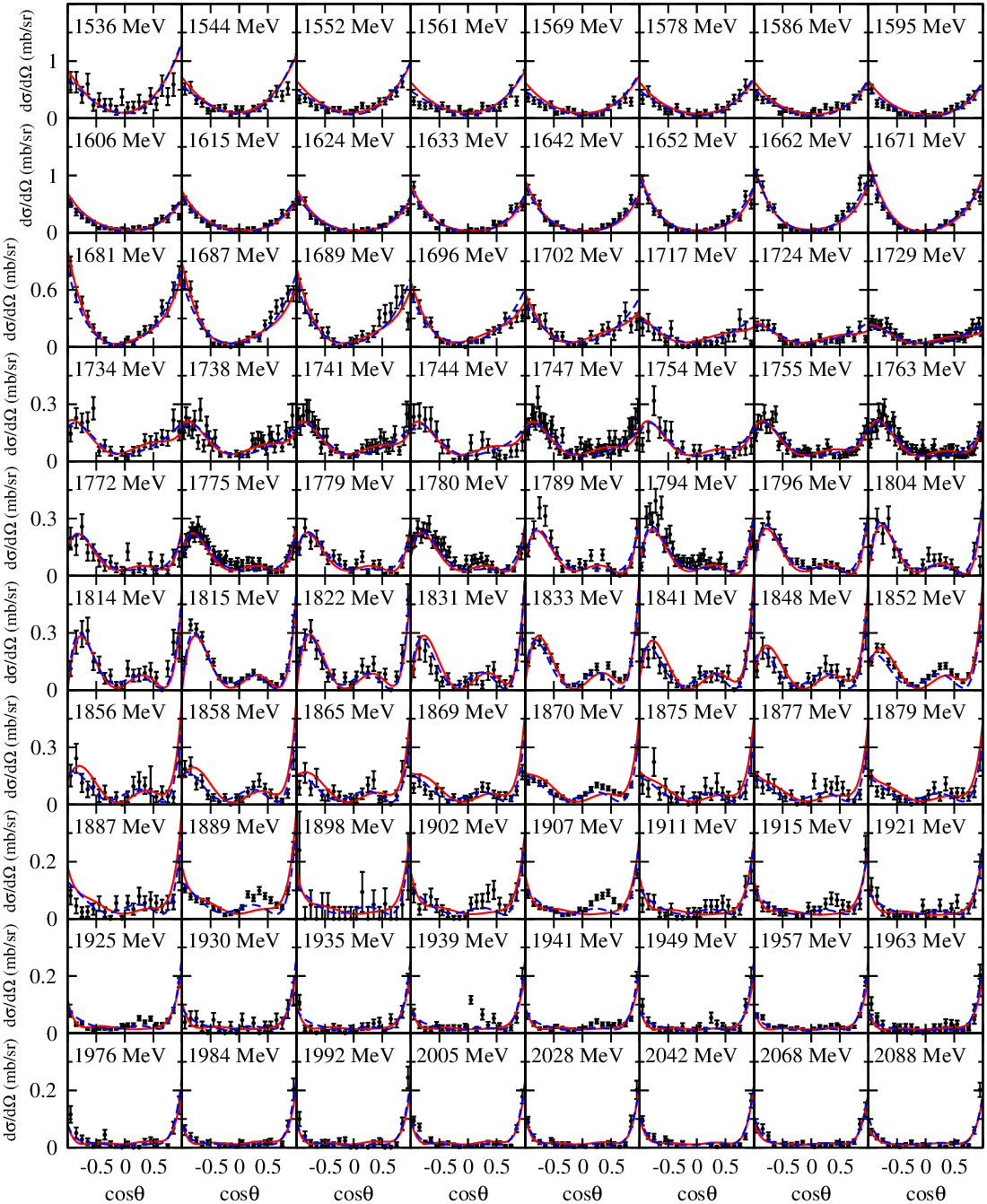}
\caption{(Color online)
$d\sigma/d\Omega$ of $K^-p \to \pi^+ \Sigma^-$.
}
\label{fig:pipsm-dc}
\end{figure}

Our fits to the differential cross sections are shown in 
Fig.~\ref{fig:pimsp-dc} for $K^-p \to \pi^-\Sigma^+$,
Fig.~\ref{fig:pi0s0-dc} for $K^-p \to \pi^0\Sigma^0$, and
Fig.~\ref{fig:pipsm-dc} for $K^-p \to \pi^+\Sigma^-$.
Both models can give good fits to the data.
At present, no data exist for these differential cross sections below $W = 1536$ MeV.
Furthermore, the data for $K^-p \to \pi^0\Sigma^0$ are available only up to $W = 1763$ MeV
with relatively large statistical errors.

\begin{figure}
\includegraphics[clip,width=\textwidth]{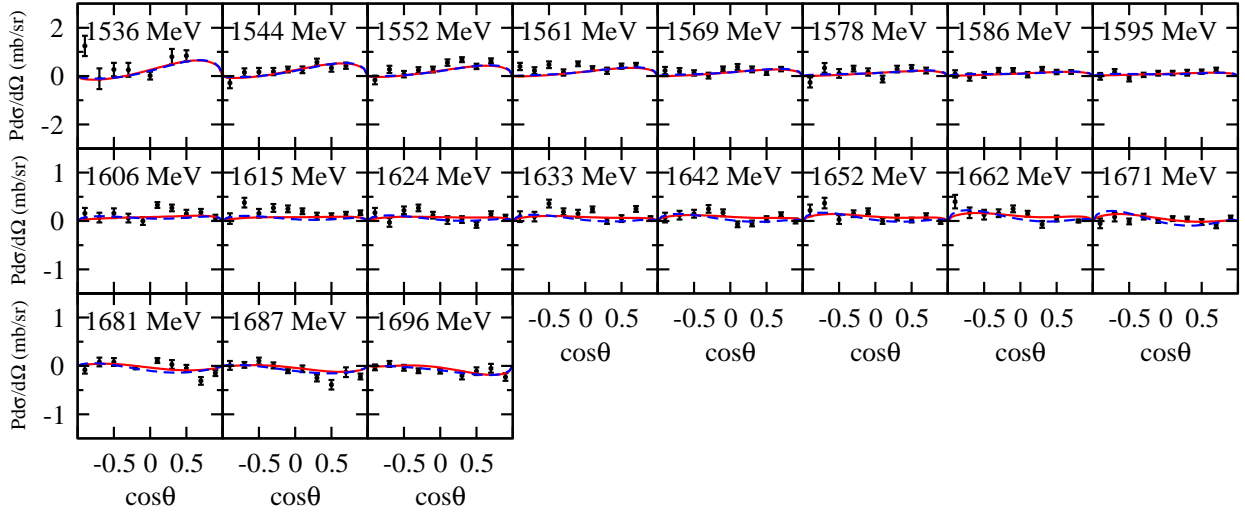}
\caption{(Color online)
$P\times d\sigma/d\Omega$ of $K^-p \to \pi^- \Sigma^+$.
}
\label{fig:pimsp-pd}
\end{figure}

\begin{figure}
\includegraphics[clip,width=\textwidth]{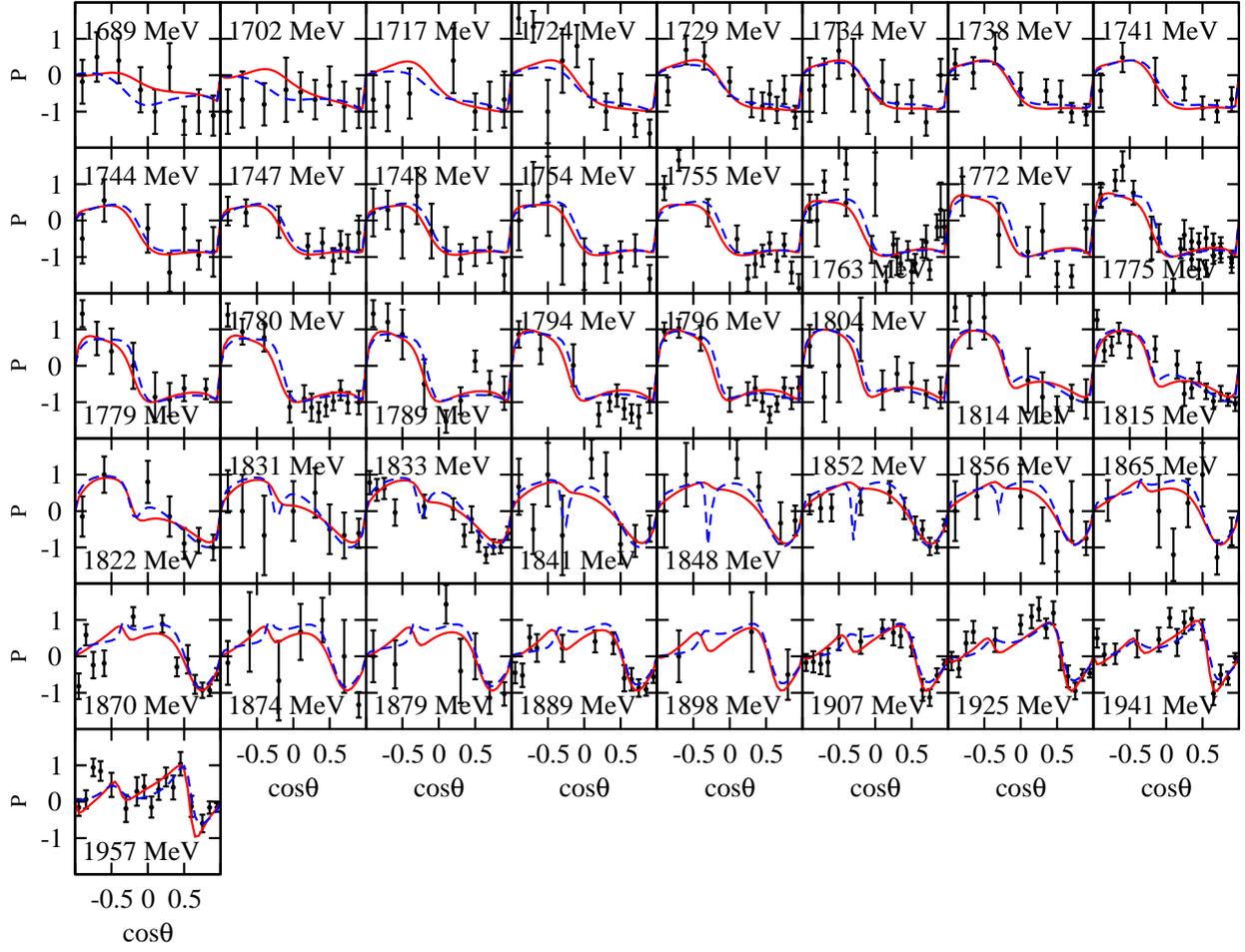}
\caption{(Color online)
$P$ of $K^-p \to \pi^- \Sigma^+$.
}
\label{fig:pimsp-p}
\end{figure}

The high precision data of $P\times d\sigma/d\Omega$ 
for $K^-p \to \pi^-\Sigma^+$ in the low energy region
can be fitted well (Fig.~\ref{fig:pimsp-pd}). 
The data of $P$ for $K^-p \to \pi^-\Sigma^+$ are available in the higher $W$ region, 
although they are very qualitative, as seen in Fig.~\ref{fig:pimsp-p}.
Both models follow well the general trend of the data of $P$,
but there is a significant difference between them at some energies:
sharp dips appear in Model B at $1831 \leq W \leq 1856$ MeV, but do not in Model A. 
This difference appears at the angles where no data exist, and thus
the precise measurements of this observable densely covering the angles 
are highly desirable for constraining the models.

\begin{figure}
\includegraphics[clip,width=\textwidth]{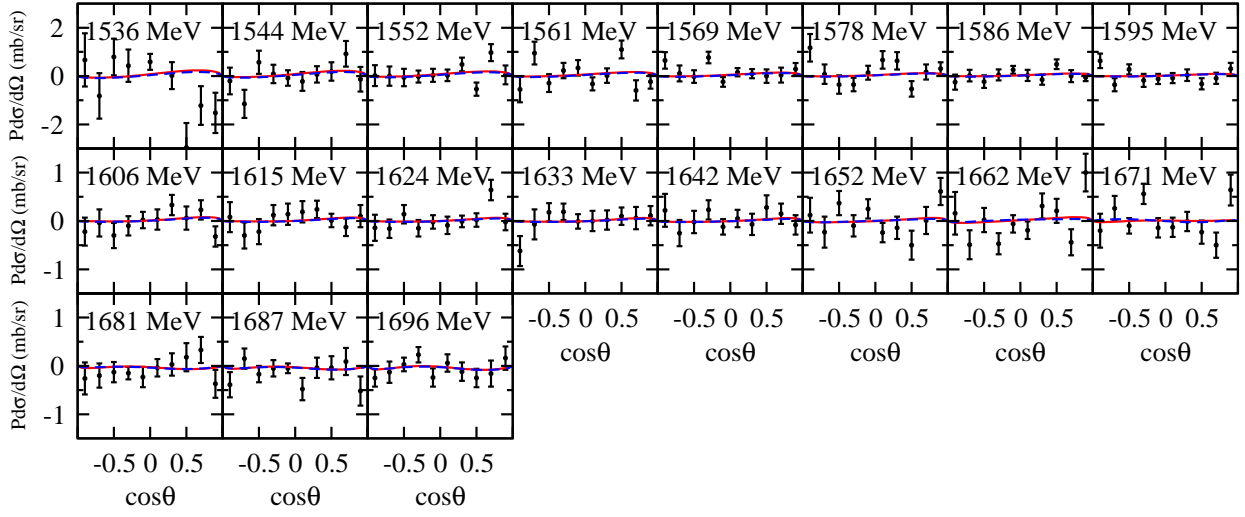}
\caption{(Color online)
$P\times d\sigma/d\Omega$ of $K^-p \to \pi^0 \Sigma^0$.
}
\label{fig:pi0s0-pd}
\end{figure}

\begin{figure}
\includegraphics[clip,width=\textwidth]{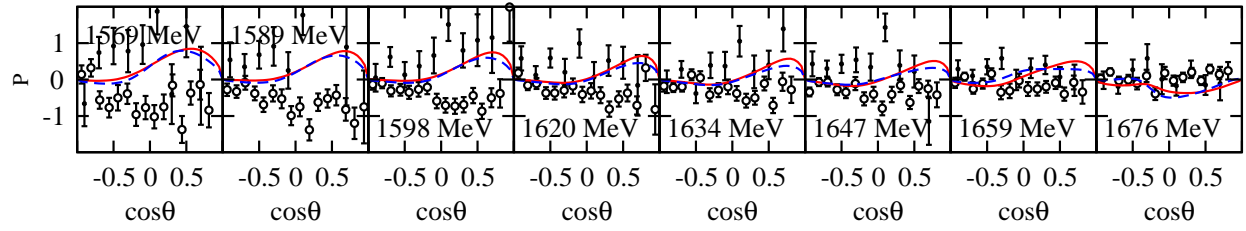}
\caption{(Color online)
$P$ of $K^-p \to \pi^0 \Sigma^0$.
Filled (open) circles are the data from Ref.~\cite{manweiler} (Ref.~\cite{prakhov}).
}
\label{fig:pi0s0-p}
\end{figure}

The polarization observables, $P\times d\sigma/d\Omega$ and $P$,
for $K^-p \to \pi^0\Sigma^0$ are presented in Figs.~\ref{fig:pi0s0-pd} and~\ref{fig:pi0s0-p}, respectively.
Although the data of $P$ are very limited, there are some 
from the recent Crystal Ball experiment~\cite{manweiler,prakhov}.
As seen in Fig.~\ref{fig:pi0s0-p}, the data from
Ref.~\cite{manweiler} and Ref.~\cite{prakhov} seem inconsistent at low energies. 
As explained in Ref.~\cite{prakhov}, this inconsistency could be from
differences in analysis methods taken by the two analysis groups, even
though they used the same data sample. 
Here we included both in our dataset, and fitted them along with other
data simultaneously. 
Our fits shown in Fig.~\ref{fig:pi0s0-p} are relatively closer to the data of 
Ref.~\cite{manweiler}, not necessarily supporting them. 
As already mentioned, this inconsistency of the data results in the large $\chi^2/$data value
of $P$ for $K^-p \to \pi^0\Sigma^0$ as shown in Table~\ref{tab:data-chi2}.

\subsection{$K^- p \to \pi^0\Lambda$}

\begin{figure}
\includegraphics[clip,width=\textwidth]{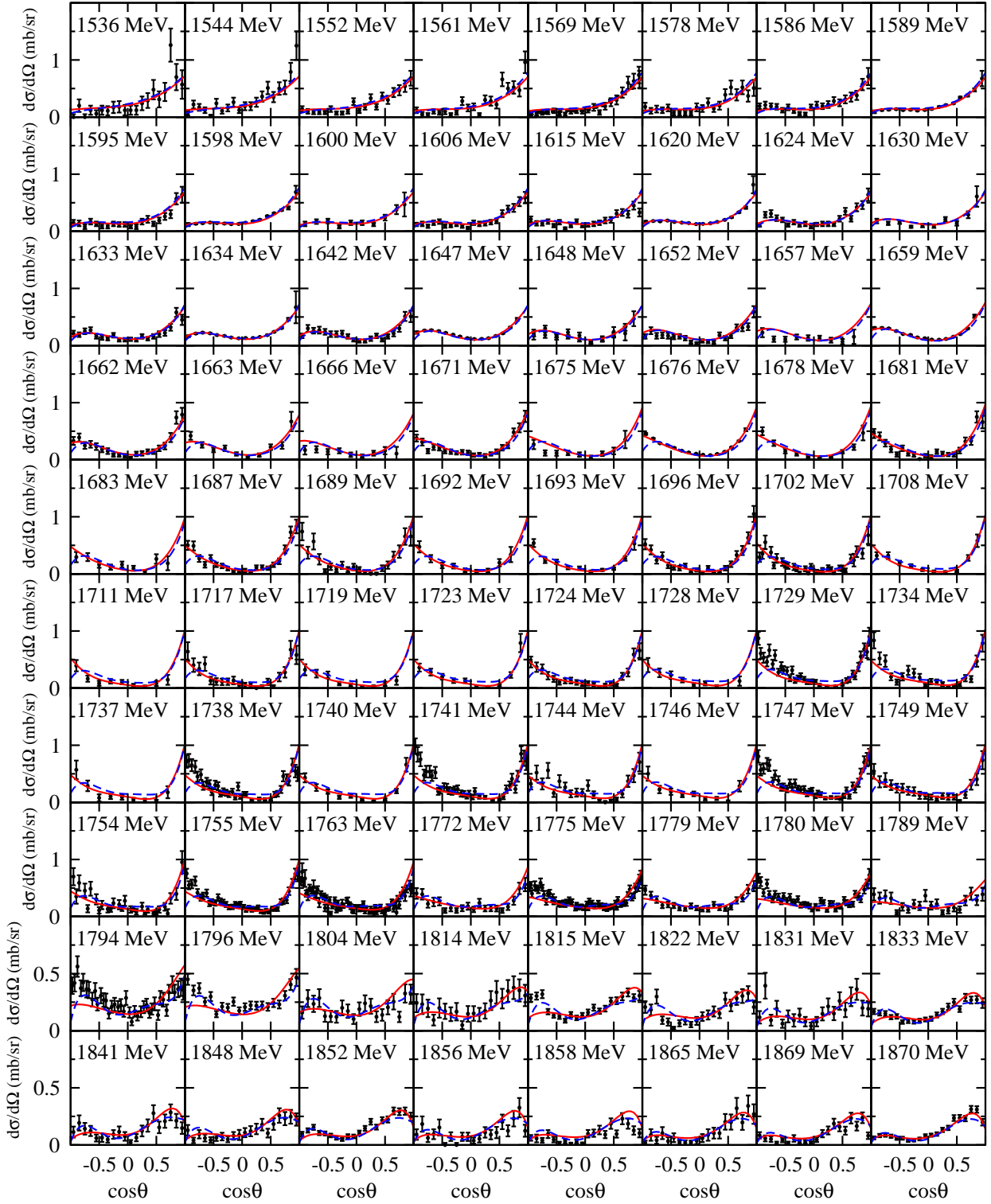}
\caption{(Color online)
$d\sigma/d\Omega$ of $K^-p \to \pi^0 \Lambda$.
}
\label{fig:pi0l0-dc-1}
\end{figure}

\begin{figure}
\includegraphics[clip,width=\textwidth]{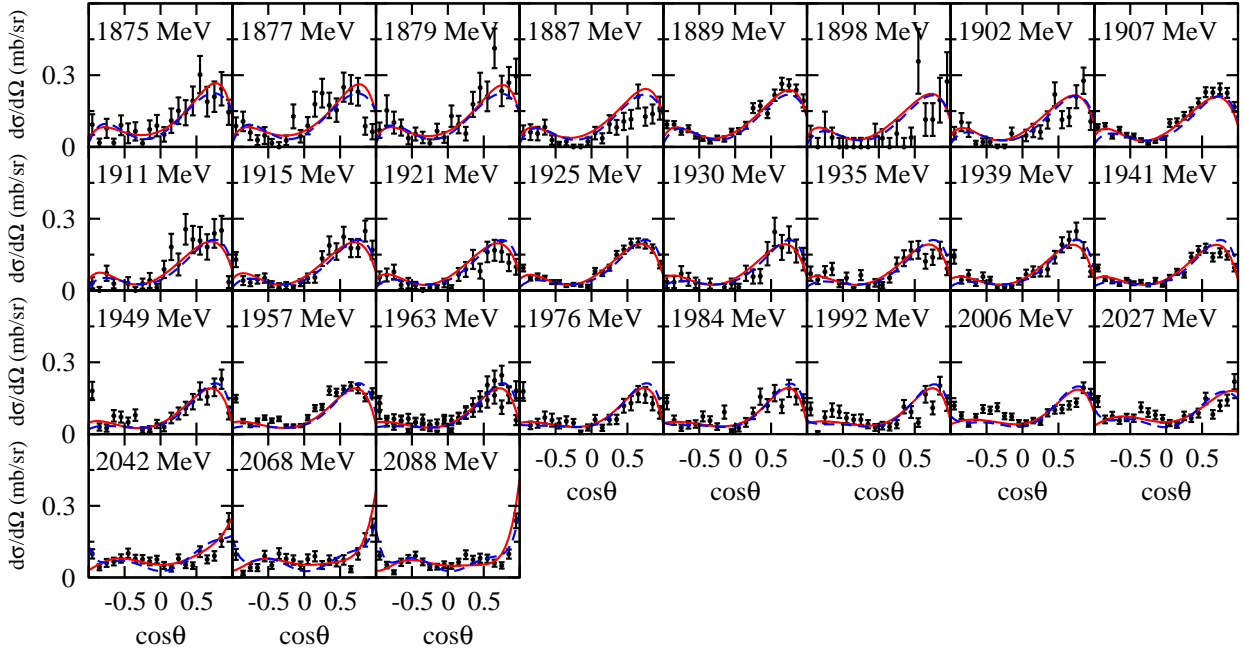}
\caption{(Color online)
$d\sigma/d\Omega$ of $K^-p \to \pi^0 \Lambda$ (continued).
}
\label{fig:pi0l0-dc-2}
\end{figure}

\begin{figure}
\includegraphics[clip,width=\textwidth]{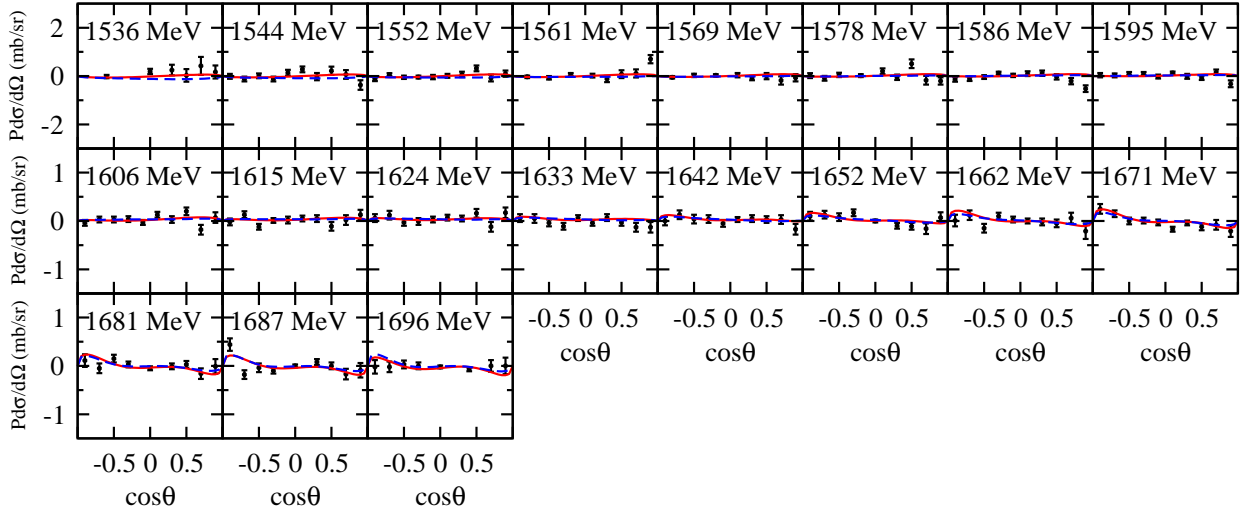}
\caption{(Color online)
$P\times d\sigma/d\Omega$ of $K^-p \to \pi^0 \Lambda$.
}
\label{fig:pi0l0-pd}
\end{figure}

\begin{figure}
\includegraphics[clip,width=\textwidth]{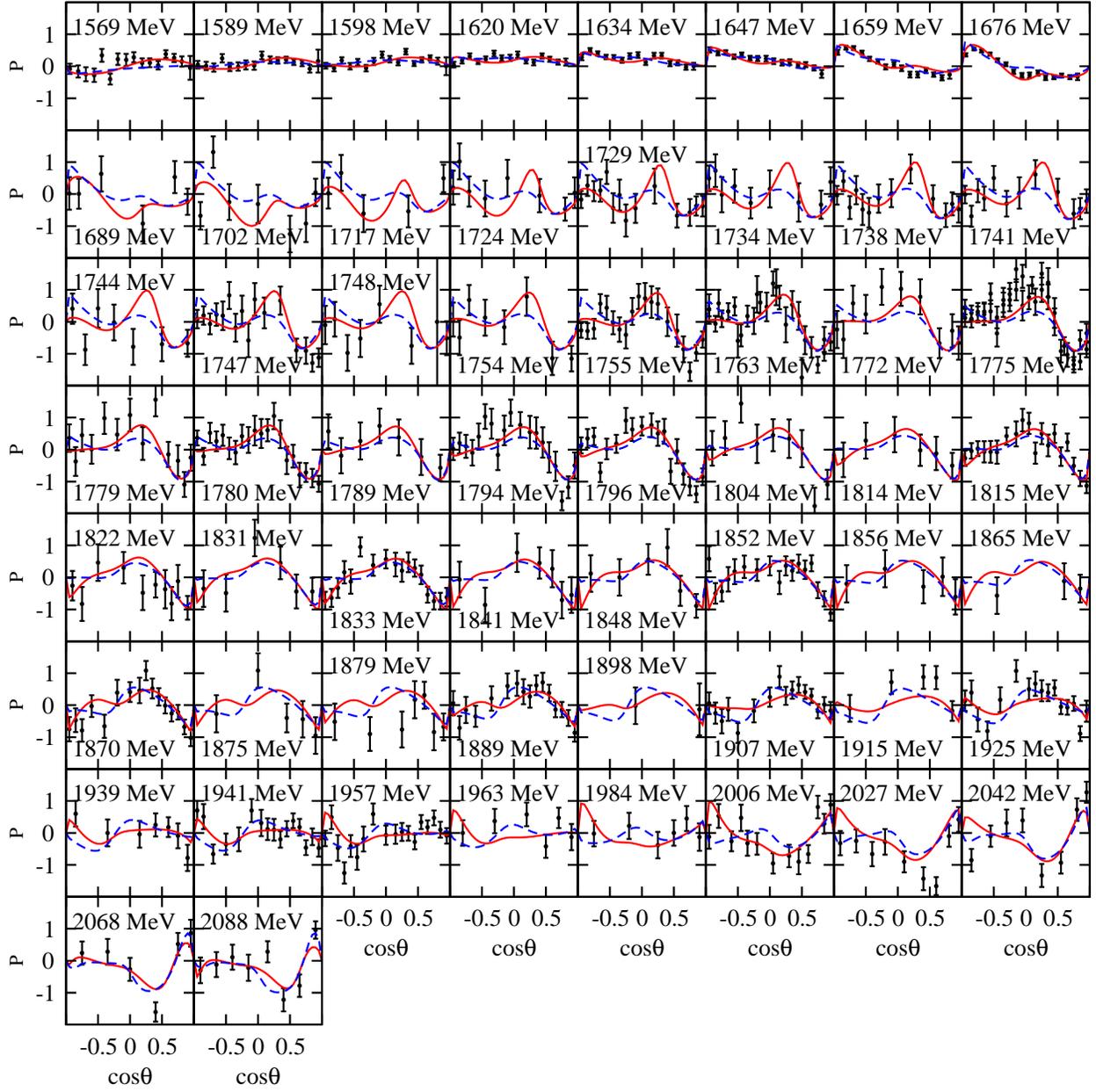}
\caption{(Color online)
$P$ of $K^-p \to \pi^0 \Lambda$.
}
\label{fig:pi0l0-p}
\end{figure}

Our fits to the differential cross section data of $K^- p \to \pi^0 \Lambda$ are shown in 
Figs.~\ref{fig:pi0l0-dc-1} and~\ref{fig:pi0l0-dc-2}. 
As is the case for $K^- p \to \pi \Sigma$, at present there are no data below $W = 1536$ MeV.
The fits from Models A and B are equally good. 
The high precision data of $P\times d\sigma/d\Omega$ 
at low $W$ can also be fitted well, as shown in Fig.~\ref{fig:pi0l0-pd}.
The data for $P$ at higher $W$ are very qualitative. 
It is seen in Fig.~\ref{fig:pi0l0-p} that our fits can reproduce the general trend of the data
and it is hard to judge our two models with the quality of the current data, 
even though there are visible differences in $P$ between them at most energies.
The more precise and extensive data of polarization observables of this reaction would be helpful
to establish the $\Sigma^*$ mass spectrum since only the $\Sigma^*$ resonances with $I=1$ 
can contribute to the $s$-channel processes.

\subsection{$K^- p \to \eta\Lambda$}
\label{sec:etl}
\begin{figure}
\includegraphics[clip,width=\textwidth]{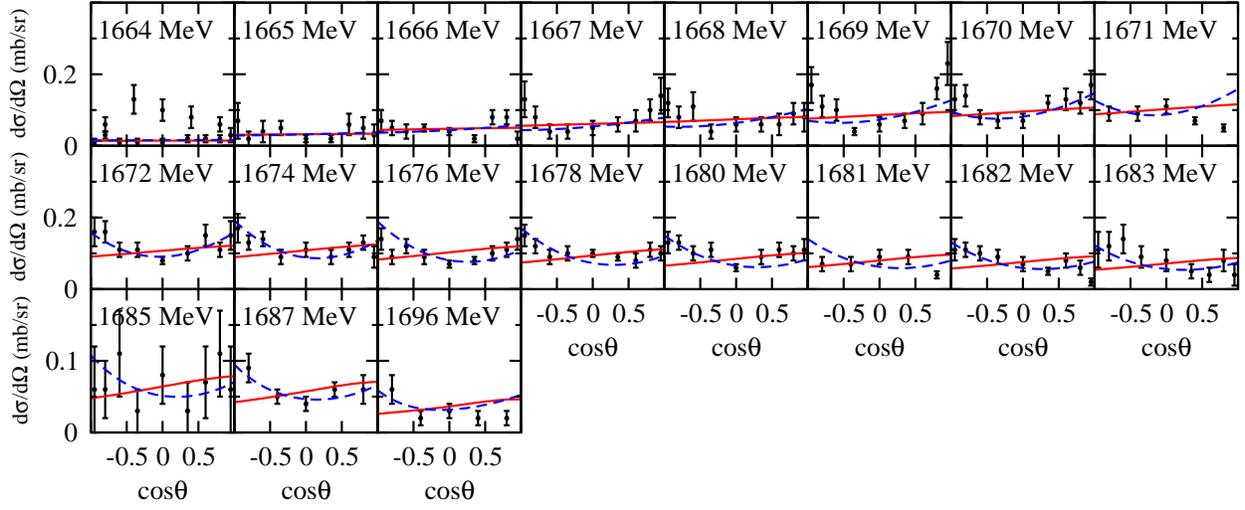}
\caption{(Color online)
$d\sigma/d\Omega$ of $K^-p \to \eta \Lambda$.
}
\label{fig:et0l0-dc}
\end{figure}

\begin{figure}
\includegraphics[clip,width=0.3\textwidth]{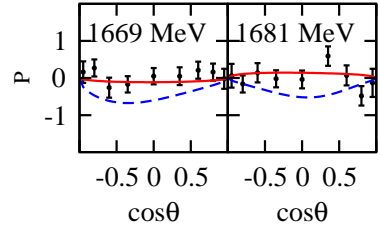}
\caption{(Color online)
$P$ of $K^-p \to \eta \Lambda$.
}
\label{fig:et0l0-p}
\end{figure}

Currently, the data of differential cross sections (Fig.~\ref{fig:et0l0-dc})
for $K^- p \to \eta\Lambda$ are limited in the near-threshold region.
We observe visible differences between Models A and B in
the differential cross sections.
Although the Model B shows a better $\chi^2/$data value for this observable,
it is not easy to judge the models within the current limited data.
There is an inconsistency in the data at $W = 1664$ MeV.
The lower set of the data comes from the recent BNL data~\cite{el-data}, 
while the higher set comes from an old bubble chamber experiment
at CERN~\cite{Arm70}.
Our model curves seem to follow the BNL data.
Also, we observe that the differential cross section data show a concave-up angular dependence at most $W$,
which suggests a possibility of the sizable contribution from higher partial waves even in the very threshold region.
In fact, Model B captures this angular dependence better than Model A
and this is due to the large contribution from $P_{03}$ wave near the threshold, 
as shown later in Sec.~\ref{sec:pwa}.
As for the polarization $P$ (Fig.~\ref{fig:et0l0-p}),
the data are available only at two $W$ points near the threshold.
Model B is found to be a bit off the data.


\subsection{$K^- p \to K\Xi$}

\begin{figure}
\includegraphics[clip,width=0.425\textwidth]{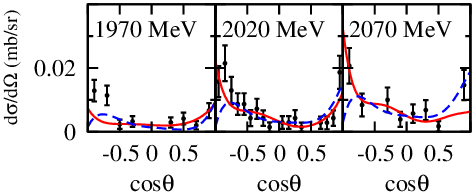}
\caption{(Color online)
$d\sigma/d\Omega$ of $K^-p \to K^0 \Xi^0$.
}
\label{fig:k0x0-dc}
\end{figure}

\begin{figure}
\includegraphics[clip,width=0.425\textwidth]{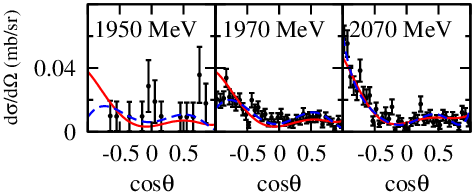}
\caption{(Color online)
$d\sigma/d\Omega$ of $K^-p \to K^+ \Xi^-$.
}
\label{fig:kpxm-dc}
\end{figure}

The results of the differential cross sections 
for the $K^- p \to K^0 \Xi^0$ and $K^- p \to K^+ \Xi^-$ reactions
are presented in Figs.~\ref{fig:k0x0-dc} and~\ref{fig:kpxm-dc}, respectively.
In both reactions, the data are only available at three $W$ points
in the considered energy region up to $W=2.1$ GeV, and
our models reproduce the data reasonably well.
However, definitely much more data are required to constrain the models,
particularly the model parameters associated with the $K\Xi$ channel.

\section{Discussions}
\label{sec:discussion}

\subsection{Comparison of partial-wave amplitudes}
\label{sec:pwa}

With the good fits to the available data of $K^-p$ reactions, 
as shown in the previous section, 
the partial-wave amplitudes from the models (Models A and B)
can be used to extract the $S=-1$ hyperon resonance parameters. 
The partial-wave amplitudes
are also essential in theoretical calculations of the production of hypernuclei from kaon-induced nuclear reactions
within the well-studied multiple scattering theory. 
It is therefore interesting to compare our resulting partial-wave amplitudes with those determined 
in the recent single-energy partial-wave analysis performed by the KSU group~\cite{zhang2013}.
There exist several previous partial-wave analyses~\cite{pwa-1,pwa-2,pwa-3,pwa-4,pwa-5,pwa-6,pwa-7} 
of $\bar KN$ reactions. 
However, these earlier works only account for limited data and are based
on simple Breit-Wigner parametrizations that do not account for 
the complex coupled-channels effects as done in the KSU analysis and in this work. 
We thus will not include those earlier partial-wave analyses in the discussions.
In our notation, the partial-wave amplitudes $F_{M'B',MB}(W)$ are given by
\begin{equation}
F_{M'B',MB}(W) = -[\rho_{M'B'}(k'_{\text{on}};W)\rho_{MB}(k_{\text{on}};W)]^{1/2} 
T_{M'B',MB}(k'_{\text{on}},k_{\text{on}};W),
\label{eq:pwa}
\end{equation}
where $\rho_{MB}(k,W) = \pi k E_M(k)E_B(k)/W$, and 
$k_{\text{on}}$ [$k'_{\text{on}}$] is the on-shell momentum defined by
$W = E_M(k_{\text{on}}) + E_B(k_{\text{on}})$ [$W = E_{M'}(k'_{\text{on}}) + E_{B'}(k'_{\text{on}})$].

\begin{figure}
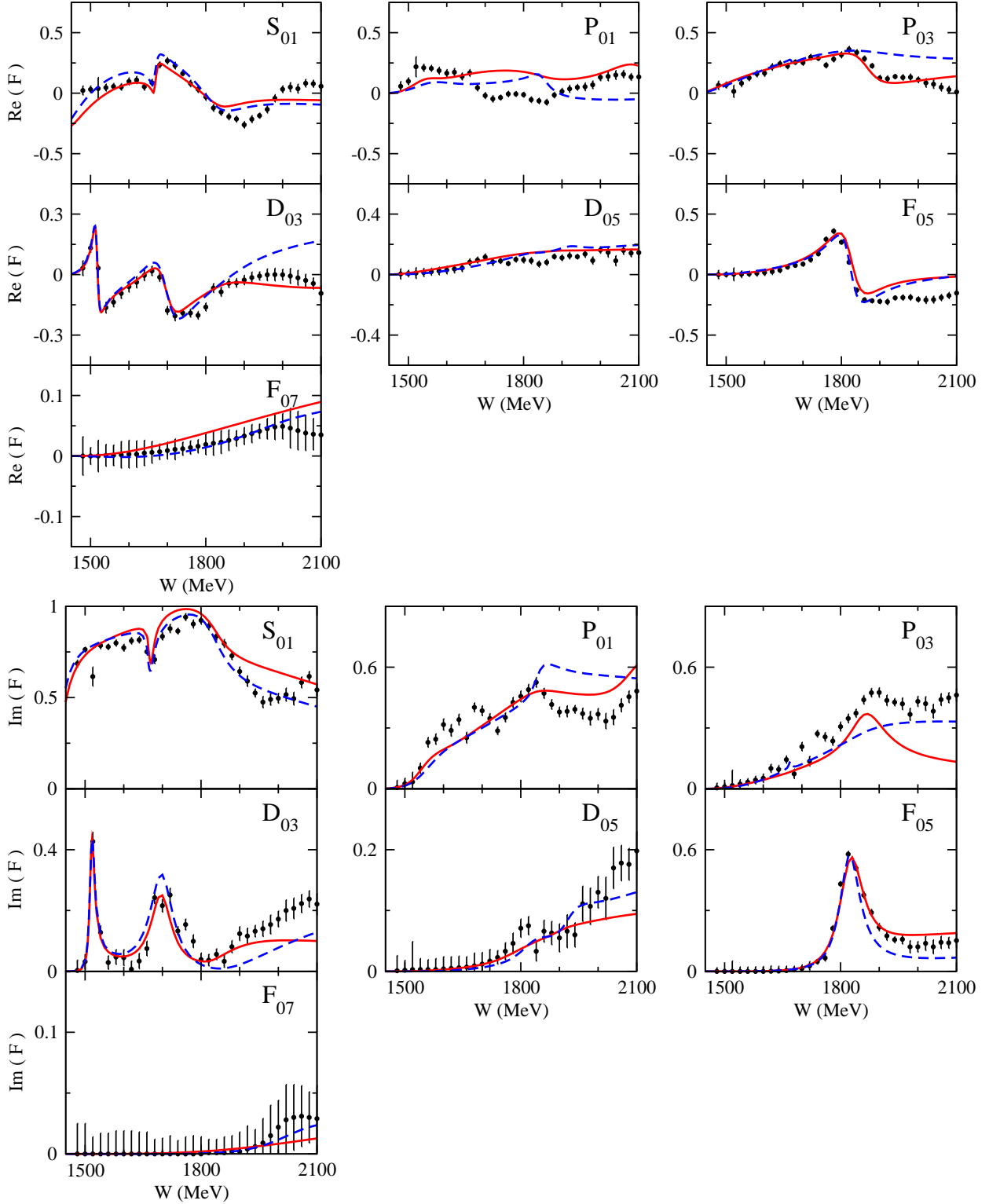

\includegraphics[clip,width=\textwidth]{kbn-re0}
\includegraphics[clip,width=\textwidth]{kbn-im0}
\caption{(Color online)
Determined partial-wave amplitudes of $\bar K N \to \bar K N$ with isospin $I=0$.
Upper (lower) panels are for real (imaginary) parts of the amplitudes.
Results of Model A (Model B) are shown in red solid (blue dashed) curves.
Our results are compared with the single-energy solution (filled circles) 
given in Ref.~\cite{zhang2013}.
}
\label{fig:kbn-amp-0}
\end{figure}

\begin{figure}
\includegraphics[clip,width=\textwidth]{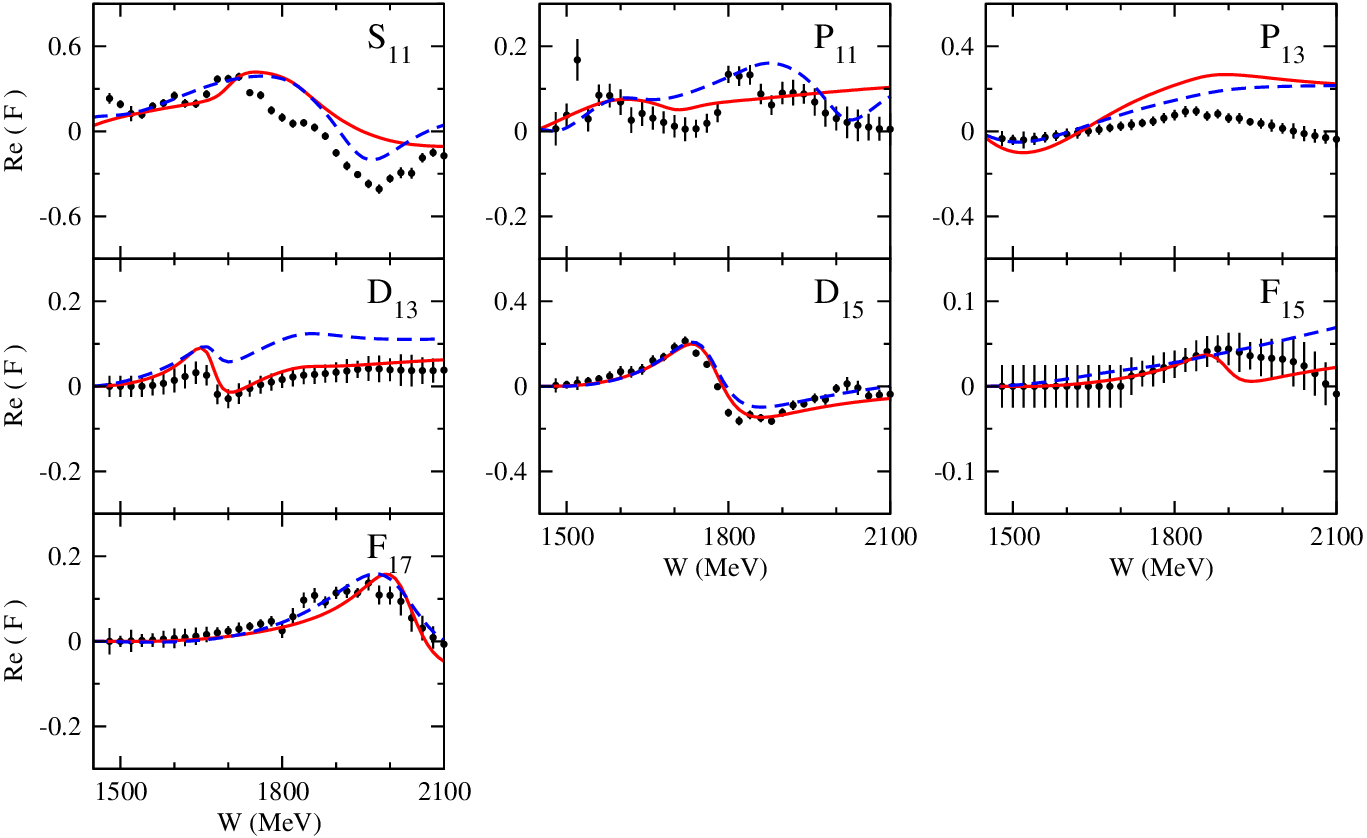}
\includegraphics[clip,width=\textwidth]{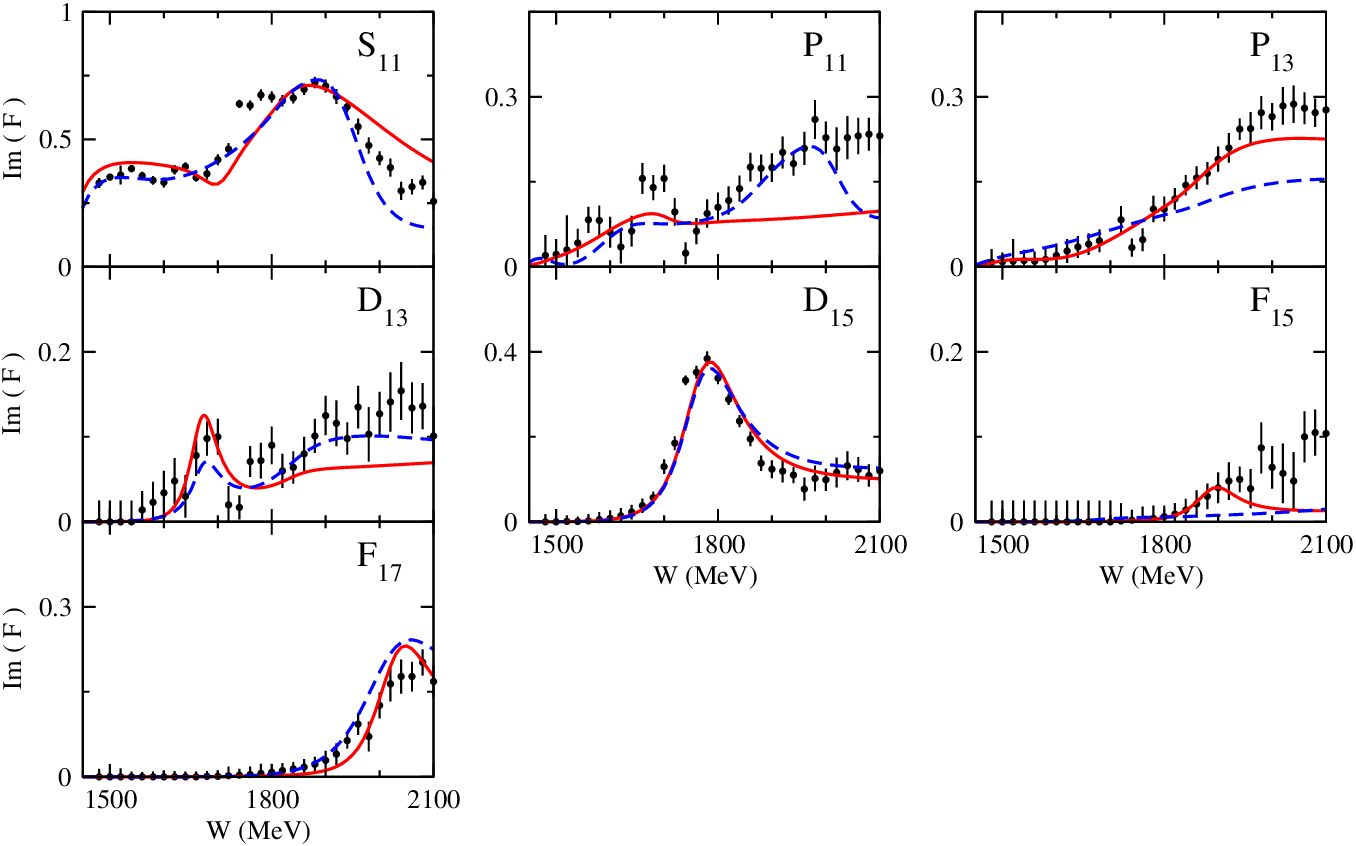}
\caption{(Color online)
Determined partial-wave amplitudes of $\bar K N \to \bar K N$ with isospin $I=1$.
See the caption of Fig.~\ref{fig:kbn-amp-0} for the description of the figure.
}
\label{fig:kbn-amp-1}
\end{figure}

\begin{figure}
\includegraphics[clip,width=\textwidth]{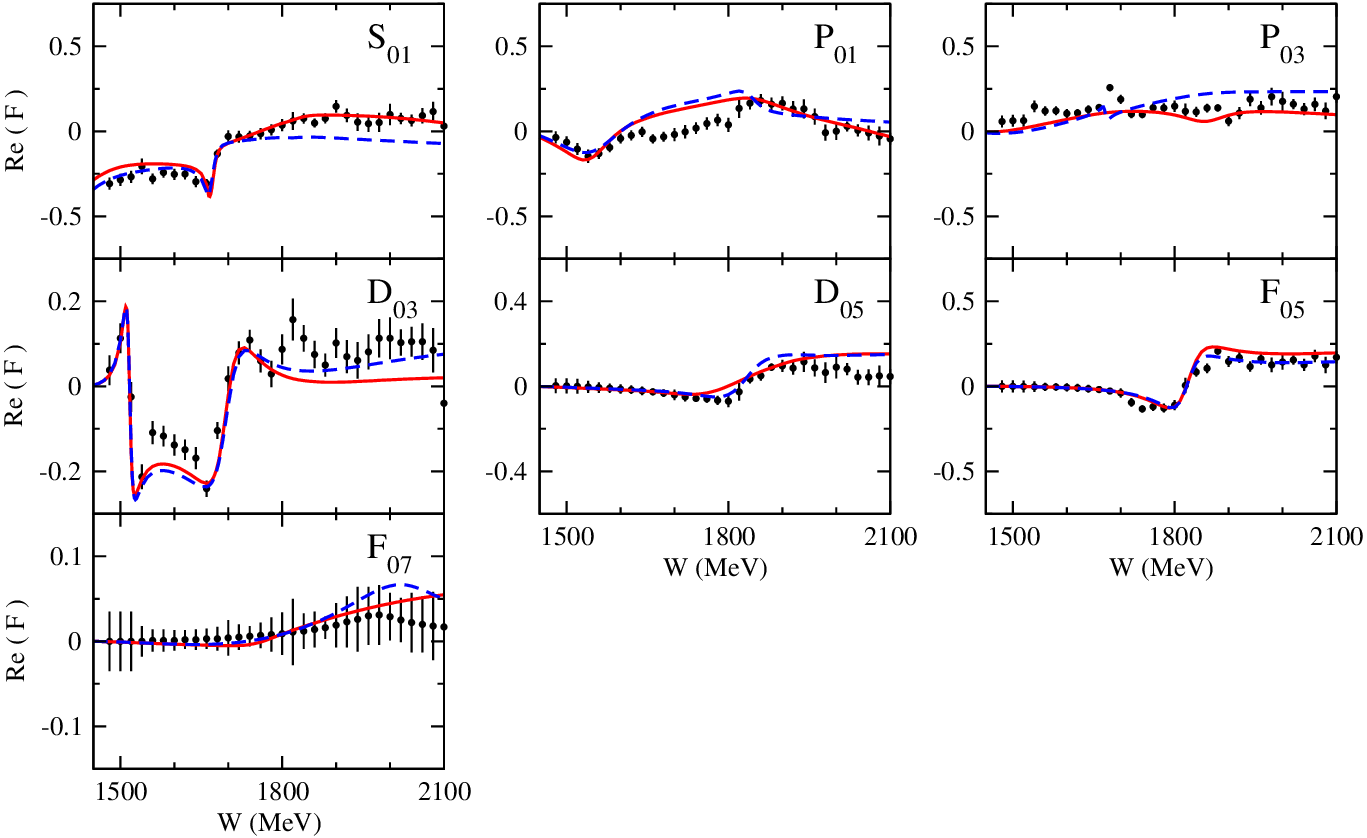}
\includegraphics[clip,width=\textwidth]{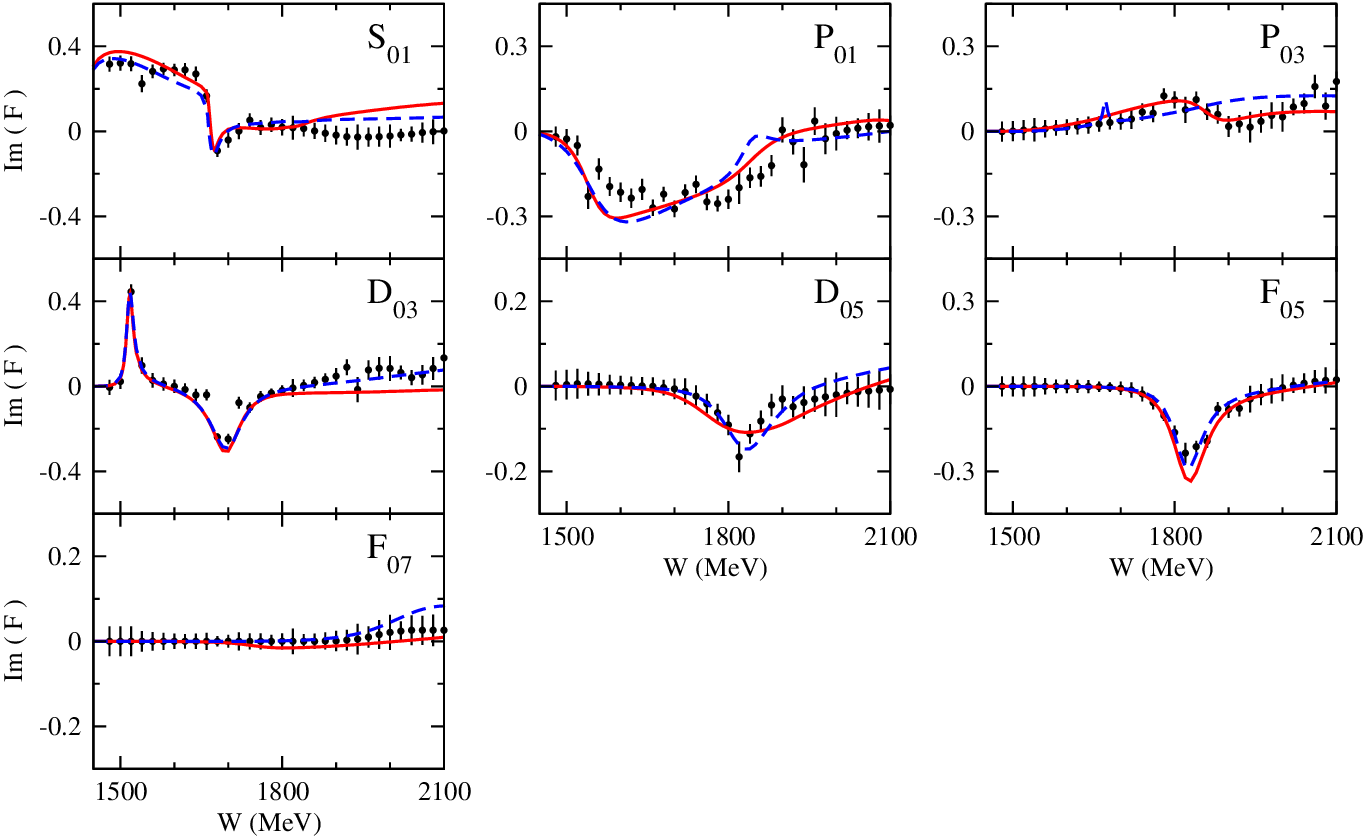}
\caption{(Color online)
Determined partial-wave amplitudes of $\bar K N \to \pi \Sigma$ with isospin $I=0$.
See the caption of Fig.~\ref{fig:kbn-amp-0} for the description of the figure.
}
\label{fig:pis-amp-0}
\end{figure}

\begin{figure}
\includegraphics[clip,width=\textwidth]{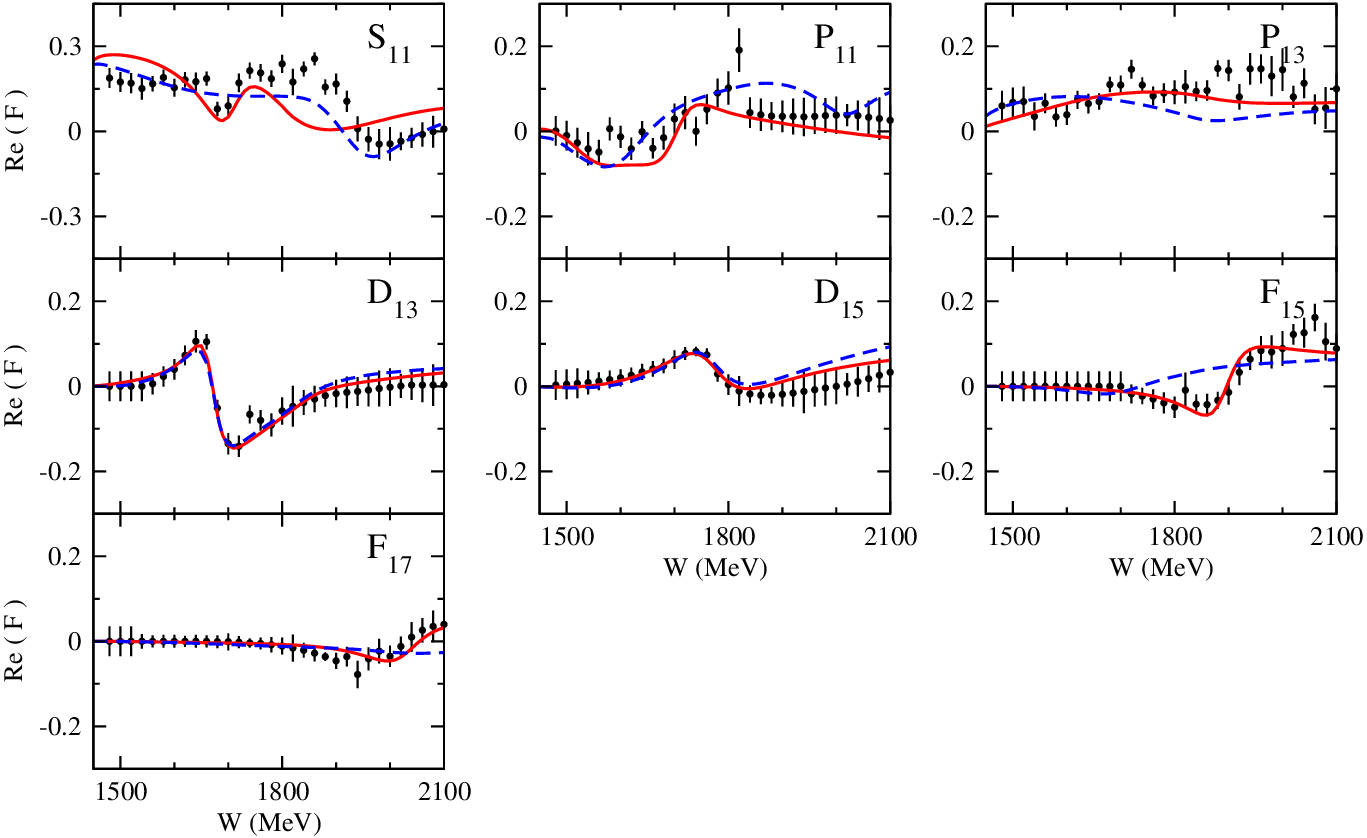}
\includegraphics[clip,width=\textwidth]{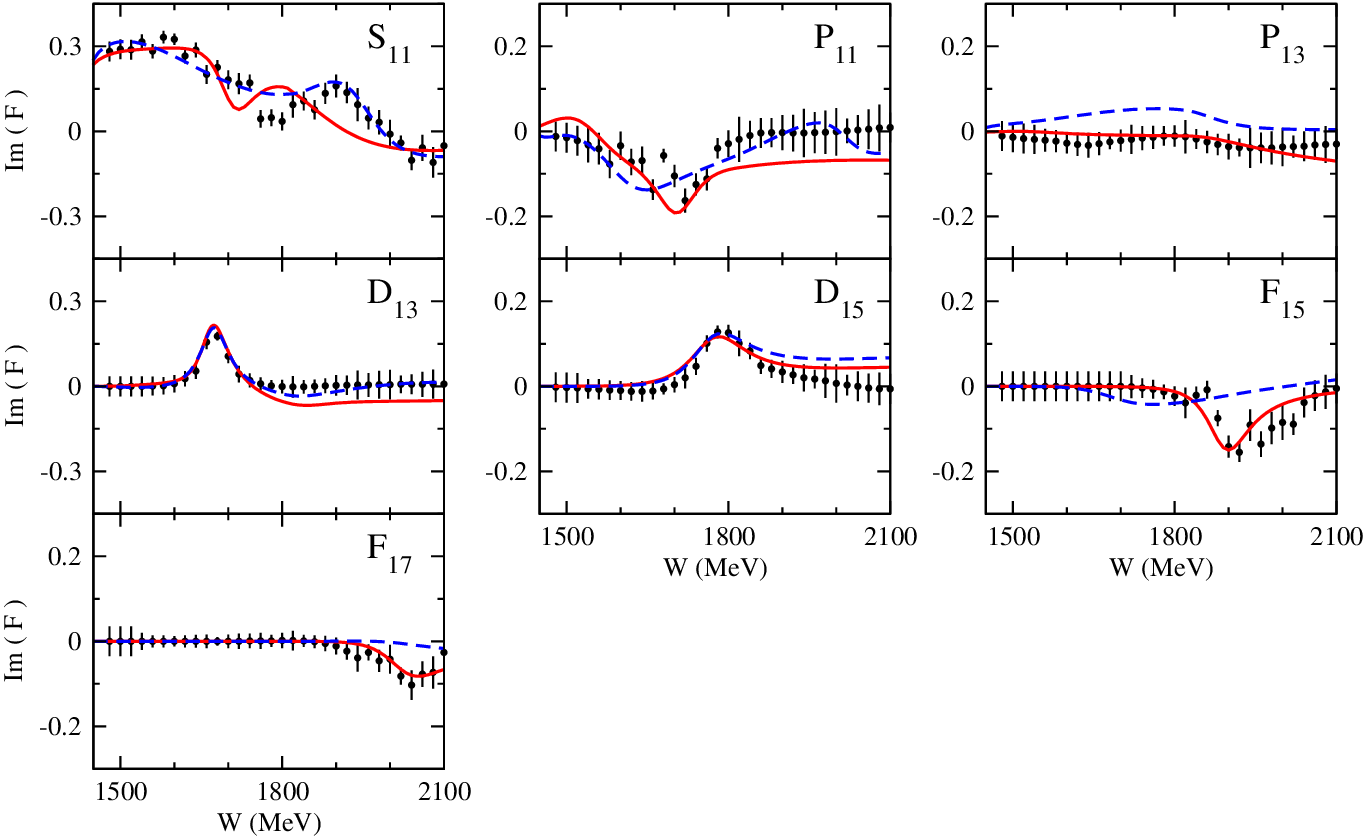}
\caption{(Color online)
Determined partial-wave amplitudes of $\bar K N \to \pi \Sigma$ with isospin $I=1$.
See the caption of Fig.~\ref{fig:kbn-amp-0} for the description of the figure.
}
\label{fig:pis-amp-1}
\end{figure}

\begin{figure}
\includegraphics[clip,width=\textwidth]{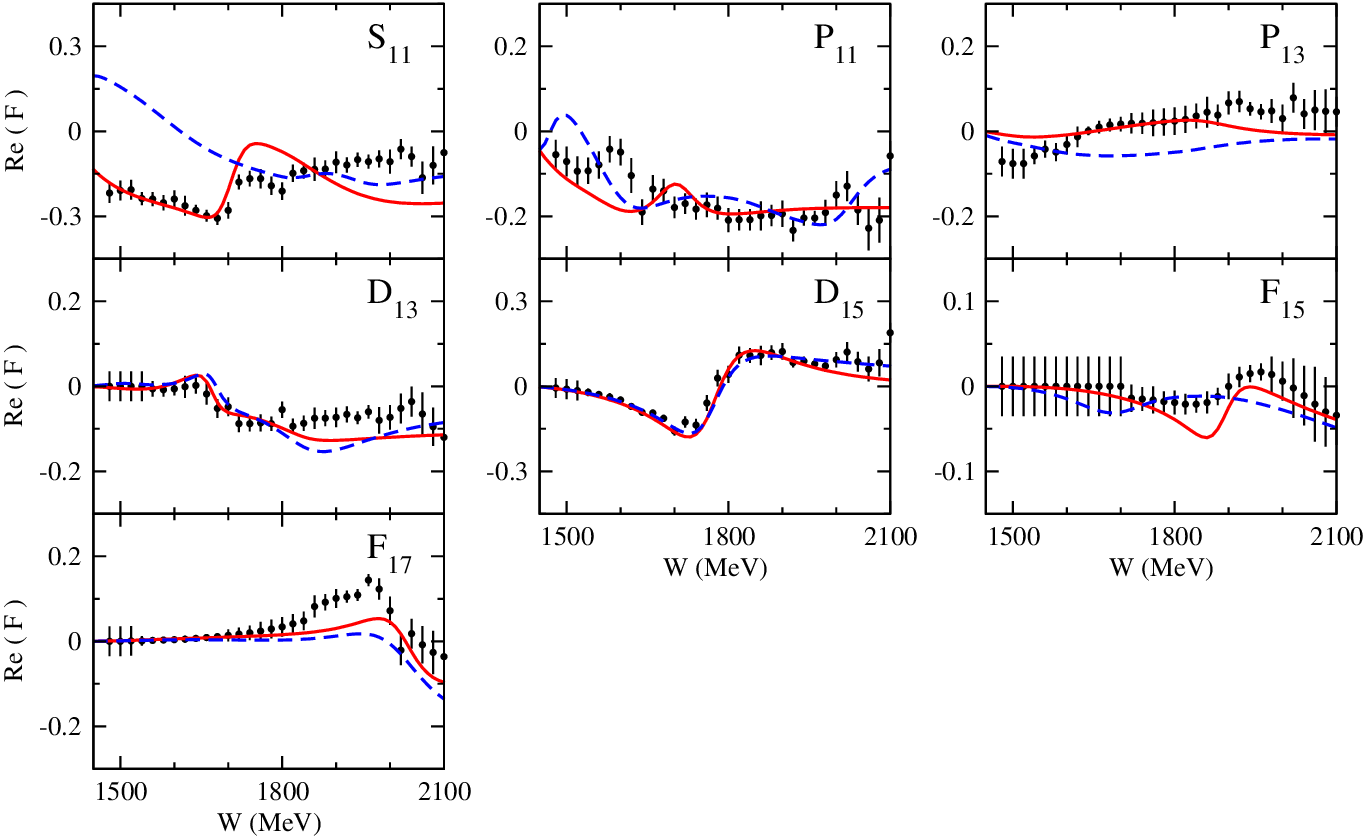}
\includegraphics[clip,width=\textwidth]{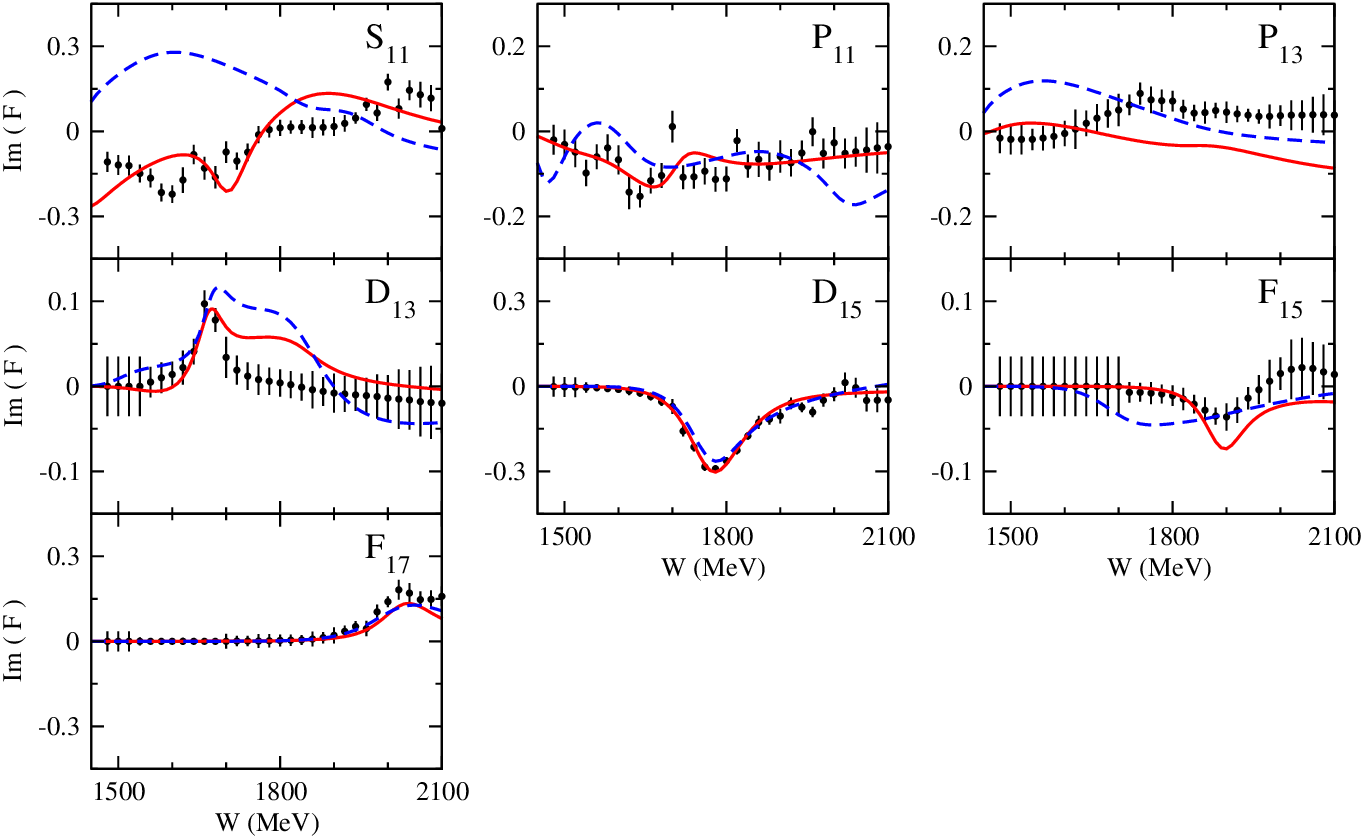}
\caption{(Color online)
Determined partial-wave amplitudes of $\bar K N \to \pi \Lambda$.
See the caption of Fig.~\ref{fig:kbn-amp-0} for the description of the figure.
}
\label{fig:pil-amp-1}
\end{figure}

In Figs.~\ref{fig:kbn-amp-0}-\ref{fig:pil-amp-1}, 
the $\bar K N \to \bar K N, \pi \Sigma, \pi \Lambda$ partial-wave amplitudes
obtained from our two models (solid red for Model A and dashed blue for Model B) 
are compared with those (solid circles with errors)
determined by the single-energy partial-wave analysis of KSU~\cite{zhang2013}. 
Overall, the results from the three analyses agree qualitatively.
In particular, some of the amplitudes, e.g.,
$S_{01}$, $D_{03}$, $F_{05}$, $D_{15}$, and $F_{17}$
of $\bar K N \to \bar K N$,
$S_{01}$, $D_{03}$, $D_{05}$, $F_{05}$, $D_{13}$, and $D_{15}$
of $\bar K N \to \pi \Sigma$, and
$D_{15}$ of $\bar K N \to \pi \Lambda$,
show  good agreements between the three analyses.
It is interesting to see that most of these ``stable'' amplitudes show a clear resonance behavior.
For example, the zero (peak) of $\mathrm{Re}~F$ ($\mathrm{Im}~F$) for $D_{03}$
at $W\sim 1520$ MeV is due to the existence of the well-established $\Lambda$ resonance with spin-parity $J^P=3/2^-$,
known as $\Lambda(1520)3/2^- $ in the notation of PDG~\cite{pdg2012}.
On the other hand, visible discrepancies can be seen in $S_{11}$ and most $P$-wave amplitudes
of all three reactions, where the discrepancy in $S_{11}$ of $\bar K N \to \pi \Lambda$ is sizable.
Such discrepancies are not surprising since the database used in the analyses
is far from \textit{complete}.
It is known that for the considered pseudoscalar-baryon scattering, 
the \textit{complete} data should include three observables, such as the
differential cross section ($d\sigma/d\Omega$), 
polarization ($P$), and the spin rotations.
From Table~\ref{tab:data-chi2} and the fits presented in Sec.~\ref{sec:results}, 
we see that no data of spin rotations are available. 
Furthermore, the number of
data points for the polarization $P$ are not sufficiently large.
The discrepancies seen here will lead to the differences of the hyperon resonances extracted
from the three partial-wave amplitudes displayed in Figs.~\ref{fig:kbn-amp-0}-\ref{fig:pil-amp-1}, 
as will be presented in our separate paper~\cite{knls14a}.


\begin{figure}
\includegraphics[clip,width=\textwidth]{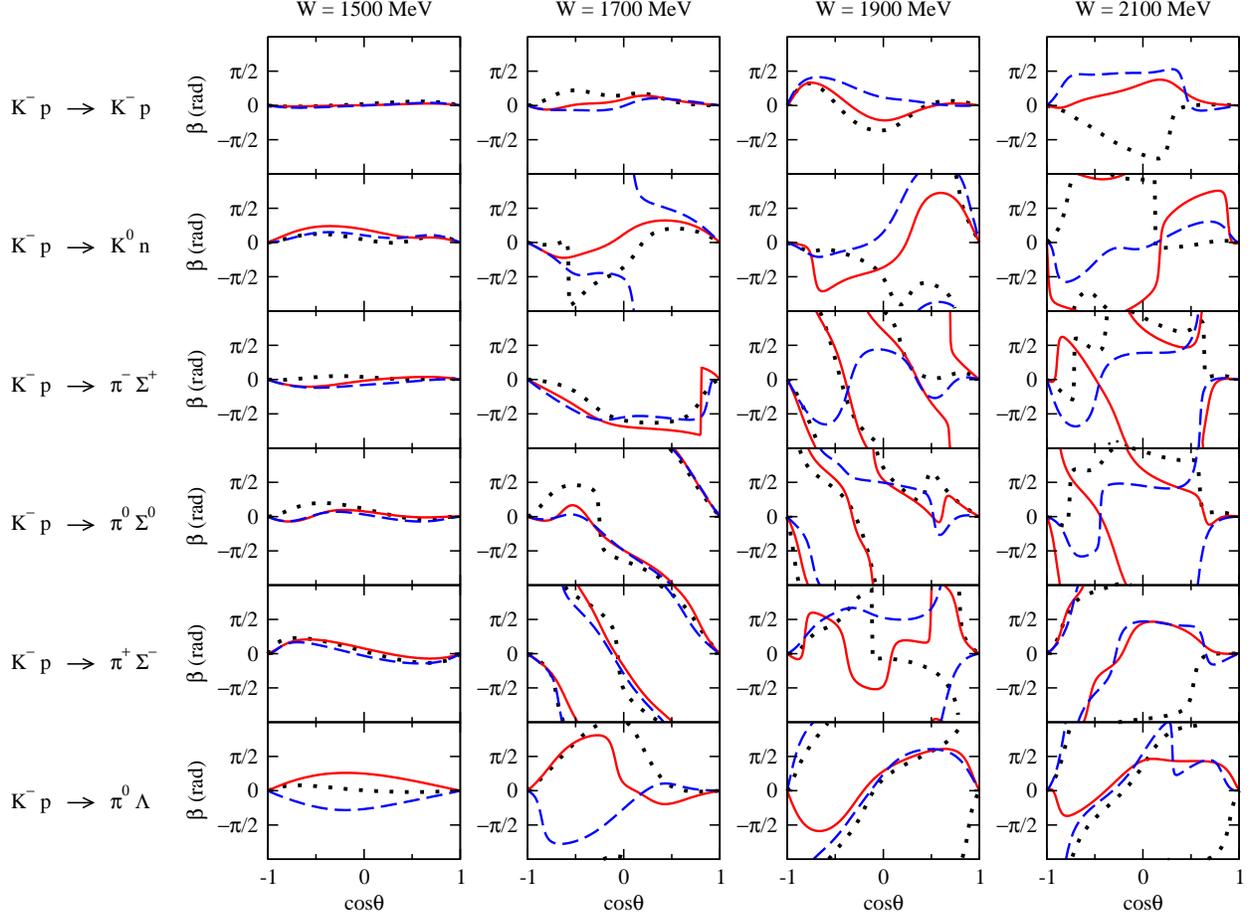}
\caption{(Color online)
Spin-rotation angle $\beta$ predicted from Model A (red solid curves) and Model B (blue dashed curves). 
The results are shown for the $K^-p \to \bar K N, \pi\Sigma, \pi\Lambda$ reactions.
Our predictions are compared with the $\beta$ calculated by using the partial-wave amplitudes of 
the KSU single-energy solution~\cite{zhang2013} (black dotted curves).
Note that $\beta$ is modulo $2\pi$.
}
\label{fig:beta}
\end{figure}

It is therefore important to obtain more high precision data of 
the $K^- p$ reactions from hadron beam facilities such as J-PARC, 
in particular for the polarization $P$ and spin rotations.
To motivate future experimental efforts, we compare in Fig.~\ref{fig:beta} 
the spin-rotation angle $\beta$ for $K^-p \to \bar KN,\pi\Sigma,\pi\Lambda$,
calculated from the considered three partial-wave amplitudes.
We observe that, 
except for $K^- p \to \pi \Lambda$,
the results agree qualitatively at low energies $W \lesssim 1600$ MeV,
while
the discrepancy becomes visible at $W\sim 1700$ MeV and sizable at higher energies.
As for $K^- p \to \pi \Lambda$ (bottom-row panels of Fig.~\ref{fig:beta}),
however, a clear discrepancy is already seen at low $W = 1500$ MeV.
This trend of the discrepancy is consistent with that of partial-wave amplitudes
shown in Figs.~\ref{fig:kbn-amp-0}-\ref{fig:pil-amp-1}. 
We expect that the discrepancy in the predicted $\beta$ can be distinguished
by experiments and thus the spin-rotation data can play a crucial role for eliminating the 
discrepancies in the determined partial-wave amplitudes particularly at higher energies.

\begin{figure}
\includegraphics[clip,width=\textwidth]{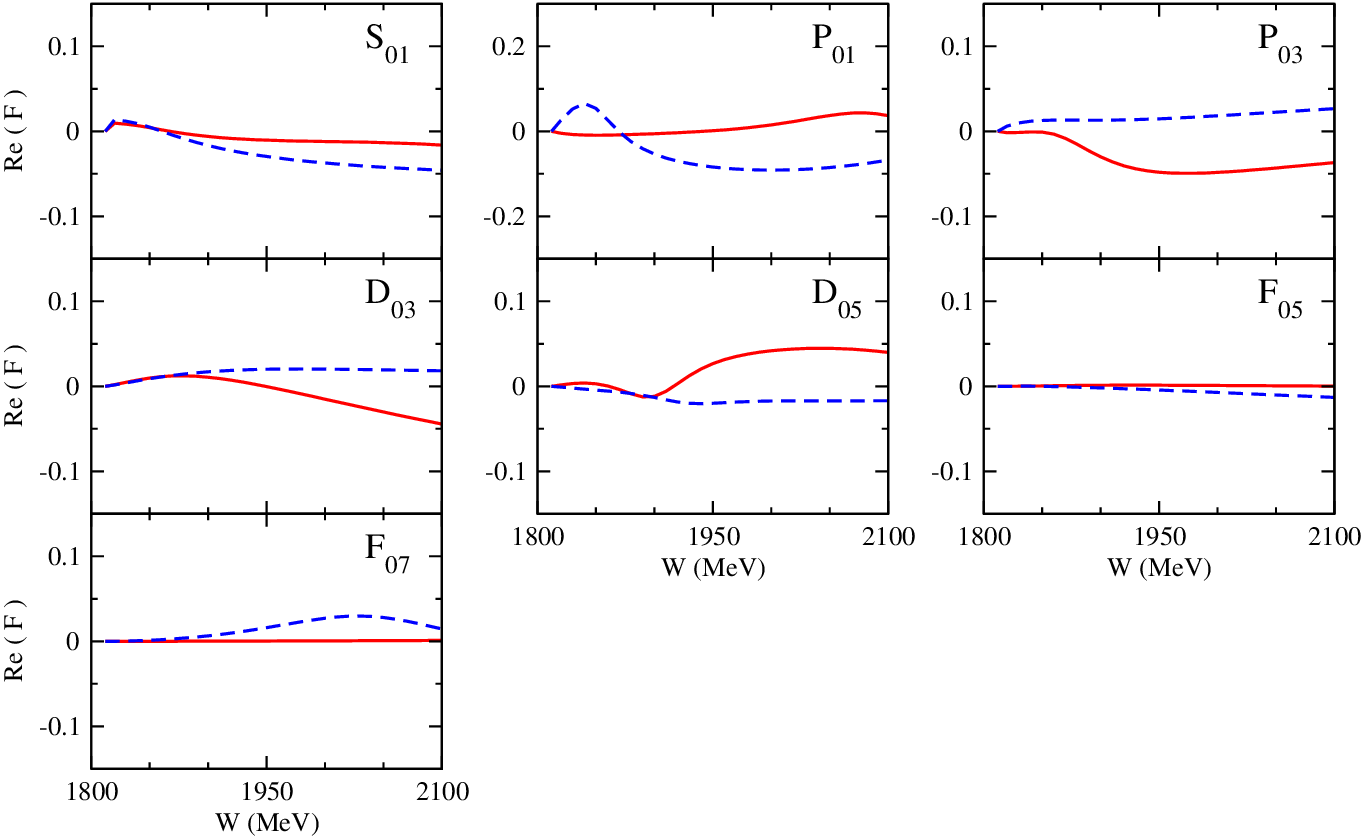}
\includegraphics[clip,width=\textwidth]{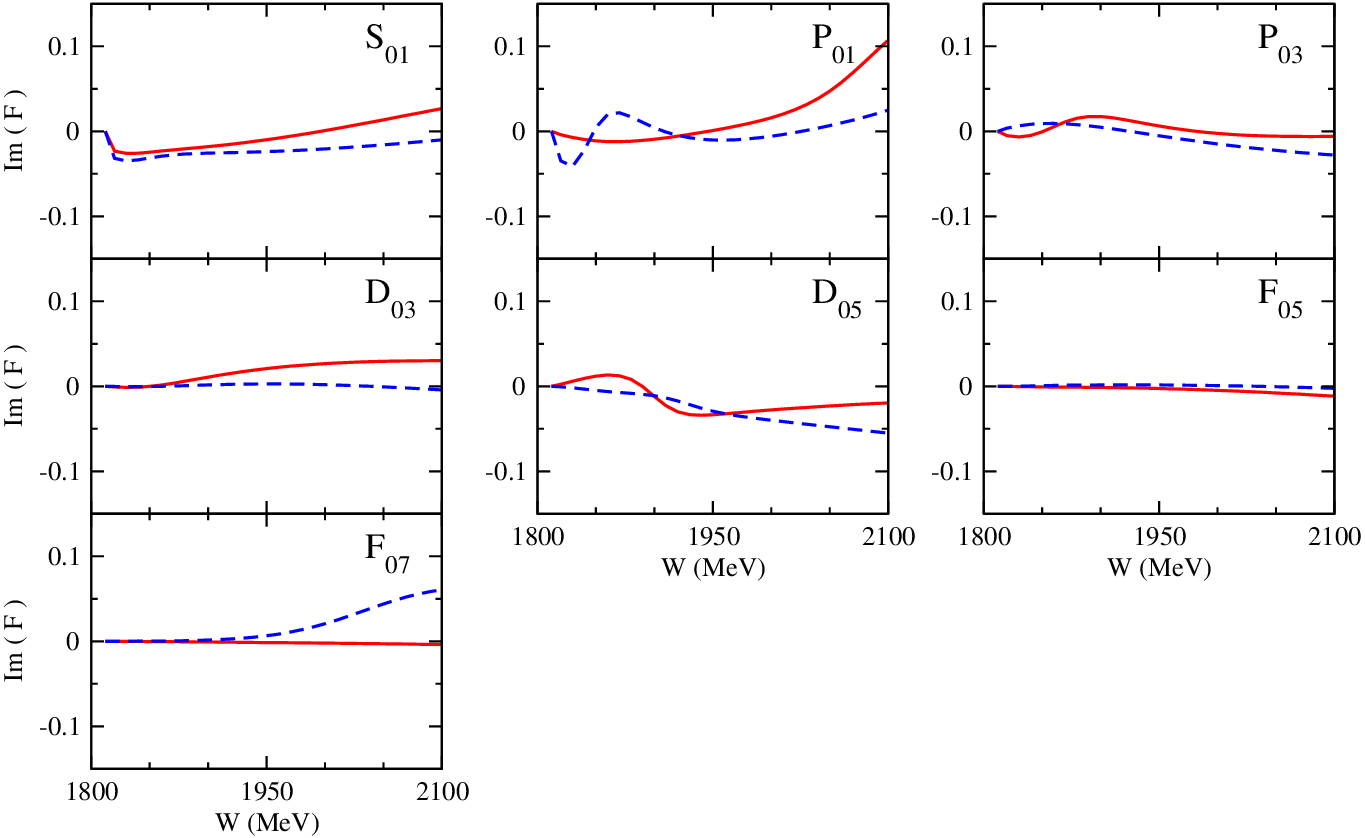}
\caption{(Color online)
Determined partial-wave amplitudes of $\bar K N \to K \Xi$ with isospin $I=0$.
See the caption of Fig.~\ref{fig:kbn-amp-0} for the description of the figure.
}
\label{fig:kxi-amp-0}
\end{figure}

\begin{figure}
\includegraphics[clip,width=\textwidth]{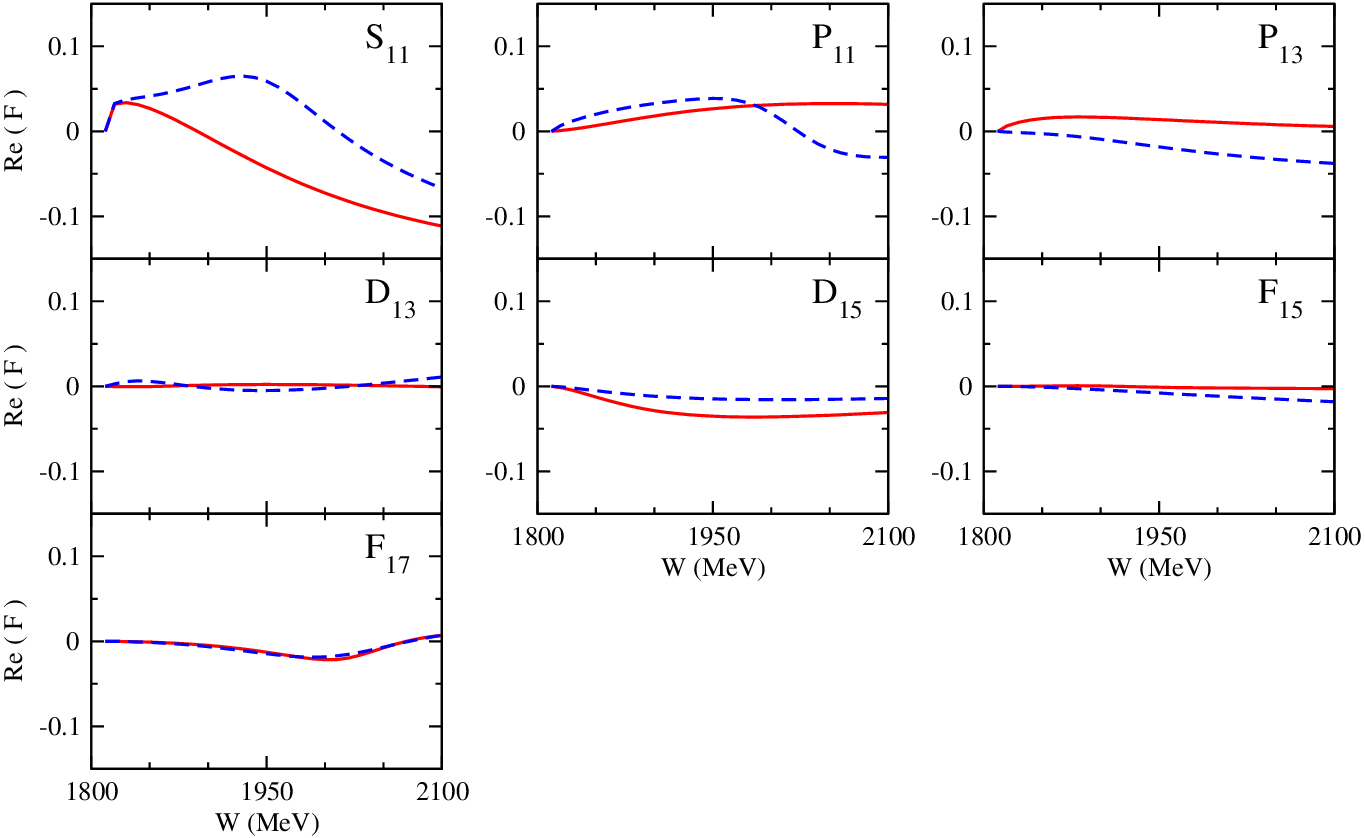}
\includegraphics[clip,width=\textwidth]{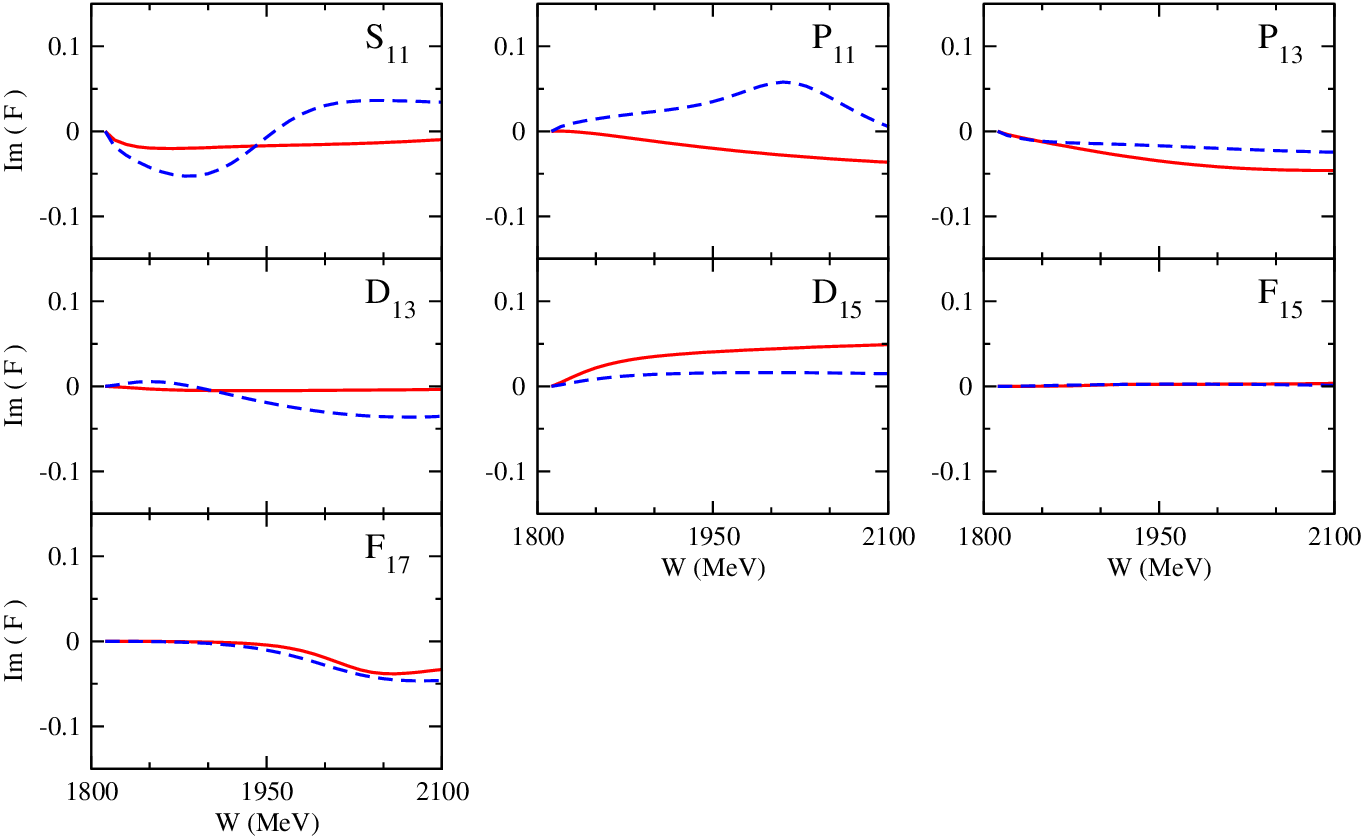}
\caption{(Color online)
Determined partial-wave amplitudes of $\bar K N \to K \Xi$ with isospin $I=1$.
See the caption of Fig.~\ref{fig:kbn-amp-0} for the description of the figure.
}
\label{fig:kxi-amp-1}
\end{figure}

\begin{figure}
\includegraphics[clip,width=\textwidth]{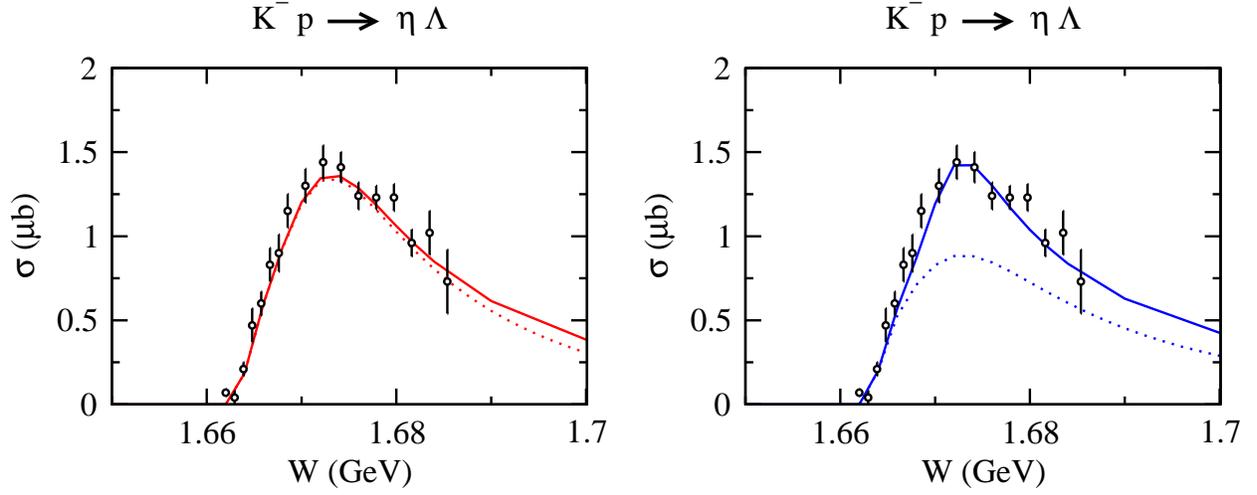}
\caption{(Color online)
The $S$-wave contribution to the $K^- p \to \eta \Lambda$ total cross sections near the threshold.
Left (right) panel is the result of Model A (Model B).
The solid curves are the full results, while the dotted curves are the contribution
of the $S_{01}$ partial wave only.
The data are taken from the latest BNL result~\cite{el-data}.
}
\label{fig:kmpetl-tcs}
\end{figure}

\begin{figure}
\includegraphics[clip,width=0.25\textwidth]{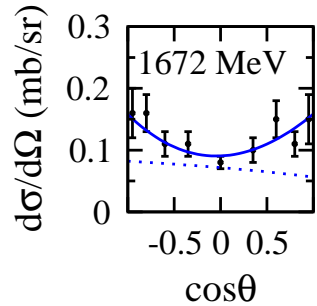}
\caption{(Color online)
The angular dependence of the near-threshold $K^- p \to \eta \Lambda$ differential cross section 
for Model B.
The result at $W=1672$ MeV is presented.
The solid curve is the full result, while the dotted curve is the result
in which the contribution of the $P_{03}$ partial wave is turned off.
}
\label{fig:e0l0-dc-p03}
\end{figure}

The $\bar K N \to K \Xi,\eta \Lambda$ partial-wave amplitudes 
are presented in Figs.~\ref{fig:kxi-amp-0}-\ref{fig:etl-amp-0}. 
Here we only compare the results from Models A and B
because KSU~\cite{zhang2013} did not provide them
(in particular, they did not include the $K\Xi$ channel in their analysis).
Overall, the amplitudes from Models A and B do not converge,
and this can be understood given the tiny amount of the data for $\bar K N \to K \Xi,\eta \Lambda$
as shown in Table~\ref{tab:data-chi2}.
As for $\bar K N \to K \Xi$, 
the $P_{01}$ amplitude exhibits a clear difference between the two models 
in the threshold region, and it is responsible for
the model dependence of the fitted results for the $K^- p \to K^0\Xi^0, K^+\Xi^-$ 
total cross sections below $W\sim$ 1.85 GeV (Fig.~\ref{fig:tcs}).
There are two characteristics for the $\bar K N \to \eta \Lambda$ amplitudes.
One is the near threshold behavior of the $S_{01}$ amplitude.
The rapid change of the amplitude is seen in both models and found to be due to 
the existence of a narrow $J^P = 1/2^-$ $\Lambda$ resonance, which would correspond to
$\Lambda(1670)1/2^-$ assigned as a four-star resonance by PDG~\cite{pdg2012}.
We find that this rapid change of the amplitude is necessary for reproducing the sharp peak 
in the $K^- p \to \eta \Lambda$ total cross section
near the threshold (Fig.~\ref{fig:tcs}). 
In other words, the inclusion of the $\eta \Lambda$ channel and the 
$K^- p \to \eta \Lambda$ data into the analysis is likely to make
the appearance of the narrow $J^P = 1/2^-$ $\Lambda$ resonance inevitable.
It is worthwhile to mention that for the $\bar K N \to \bar K N, \pi \Sigma$ reactions,
this resonance appears as a dip at $W\sim 1670$~MeV in the $S_{01}$ amplitudes. 
Another characteristics is the near threshold behavior of the $P_{03}$ amplitude,
which is one example indicating the model/analysis dependence of the extracted
amplitudes.
We see that Model A shows a smooth behavior for the $P_{03}$ amplitude, while
Model B shows a rapid change similar to the case of the $S_{01}$ amplitude.
In fact, we find this rapid change in Model B originates from the existence of 
a narrow $\Lambda$ resonance with $J^P = 3/2^+$ near the $\eta \Lambda$ threshold, 
which is not seen in Model A.
We show in Fig.~\ref{fig:kmpetl-tcs} the contribution of the $S_{01}$ partial wave to the 
$K^- p \to \eta \Lambda$ total cross section near the threshold.
For Model A, the total cross section is dominated by the $S_{01}$ partial wave.
However, for Model B, the contribution of the $S_{01}$ partial wave is just about 60~\%,
and the remaining 40~\% is found to almost come from the $P_{03}$ partial wave.
It is hard to judge the two models only from the total cross section,
but their difference should be enhanced in the angular dependence of 
the differential cross sections.
As mentioned in Sec.~\ref{sec:etl}, the near-threshold $K^- p \to \eta \Lambda$ 
differential cross section data (Fig.~\ref{fig:et0l0-dc}) 
show a concave-up angular dependence, which is reproduced well in Model B, but not in Model A.
This can be understood because the $S$-wave cannot produce any angular dependence
in the differential cross section.
In fact, we find that such a concave-up angular dependence of Model B
comes from the $P_{03}$ partial wave (Fig.~\ref{fig:e0l0-dc-p03}).
The current data therefore suggest that the non-negligible contribution from higher partial waves
is required even at the very low energies.
However, such contributions may not necessarily be originated from the resonance as in Model B.
The $J^P = 3/2^+$ $\Lambda$ resonance seen in Model B gives just a tiny contribution
to the $\bar K N \to \bar K N, \pi \Sigma$ reactions 
[e.g., a very small peak in the imaginary part of the $\bar K N \to \pi \Sigma$ $P_{03}$ 
amplitude at $W\sim 1670$~MeV (Fig.~\ref{fig:pis-amp-0})]
in contrast to the $J^P = 1/2^-$ $\Lambda$ resonance.
To judge the existence of the $J^P = 3/2^+$ $\Lambda$ resonance, more data of
the polarization observables would be desirable.


\begin{figure}
\includegraphics[clip,width=\textwidth]{etl-re0}
\includegraphics[clip,width=\textwidth]{etl-im0}
\caption{(Color online)
Determined partial-wave amplitudes of $\bar K N \to \eta \Lambda$.
See the caption of Fig.~\ref{fig:kbn-amp-0} for the description of the figure.
}
\label{fig:etl-amp-0}
\end{figure}

We should emphasize here that it may be
unlikely that the \textit{high} precision data from
\textit{complete} experiments can be realized in practice. 
Furthermore, it is not clear that one really can determine partial-wave amplitudes model-independently 
even if the data of complete experiments are available. 
This was examined~\cite{shkl11} carefully for the pion photoproduction data.
Thus, as mentioned in the introduction, it is advantageous to determine the partial-wave amplitudes
using a model within which the well-established 
physics is used to \textit{extrapolate} the available data to the region where the measurements
are difficult. 
This was done~\cite{hohler,cutkosky} for determining the partial-wave amplitudes 
of $\pi N$ scattering using the dispersion relations. 
Also, the very well-established $NN$ amplitudes at low energies
were obtained by imposing one-pion-exchange tails in all partial-wave analyses.
Here we follow the same approach by also making use of the hadron-exchange mechanisms. 
The purely phenomenological $K$-matrix analysis of the KSU group 
does not have such theoretical constraints, and the accuracy of their partial-wave amplitudes  
totally depends on the amount and quality of the data.
On the other hand, they are much more flexible in fitting the data, 
while the dynamical model may lead to large errors in the region where 
the hadron-exchange picture of reactions is not valid. 
Therefore the cross checks of results from two different approaches are
essential to pin down the resonance parameters.

\subsection{Threshold behavior of the total cross sections}

\begin{figure}
\includegraphics[clip,width=\textwidth]{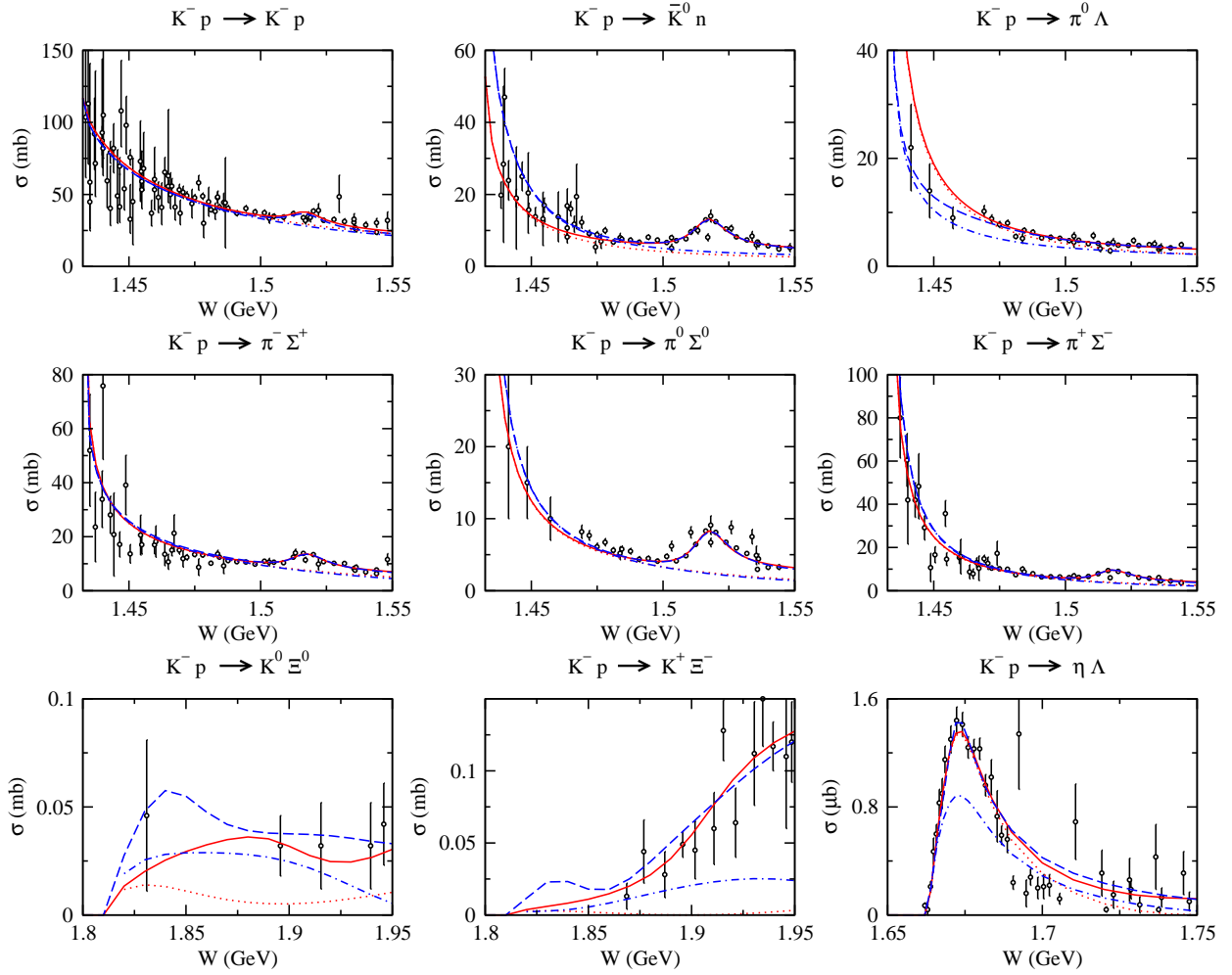}
\caption{(Color online)
The $K^- p$ reaction total cross sections in the threshold region.
Solid (dashed) curves are the full results from Model A (Model B), which are
the same as shown in Fig.~\ref{fig:tcs}, while dotted (dashed-dotted)
curves are the $S$-wave contribution from Model A (Model B).
}
\label{fig:kmp-thre}
\end{figure}

In this subsection, we discuss the threshold behavior of the total cross sections.
It is naively expected that the total cross sections near the threshold 
will be dominated by the $S$ wave. 
On the basis of this expectation,
a number of theoretical studies of threshold phenomena
have been performed by using various reaction models such as chiral unitary models~\cite{ucm-1},
where their model parameters are determined by fitting their $S$-wave cross sections
to the total cross section data near the threshold.
In Fig.~\ref{fig:kmp-thre}, we present the $K^- p$ reaction total cross sections near the threshold
obtained by Models A and B and their $S$-wave contributions.
As for the $K^-p \to \bar K N, \pi\Sigma$ reactions, the total cross sections
are dominated by the $S$-wave contributions up to $W\sim 1.5$~GeV, which seems consistent with the naive expectation.
Above that energy, however, the $S$-wave contributions start to deviate from the full results and
underestimate the data. 
This is mainly because of the large contribution from the $J^P = 3/2^-$ $\Lambda$ resonance in the $D_{03}$ partial wave
that is responsible for the peak of the cross sections at $W\sim 1520$ MeV.
In contrast with the $K^-p \to \bar K N, \pi\Sigma$ reactions, the 
$K^-p \to \pi\Lambda,\eta\Lambda,K\Xi$ reactions seem not to follow the naive expectation.
For these reactions, the higher partial waves contribute already at the energies very close to the threshold.
The appearance of the $S$-wave dominance in the $K^-p \to \bar K N, \pi\Sigma$ reactions
over the relatively wide energy region, i.e., up to about 70 MeV above the threshold,
would be due to the special circumstance of the existence of $\Lambda(1405)1/2^-$
lying just below the $\bar K N$ threshold. 
The results of Fig.~\ref{fig:kmp-thre} suggest that 
theoretical studies neglecting higher partial waves 
might fail to extract correct information from the observables even in the threshold region.

\begin{figure}
\includegraphics[clip,width=0.75\textwidth]{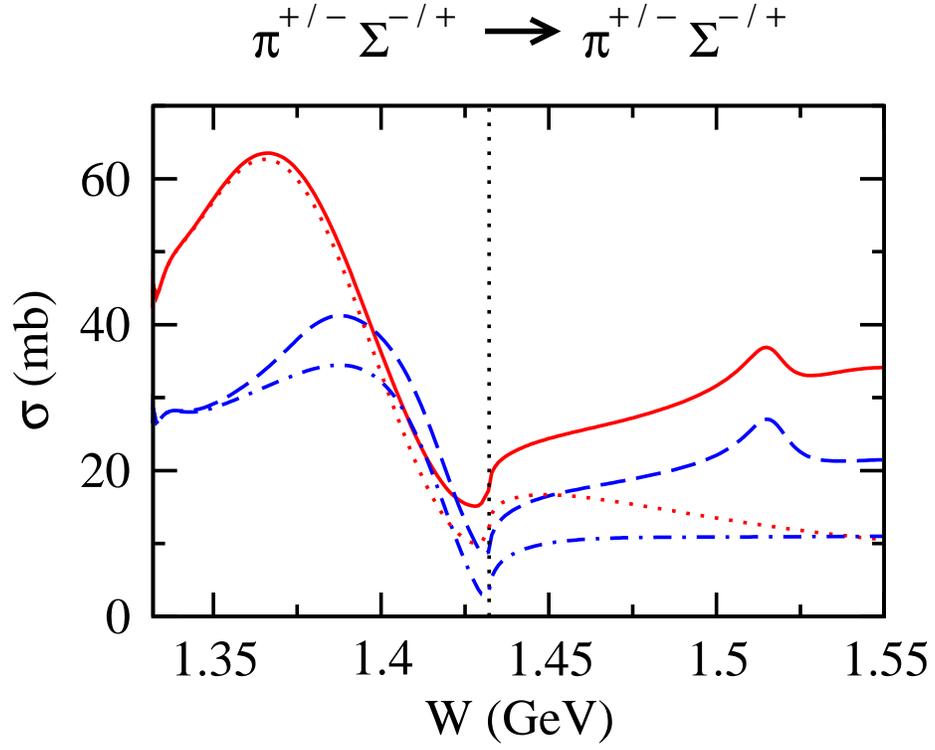}
\caption{(Color online)
The predicted $\pi^{\pm} \Sigma^{\mp} \to \pi^{\pm} \Sigma^{\mp}$
total cross sections from the threshold up to $W=1.55$ GeV.
The vertical dotted line indicates the $\bar K N$ threshold.
The meaning of each predicted curve is the same as in Fig.~\ref{fig:kmp-thre}
}
\label{fig:pis-pis}
\end{figure}

Before closing this subsection, we present the predicted total cross sections for
the $\pi^{\pm} \Sigma^{\mp} \to \pi^{\pm} \Sigma^{\mp}$ scatterings as another example (see Fig.~\ref{fig:pis-pis}).
Above $W \sim 1.42$~GeV, the contribution from higher partial waves
are found to be comparable with the $S$ wave for both Models A and B.
Furthermore, it is found that for Model B,
20 \% of the total cross sections in the $\Lambda(1405)1/2^-$ region,
namely at the energies near $W\sim 1.4$ GeV, come from the higher partial waves.
These observations are not surprising because the $\Lambda(1405)1/2^-$ region 
locates high ($\sim$ 80 MeV) above the $\pi \Sigma$ threshold,
and suggest that the higher partial waves should be treated on
the same footing as the $S$ wave if one wants to make a quantitative understanding of 
the nature of $\Lambda(1405)1/2^-$ using the data below the $\bar K N$ threshold, 
as will be provided by the J-PARC E31 experiment~\cite{noumi}.

\subsection{Threshold parameters}

\begin{table}
\caption{\label{tab:length}
Scattering length ($a_{MB}$) and effective range ($r_{MB}$) extracted from our analysis.
The results are shown in the isospin basis.
The sign convention of these threshold parameters is taken to be the same as that in Ref.~\cite{ihjksy11}.
}
\begin{ruledtabular}
\begin{tabular}{ccccc}
& \multicolumn{2}{c}{Model A} & \multicolumn{2}{c}{Model B} \\
\cline{2-3} \cline{4-5}
                      & $I=0$ & $I=1$ & $I=0$ & $I=1$ \\
\hline
$a_{\bar KN}$ (fm)    & $-1.37+i0.67$ & $~~\,0.07+i0.81$& $-1.62+i1.02$   & $~~\,0.33+i0.49$\\
$a_{\eta\Lambda}$ (fm)  & $~~\,1.35+i0.36$ &  - &  $~~\,0.97+i0.51$& - \\
$a_{K\Xi}$ (fm)       & $-0.81+i0.14$ & $-0.68+i0.09$ &$-0.89+i0.13$& $-0.83+i0.03$ \\
\\
$r_{\bar KN}$ (fm)    & $~~\,0.67-i0.25$ & $~~\,1.01-i0.20$& $~~\,0.74-i0.25$   & $-1.03+i0.19$\\
$r_{\eta\Lambda}$ (fm)    & $-5.67-i2.24$ & - & $-5.82-i3.32$  & - \\
$r_{K\Xi}$ (fm)       & $-0.01-i0.33$ & $-0.42-i0.49$ &$~~\,0.13-i0.20$& $-0.22-i0.11$ \\
\end{tabular}
\end{ruledtabular}
\end{table}
Threshold parameters such as the scattering length and effective range
are important quantities that characterize interacting systems.
Those parameters can also provide information on resonances existing near the threshold
(see, e.g., Ref.~\cite{ihjksy11}).
It is therefore interesting to see the threshold parameters given
by our constructed models.
In Table~\ref{tab:length}, we present the scattering lengths ($a_{MB}$)
and effective ranges ($r_{MB}$) for the $MB \to MB$ scattering 
($MB = \bar K N, \eta \Lambda, K\Xi$) above $\bar KN$ threshold.
The results are listed in the isospin basis.
Although we have not found a previous work that gave the effective
range, we present it for a future reference.

It is not an easy task to extract the threshold parameters precisely
because currently just a few data with large errors are available for the $\bar K N$ scattering near the threshold
and direct measurements of the $\eta \Lambda$ and $K\Xi$ scatterings are practically not possible.
In fact, some of the extracted threshold parameters exhibit significant model dependences.
As for the $\bar K N$ channel, the threshold parameters are found to present sizable differences
between Models A and B except for the $I=0$ effective range.
The difference in the $I=0$ $\bar K N$ scattering length would strongly affect
the analytic structure of the $S_{01}$ $\bar K N$ scattering amplitude
near the threshold and accordingly the pole position of the well-known first $J^P=1/2^-$ $\Lambda$ resonance, $\Lambda(1405)1/2^-$~\cite{knls14a}.
Although the $\bar K N$ scattering lengths in the isospin basis exhibits large model dependences,
the resulting $K^- p$ scattering length given by $a_{K^- p} = (a_{\bar KN}^{I=0} + a_{\bar KN}^{I=1} )/2$ seems rather stable,
$a_{K^- p} =-0.65 + i0.74$ fm for Model A and $a_{K^- p} = -0.65 +i0.76$ fm for Model B, and 
are consistent with previous works, e.g., Refs.~\cite{bmn06,dm11,ihw12}.
The $K^- n$ scattering length that contains purely $I=1$ contribution, $a_{K^- n} = a_{\bar KN}^{I=1}$, however,
presents a significant difference between Models A and B as shown in Table~\ref{tab:length}, 
and the latter seems consistent with a theoretical evaluation in Ref.~\cite{ihw12}.

As for the $\eta \Lambda$ channel, the model dependence in the threshold parameters is significant.
As for the $K\Xi$ channel, however, the scattering lengths agree
reasonably well between Models A and B, while significant model dependences are seen for the effective ranges.
The model dependence in the $\eta\Lambda$ and $K\Xi$ threshold parameters are understandable, 
because they are constrained indirectly by the data of $K^- p$ reactions,
in particular by those of $K^- p\to \eta \Lambda$ and $K^- p\to K\Xi$, respectively, through coupled-channels effects.
Because the quality of the data for $K^- p\to \eta \Lambda$ and $K^- p\to K\Xi$ is still poor, it is
difficult to strongly constrain the threshold parameters for $\eta \Lambda$ and $K\Xi$.

Here it is noted that we extracted the threshold parameters
from the $K^- p$ reaction data above the $\bar KN$ threshold 
through the analysis in which the isospin symmetry is assumed.
For a more precise determination, we would need to take into account the
isospin breaking effect, and also need to use data that provide accurate threshold 
information such as the kaonic hydrogen spectrum~\cite{ihw12}. 
Furthermore, the Coulomb corrections, which are expected to negligibly affect
hyperon resonance parameters and thus not considered in this work, should also be included.

The threshold parameters for channels below the $\bar KN$ threshold 
($\pi\Sigma$, $\pi\Lambda$) are difficult to extract unambiguously 
because of the lack of data below the $\bar KN$ threshold.
Analyses including only data above the $\bar KN$ threshold, as done in this work,
would result in obtaining rather model-dependent
threshold parameters for the subthreshold channels.
This was demonstrated in Ref.~\cite{ihjksy11} within various chiral unitary models 
and phenomenological potential approaches.
Thus, at this moment we refrain from presenting the threshold parameters 
for the subthreshold channels.

\subsection{$K^-p$ reaction total cross section}

Finally, we present the $K^- p$ reaction total cross sections predicted from our models.
Within our current framework, the contributions from the considered reaction channels, i.e.,
$K^- p \to MB$ with $MB = \bar K N,\pi\Sigma,\pi\Lambda,\eta \Lambda, K\Xi, \pi \Sigma^*, \bar K^* N$,
to the $K^- p$ reaction total cross section are expressed as 
\begin{equation}
\sigma^{\text{tot}}_{K^- p}(W) = 
\sum_{M'B' = \bar K N,\pi\Sigma,\pi\Lambda,\eta\Lambda,K\Xi} \sigma_{K^- p \to M'B'}(W) 
+
\sum_{M'B' = \pi \Sigma^* \bar K^* N} \bar \sigma_{K^- p \to M'B'}(W) .
\label{eq:tcs-all}
\end{equation}
Here, the first (second) term in the right-hand side represents
the contribution from the two-body $\bar K N$, $\pi\Sigma$, $\pi\Lambda$, $\eta\Lambda$, and $K\Xi$ channels
(the quasi-two-body $\pi\Sigma^*$ and $\bar K^* N$ channels)
to the cross section.

The contribution from a stable two-body channel $M'B'$, $\sigma_{K^-p \to M'B'}$, is calculated with
\begin{eqnarray}
\sigma_{K^-p \to M'B'}(W) 
&=& 
\frac{4\pi}{k^2} \rho_{M'B'}(k';W) \rho_{\bar K N}(k;W)
\nonumber\\
&&
\times
\frac{1}{2}\sum_{LS,JI}(2J+1)
|C_{K^- p}^{I}\times T^{IJ}_{M'B'(LS),\bar KN(LS)}(k',k;W)|^2 ,
\label{eq:tcs-2body}
\end{eqnarray}
where the explicit form of $\rho_{MB}(k;W)$ is given just below Eq.~(\ref{eq:pwa});
$k$ and $k'$ are defined by $W = E_{\bar K}(k) + E_{N}(k) = E_{M'}(k') + E_{B'}(k')$; 
the factor $1/2$ comes from the spin average of the initial proton;
$C_{K^- p}^{I}$ is the isospin factor with $C_{K^- p}^{I=0} =C_{K^- p}^{I=1} = 1/2$; 
and all indices of $T_{M'B',\bar KN}(k',k;W)$ are explicitly shown.
In this formula, all allowed charge states of the final $M'B'$ channel are summed up.
Multiplying Eq.~(\ref{eq:tcs-2body}) by an appropriate isospin factor 
for the final $M'B'$ channel, we can have the $K^- p \to M'B'$ total cross section 
for a specific charge state, which has been shown in Fig.~\ref{fig:tcs}.

The explicit form of $\bar \sigma_{K^- p \to \pi\Sigma^*}(W)$ is given by
\begin{eqnarray}
\bar \sigma_{K^- p \to \pi\Sigma^*}(W)
&=&
\int^{W-m_\pi}_{m_\pi+m_\Lambda} dM_{\pi\Lambda} \frac{M_{\pi\Lambda}}{E_{\Sigma^*}(k')}
\nonumber\\
&&
\times
\frac{1}{2\pi}\frac{\Gamma_{\pi\Sigma^*}(k';W)}{|W-E_\pi(k')-E_{\Sigma^*}(k')-\Sigma_{\pi\Sigma^*}(k';W)|^2}
\nonumber\\
&&
\times
\sigma_{K^-p \to \pi\Sigma^*}(k';W) ,
\label{eq:tcs-23body}
\end{eqnarray}
where $k'$ is defined by $W = E_{\pi}(k') + E_{\pi\Lambda}(k')$ with  
$E_{\pi\Lambda}(k') = \sqrt{M_{\pi\Lambda}^2+(k')^2}$;
$\Sigma_{\pi\Sigma^*}(k';W)$ is the self-energy of the $\pi\Sigma^*$ Green's function
given in Eq.~(\ref{eq:self-pisstar}); 
$\Gamma_{\pi \Sigma^*}(k';W) = -2 \mathrm{Im}[\Sigma_{\pi\Sigma^*}(k';W)]$;
and $\sigma_{K^-p \to \pi\Sigma^*}(k';W)$ is the total cross section for 
the half-off-shell $K^-p \to \pi\Sigma^*$ reaction,
\begin{eqnarray}
\sigma_{K^-p \to \pi\Sigma^*}(k';W) 
&=& 
\frac{4\pi}{k^2}
\rho_{\pi\Sigma^*}(k';W) \rho_{\bar K N}(k;W)
\nonumber\\
&&
\times
\frac{1}{2}\sum_{L'S',LS,JI}(2J+1)
|C_{K^- p}^{I}\times T^{IJ}_{\pi\Sigma^*(L'S'),\bar KN(LS)}(k',k;W)|^2 ,
\label{eq:tcs-q2body}
\end{eqnarray}
with $k$ defined by $W = E_{\bar K}(k) + E_{N}(k)$.
Comparing Eqs.~(\ref{eq:tcs-2body}) and~(\ref{eq:tcs-q2body}), 
we see that in Eq.~(\ref{eq:tcs-q2body})
the summation of total spin ($S$) and angular momentum ($L$)
is taken independently for the initial $\bar K N$ and final $\pi\Sigma^*$ channels.
This is because $\Sigma^*$ has the spin $3/2$
and allowed $LS$ quantum numbers for a given total $J^P$ are
different between the $\bar K N$ and $\pi\Sigma^*$ channels as shown in Table~\ref{tab:pw}.
Also, it should be emphasized that 
the decay of $\Sigma^*$ to the $\pi \Lambda$ state, 
which subsequently occurs after the $K^- p \to \pi\Sigma^*$ process,
is appropriately taken into account in Eq.~(\ref{eq:tcs-23body}).
The corresponding expression for $\bar\sigma_{K^- p \to \bar K^* N}(W)$
can be obtained from Eqs.~(\ref{eq:tcs-23body}) and~(\ref{eq:tcs-q2body})
by changing the channel labels.

\begin{figure}
\includegraphics[clip,width=\textwidth]{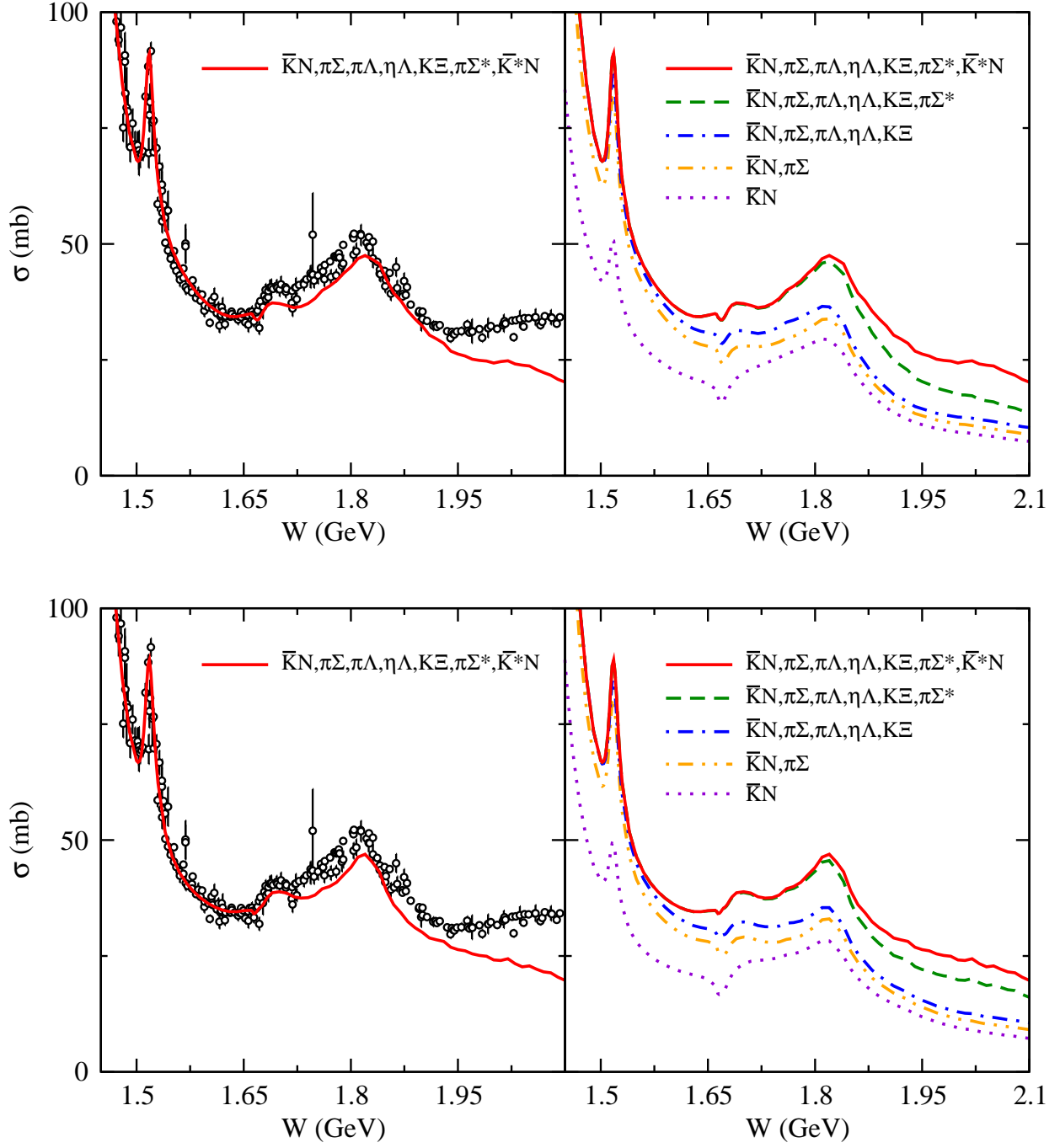}
\caption{\label{fig:tcs-all}
(Color online)
Predicted $K^- p $ reaction total cross section.
The upper (lower) row is the results of Model A (Model B).
(Left) Comparison of our predicted $\sigma_{K^-p}^{\text{tot}}$ (solid curve) 
with the data (open circles). 
The data are taken from Ref.~\cite{pdg2012}.
(Right) The curves showing how the predicted contributions from each channel
are added up to the total $\sigma_{K^-p}^{\text{tot}}$.
}
\end{figure}

The resulting $K^- p$ reaction total cross sections are shown in Fig.~\ref{fig:tcs-all}.
In the left panels, the comparison with the experimental data is presented.
Our results agree well with the data up to $W \sim 1.7$ GeV,
and then start to underestimate the data. 
Since the contributions from the two-body $\bar K N$, $\pi\Sigma$, $\pi\Lambda$, $\eta\Lambda$, and $K\Xi$ channels
are well fixed by the data as shown in Fig.~\ref{fig:tcs}, the difference 
in $\sigma_{K^-p}^{\text{tot}}$ between Models A and B arises from
the predicted contributions from the $\pi\Sigma^*$ and $\bar K^* N$ channels.
The underestimation of our results becomes significant above $W \sim 1.9$ GeV,
which is expected to be mainly because other inelastic channels that are not included
in this work also become relevant.

It is seen from the right panels of Fig.~\ref{fig:tcs-all} 
that the $\bar K N$ channel have the largest contribution in the considered energy region.
The $\pi \Sigma$ channel also gives a sizable contribution at low energies, while
it becomes very small above $W\sim 1.85$ GeV.
The contributions from the $\pi\Lambda$, $\eta\Lambda$, and $K\Xi$ channels
are rather small in the entire energy region considered.
The contributions from the quasi-two-body $\pi\Sigma^*$ and $\bar K^* N$ channels become 
visible at $W \sim 1.65$ GeV and at $W \sim 1.85$ GeV, respectively, and
those become comparable with 
the two-body $\bar K N$, $\pi\Sigma$, $\pi\Lambda$ and $K\Xi$ contributions
above $W\sim 1.9$ GeV.

\section{Summary and future developments}
\label{sec:summary}

In this work, we have constructed a dynamical coupled-channels model of $K^- p$ reactions
within the Hamiltonian formulation developed in 
Refs.~\cite{msl07,jlms07,jlmss08,jklmss09,kjlms09-1,kjlms09-2,knls10,knls13,kpi2pi13}.
The model consists of meson-baryon potentials $v_{M'B',MB}$
derived from the phenomenological SU(3) Lagrangian, and
vertex interactions $\Gamma_{MB,Y^*}$ describing the decays of the bare excited
hyperon states $Y^*$  into $MB$ states.
The parameters of the model
are determined by fitting the data of the unpolarized and polarized observables 
of the $K^- p \to \bar K N, \pi \Sigma, \pi \Lambda, \eta\Lambda, K\Xi$ reactions
from the threshold up to $W=2.1$ GeV. 
Practically, we have constructed two models, Models A and B, for which 
we used different vector-meson-exchange mechanisms in $v_{M'B',MB}$,
yet both reproduce equally well 
the currently available data of $K^- p$ reactions within their uncertainties.
Once a model is constructed, we can extract various physical parameters associated with
$Y^*$ resonances,
e.g., complex resonance masses and coupling constants defined by poles and residues
of the scattering amplitudes, from the model by performing the
analytic continuation to the complex energy plane.
The extracted $Y^*$ resonance parameters
will be presented in detail in a separate paper~\cite{knls14a}.
Although we have employed different vector-meson-exchange mechanisms for Models A and B,
it seems difficult to figure out its consequence on the dynamical content of 
the determined partial-wave amplitudes, given the incompleteness of the available data.
However, such a difference could become more visible when one investigates
the role of reaction dynamics in understanding $Y^*$ resonances.

We found that the determined partial-wave amplitudes depend rather strongly on the analysis
methods, owing to the fact that the available data of $K^- p$ reactions are far from complete.
With comparable quality of the fits to the data,
two models constructed in our fits  give rather different results in several partial waves.
Similar large differences are also found with the results from the recent single-energy analysis
by the KSU group~\cite{zhang2013}. 
More high precision data on $K^- p$ reactions, in particular for
the spin-dependent observables, $P$ and spin rotations ($\beta$, $A$, or $R$),
from J-PARC will be highly desirable to pin down the partial-wave amplitudes 
for high precision extractions of hyperon resonances.

The antikaon-nucleon scattering amplitudes obtained in this work can be used to
investigate various phenomena in the strangeness sector such as hypernuclei
and kaonic nuclei production reactions, as being actively pursued at J-PARC.
Also, the amplitudes set a basis to explore the dynamics below the 
$\bar K N$ threshold where $\Lambda (1405)1/2^-$ is expected to play an important role.
Previously, this interesting region and $\Lambda (1405)1/2^-$
have been studied with an assumption
that a $S$-wave interaction dominates
the $\bar K N$-$\pi\Sigma$ coupled-channels system~\cite{ucm-1}.
However, $\Lambda (1405)1/2^-$ mass is above the $\pi\Sigma$ threshold by
$\sim$ 80 MeV, and there is no reason to ignore the dynamics in partial
waves higher than the $S$ wave.
With the DCC model that treats all relevant partial waves on the same footing, 
we indeed have shown that contributions from higher partial waves can be sizable in
the $\pi\Sigma\to\pi\Sigma$ amplitude in the $\Lambda(1405)$ region.
Thus it will be interesting to apply the DCC model for analyzing data
not only above the $\bar K N$ threshold but also below that.  
Such combined analysis, if done, will be much more comprehensive than what has been done so
far for the $S=-1$ meson-baryon system, 
and would give a clear picture of the dynamics in the $\Lambda (1405)1/2^-$ region.
A J-PARC experiment~\cite{noumi} is expected to provide useful information for this interesting
future prospect.

The formulas presented in Sec.~\ref{sec:dcc} satisfy the two- and three-body unitarity conditions.
However, at this stage we have neglected the $Z$-diagram term, $Z^{(E)}_{M'B',MB}(k',k;W)$ 
in Eq.~(\ref{eq:veff-mbmb}), in solving the coupled-channels equations. 
While we expect from our previous investigation~\cite{knls13} of $\pi N$ reactions that
the effects of $Z^{(E)}_{M'B',MB}(k',k;W)$ have only a few percent effects
on the total cross sections, we need to include this in our next investigations.
Also, we have not included the direct 
$\pi \Sigma^* \to \pi \Sigma^*$, 
$\pi \Sigma^* \leftrightarrow \bar K^* N$, and 
$\bar K^* N \to \bar K^* N$ mechanisms
in the meson-baryon potentials $v_{M'B',MB}$.
(In the current work, those transitions occur indirectly
via the processes such as $\pi \Sigma^* \to \bar K N \to \pi \Sigma^*$.)
While in the $K^- p$ reactions the direct quasi-two-body to quasi-two-body transition mechanisms 
are expected to be less important than the two-body to two-body or two-body to quasi-two-body
processes, we must also include them in the future.

\begin{acknowledgments}
H.K. thanks Y. Ikeda for useful discussions on the scattering lengths and 
for providing the data of the $K^- p$ reaction total cross sections near the threshold.
This work was supported by the Japan Society for the Promotion of Science (JSPS) KAKENHI Grant No. 25800149 (H.K.) and 
Nos. 24540273 and 25105010 (T.S.), 
and by the U.S. Department of Energy, Office of Nuclear Physics Division, 
under Contract No. DE-AC02-06CH11357.
H.K. acknowledges the support of the HPCI Strategic Program 
(Field 5 ``The Origin of Matter and the Universe'') of 
Ministry of Education, Culture, Sports, Science and Technology (MEXT) of Japan.
This research used resources of the National Energy Research Scientific Computing Center,
which is supported by the Office of Science of the U.S. Department of Energy
under Contract No. DE-AC02-05CH11231, and resources provided on Blues and/or Fusion,
high-performance computing cluster operated by the Laboratory Computing Resource Center
at Argonne National Laboratory. 
\end{acknowledgments}

\appendix

\section{Model Lagrangian}
\label{app:lag}

Here we present the effective Lagrangian used in our model.
In this appendix and Appendix~\ref{app:pot}, the symbols
$P$, $V$, $S$, $B$, and $D$
denote pseudoscalar-octet meson, vector-octet meson, scalar-octet meson,
spin-$\frac{1}{2}$ octet baryon, and spin-$\frac{3}{2}$ decuplet baryon, 
respectively.

\subsection{$PBB'$ interaction}

The Lagrangian for the $PBB'$ interaction is expressed as
\begin{equation}
L_{PBB'} = L^{\rm L}_{PBB'} \times L^{\rm F}_{PBB'} +[{\rm H.c.~for~B \not= B'}],
\end{equation}
where the superscripts L and F indicate the Lorentz and flavor parts of 
the Lagrangian, respectively.
The Lorentz part is explicitly given by
\begin{equation}
L^{\rm L}_{PBB'} = - \bar B \gamma_\mu \gamma_5 B' \partial^\mu P .
\end{equation}
The flavor part of the $PBB'$ Lagrangian is derived from the following SU(3) 
singlet form~\cite{su3},
\begin{equation}
L^{\rm F}_{P_{\bf 8}B_{\bf 8}B'_{\bf 8}} =
g_1[[B_{\bf 8}^\dag \otimes B'_{\bf 8}]^{({\bf 8}_1)} \otimes P_{\bf 8}]^{({\bf 1})} 
+g_2[[B_{\bf 8}^\dag \otimes B'_{\bf 8}]^{({\bf 8}_2)} \otimes P_{\bf 8}]^{({\bf 1})} ,
\end{equation}
where
$B_{\bf 8}$, $B'_{\bf 8}$, and $P_{\bf 8}$ denote the SU(3) octet representations,
to which the $B$, $B'$, and $P$ hadrons belong, respectively.
The necessary information for our calculation is then the following flavor matrix elements,
\begin{equation}
\bra{B}L_{PBB'}^{\rm F}\ket{PB'}=
\bra{B}L_{P_{\bf 8}B_{\bf 8}B'_{\bf 8}}^{\rm F}\ket{PB'}=
G_{B,P B'} \times (I_P I_P^z, I_{B'} I_{B'}^z|I_B I_B^z) , 
\label{eq:pbb-mat1}
\end{equation}
\begin{equation}
\bra{PB}L_{PBB'}^{\rm F}\ket{B'}=
\bra{PB}L_{P_{\bf 8}B_{\bf 8}B'_{\bf 8}}^{\rm F}\ket{B'}=
G_{PB, B'} \times (I_P I_P^z, I_{B} I_{B}^z|I_{B'} I_{B'}^z) .
\label{eq:pbb-mat2}
\end{equation}
With the isoscalar factors~\cite{su3}, $G_{B,PB'}$ and $G_{PB,B'}$ are given by
\begin{align}
G_{B,PB'} &=
(-1)^{-(Y_{B'}/2)+I_{B'}-1}\frac{1}{\sqrt{8}}
\sqrt{\frac{2I_P+1}{2I_B+1}}
\left[
\sum_{\gamma=1,2} g_\gamma
\left(
\begin{array}{cc|c}
8              & 8                 & 8_{(\gamma)} \\
I_B -Y_B & I_{B'} Y_{B'} & I_P -Y_P
\end{array}
\right)
\right],
\label{eq:pbb-g}
\end{align}
and $G_{PB, B'}= G_{B,PB'}$.
In Eqs.~(\ref{eq:pbb-mat1})-(\ref{eq:pbb-g}),
$I_P$, $I_P^z$, and $Y_P$ denote the isospin, its $z$ component, 
and the hypercharge of a hadron $P$, respectively.
Owing to the SU(3) symmetric construction of the Lagrangian, all of the $PBB'$ coupling constants
are expressed with just two parameters, $g_1$ and $g_2$.
Introducing the notation $g_{p} = (\sqrt{30}/40)g_1 + (\sqrt{6}/24)g_2$ and
$\alpha^D_{p} = (\sqrt{30}/40)(g_1/g_{p})$,
we can relate these to $g_{\pi NN} = f_{\pi NN}/m_\pi$ and $\alpha$ appearing
in Eqs.~(B28)-(B37) of Ref.~\cite{knls13} as follows,
\begin{equation}
g_{p} = - g_{\pi NN}, \qquad \alpha^D_{p} = \alpha .
\end{equation}
In this work, we fix the value of $g_{p}$ by $f_{\pi NN} = \sqrt{4\pi \times 0.08}$
as in Ref.~\cite{knls13}, while $\alpha^D_{p}$ is varied freely in the fits.

\subsection{$VBB'$ interaction}

The Lagrangian for the $VBB'$ interaction is expressed as
\begin{equation}
L_{VBB'} = L^{\rm L}_{VBB'} \times L^{\rm F}_{VBB'} +[{\rm H.c.~for~B \not= B'}].
\end{equation}
The Lorentz part is given by
\begin{equation}
L_{VBB'}^{\rm L} = \bar B 
\left[\not\!V - \frac{\kappa_{VBB'}}{m_B+m_B'}\sigma^{\mu\nu} (\partial_\nu V_\mu)\right] B'.
\end{equation}
The flavor part of the $VBB'$ Lagrangian has the exactly same structure as that of $PBB'$,
and it is obtained by making the replacement of 
$\pi \to \rho$, $K \to K^*$, $\bar K \to \bar K^*$, $\eta \to \omega_8$, 
$g_{p} \to g_{v}$, and $\alpha^D_{p} \to \alpha^D_{v}$. 
Here $\omega_8$ is the eighth component of the octet representation of the vector mesons,
and $g_v$ and $\alpha^D_{v}$ are the counterpart of $g_{p}$ and $\alpha^D_{p}$ in 
the $PBB'$ interaction, respectively. 
Assuming the ideal mixing, $\omega_8$ is related to the physical $\omega$ and $\phi$ mesons as
\begin{equation}
\omega_8 = \frac{1}{\sqrt{3}}\omega - \sqrt{\frac{2}{3}}\phi .
\end{equation}
In this work, $g_v$, $\alpha^D_v$ and $\kappa_{VBB'}$ with 
$VBB' = \rho NN$, $\omega NN$, $\phi NN$, $\rho \Xi\Xi$, 
$\bar K^* N\Lambda$, $K^* \Xi\Lambda$, $K^* \Xi\Sigma$ are parameters determined by the fits.
The other $\kappa_{VBB'}$ are fixed by $\kappa_{VBB'}/(m_B+m_{B'}) = \kappa_{\rho NN}/(2m_N)$.
Note that $g_v$ is related to the $\rho NN$ coupling constant, $g_{\rho NN}$ in Ref.~\cite{knls13},
as $g_v = -g_{\rho NN}$.

\subsection{$SBB'$ interaction}

The Lagrangian for the $SBB'$ interaction is expressed as
\begin{equation}
L_{SBB'} = L^{\rm L}_{SBB'} \times L^{\rm F}_{SBB'} +[{\rm H.c.~for~B \not= B'}].
\end{equation}
The Lorentz part is given by
\begin{equation}
L_{SBB'}^{\rm L} = \bar B  B' S .
\end{equation}
As is the case in the $VBB'$ interaction, the flavor part of the $SBB'$ Lagrangian
also has the same structure as that of $PBB'$, and it is obtained by
making the replacement of $\pi \to S_{3,4,5}$, $K\to \kappa$, $\bar K \to \bar \kappa$,
$\eta \to S_{8}$, $g_p \to g_s$, and $\alpha^D_p\to\alpha^D_s$.
Assuming again the ideal mixing, the eighth component of 
the scalar-octet meson, $S_8$, is related to the $\sigma$ and $f_0$ mesons as
\begin{equation}
S_8 = \frac{1}{\sqrt{3}}\sigma - \sqrt{\frac{2}{3}}f_0 .
\end{equation}
The value of $g_s$ and $\alpha^D_s$, which are the counterpart of
$g_p$ and $\alpha^D_p$ in the $PBB'$ interaction, are determined by the fits.
With the above definition of $\sigma$ meson, 
the $\sigma NN$ coupling constant, $g_{\sigma NN}$ in Ref.~\cite{knls13},
is related to $g_s$ and $\alpha^D_s$ as $g_{\sigma NN} = g_{S_8 NN}/\sqrt{3} = -g_s(3-4\alpha^D_s )/3$.

\subsection{$PBD$ interaction}

The Lagrangian for the $PBD$ interaction is expressed as
\begin{equation}
L_{PBD} = L^{\rm L}_{PBD} \times L^{\rm F}_{PBD} +{\rm H.c.}.
\end{equation}
The Lorentz part is given by
\begin{equation}
L_{PBD}^{\rm L} = - \bar D^\mu B \partial_\mu P .
\end{equation}
In the same manner as the $PBB'$ interaction, the flavor part of the $PBD$ 
Lagrangian is derived from the following SU(3) singlet form,
\begin{equation}
L^{\rm F}_{P_{\bf 8}B_{\bf 8}D_{\bf 10}} =
g[[D_{\bf 10}^\dag \otimes B_{\bf 8}]^{({\bf 8})} \otimes P_{\bf 8}]^{({\bf 1})}, 
\end{equation}
The necessary information for our calculation is then the following flavor matrix elements,
\begin{equation}
\bra{D}L_{PBD}^{\rm F}\ket{PB}=
\bra{D}L_{P_{\bf 8}B_{\bf 8}D_{\bf 10}}^{\rm F}\ket{PB}=
G_{D,PB} \times (I_P I_P^z, I_{B} I_{B}^z|I_D I_D^z) , 
\label{eq:pbd-mat1}
\end{equation}
\begin{equation}
\bra{PB}L_{PBD}^{{\rm F}\dag}\ket{D}=
\bra{PB}L_{P_{\bf 8}B_{\bf 8}D_{\bf 10}}^{{\rm F}\dag}\ket{D}=
G_{PB,D} \times (I_P I_P^z, I_{B} I_{B}^z|I_D I_D^z) , 
\label{eq:pbd-mat2}
\end{equation}
\begin{equation}
\bra{PD}L_{PBD}^{\rm F}\ket{B}=
\bra{PD}L_{P_{\bf 8}B_{\bf 8}D_{\bf 10}}^{\rm F}\ket{B}=
G_{PD,B} \times (I_P I_P^z, I_{D} I_{D}^z|I_B I_B^z) , 
\label{eq:pbd-mat3}
\end{equation}
\begin{equation}
\bra{B}L_{PBD}^{{\rm F}\dag}\ket{PD}=
\bra{B}L_{P_{\bf 8}B_{\bf 8}D_{\bf 10}}^{{\rm F}\dag}\ket{PD}=
G_{B,PD} \times (I_P I_P^z, I_{D} I_{D}^z|I_B I_B^z) , 
\label{eq:pbd-mat4}
\end{equation}
with
\begin{equation}
G_{D,PB} =
(-1)^{-(Y_B/2)+I_B-1}
\frac{1}{\sqrt{8}}\sqrt{\frac{2I_P+1}{2I_D+1}}
g
\left(
\begin{array}{cc|c}
10^* & 8 & 8 \\
I_D -Y_D & I_B Y_B & I_P -Y_P
\end{array}
\right),
\end{equation}
\begin{equation}
G_{PD,B} =
(-1)^{-(Y_D/2)+I_D-1}
\frac{1}{\sqrt{8}}
\sqrt{\frac{2I_P+1}{2{I_B+1}}}
g
\left(
\begin{array}{cc|c}
10^* & 8 & 8 \\
I_D -Y_D & I_B Y_B & I_P  Y_P
\end{array}
\right),
\end{equation}
\begin{equation}
G_{PB,D} = G_{D,PB}, \qquad G_{B,PD} = G_{PD,B}.
\end{equation}
Within the above SU(3) symmetric Lagrangian,
all of the $PBD$ coupling constants are specified by the single parameter $g$.
In this work, the value of $g_{pbd}\equiv - g/(2\sqrt{5})$, which corresponds
to $f_{\pi N\Delta}/m_\pi$ in Ref.~\cite{knls13}, is determined by the fits.

\subsection{$VBD$ interaction}

The Lagrangian for the $VBD$ interaction is expressed as
\begin{equation}
L_{VBD} = L^{\rm L}_{VBD} \times L^{\rm F}_{VBD} +{\rm H.c.}.
\end{equation}
The Lorentz part is given by
\begin{equation}
L_{VBD}^{\rm L} = -i \bar D^\mu \gamma^\nu\gamma_5B (\partial_\mu V_\nu - \partial_\nu V_\mu) .
\end{equation}
As in the case of $PBB'$ and $VBB'$ interactions,
the flavor part of the $VBD$ Lagrangian is given by that of $PBD$
with the replacement of 
$\pi \to \rho$, $K \to K^*$, $\bar K \to \bar K^*$, $\eta \to \omega_8$, 
and $g \to \bar g$.
In this work, the value of $g_{vbd}\equiv -\bar g/(2\sqrt{5})$, which corresponds
to $f_{\rho N\Delta}/m_\rho$ in Ref.~\cite{knls13}, is determined by the fits.

\subsection{$PDD'$ interaction}

The Lagrangian for the $PDD'$ interaction is expressed as
\begin{equation}
L_{PDD'} = L^{\rm L}_{PDD'} \times L^{\rm F}_{PDD'} + [{\rm H.c.~for~}D\not=D'].
\end{equation}
The Lorentz part is given by
\begin{equation}
L_{PDD'}^{\rm L} = + \bar D^\mu \gamma^\nu \gamma_5 D'_\mu \partial_\nu P .
\end{equation}
The flavor part of the $PDD'$ Lagrangian is derived from the following SU(3) singlet form,
\begin{equation}
L^{\rm F}_{P_{\bf 8}D_{\bf 10}D'_{\bf 10}} =
g'[[D_{\bf 10}^\dag \otimes D'_{\bf 10}]^{({\bf 8})} \otimes P_{\bf 8}]^{({\bf 1})} .
\end{equation}
The necessary information for our calculation is then the following flavor matrix elements,
\begin{equation}
\bra{D}L_{PDD'}^{\rm F}\ket{PD'}=
\bra{D}L_{P_{\bf 8}D_{\bf 10}D'_{\bf 10}}^{\rm F}\ket{PD'}=
G_{D,PD'} \times (I_P I_P^z, I_{D'} I_{D'}^z|I_D I_D^z) , 
\label{eq:pdd-mat1}
\end{equation}
\begin{equation}
\bra{PD}L_{PDD'}^{\rm F}\ket{D'}=
\bra{PD}L_{P_{\bf 8}D_{\bf 10}D'_{\bf 10}}^{\rm F}\ket{D'}=
G_{PD,D'} \times (I_P I_P^z, I_{D} I_{D}^z|I_{D'} I_{D'}^z) , 
\label{eq:pdd-mat2}
\end{equation}
with
\begin{equation}
G_{D,PD'} =
(-1)^{-(Y_{D'}/2)+I_{D'}-1}
\frac{1}{\sqrt{8}}
\sqrt{\frac{2I_P+1}{2I_D+1}}
g'
\left(
\begin{array}{cc|c}
10^* & 10 & 8 \\
I_D -Y_D & I_{D'} Y_{D'} & I_P -Y_P
\end{array}
\right),
\end{equation}
and $G_{PD,D'}= G_{D,PD'}$.
In this work, we freely vary $g_{pdd}$, which is defined by 
$g_{pdd} = -g'/(2\sqrt{15})$
and corresponds to $f_{\pi\Delta\Delta}/m_\pi$ in Ref.~\cite{knls13}, in the fits.

\subsection{$VPP'$ interaction}

The Lagrangian for the $VPP'$ interaction is expressed as
\begin{equation}
L_{VPP'} = L^{\rm L}_{VPP'} \times L^{\rm F}_{VPP'} .
\end{equation}
The Lorentz part is given by
\begin{equation}
L_{VPP'}^{\rm L} = i P (\partial_\mu P') V^\mu .
\end{equation}
The flavor part of the $VPP'$ Lagrangian is derived from the following SU(3) singlet form,
\begin{equation}
L^{\rm F}_{V_{\bf 8}P_{\bf 8}P'_{\bf 8}} =
g''[[P_{\bf 8} \otimes P'_{\bf 8}]^{({\bf 8}_2)} \otimes V_{\bf 8}]^{({\bf 1})}. 
\end{equation}
Let us evaluate the matrix elements for the flavor part that are needed for our calculation.
We first consider the $VP_1 \to P_2$ and $P_1 \to VP_2$ transitions.
For the case that the operator $P'$ ($P$) contracts with 
the $P_1$ ($P_2$) meson in the ket (bra) state, we have
\begin{equation}
\bra{P_2}L_{VPP'}^{\rm F}\ket{VP_1}=
\bra{P_2}L_{V_{\bf 8}P_{\bf 8}P'_{\bf 8}}^{\rm F}\ket{VP_1}=
G_{P_2,VP_1} \times (I_V I_V^z, I_{P_1} I_{P_1}^z|I_{P_2} I_{P_2}^z) , 
\label{eq:vpp-mat1}
\end{equation}
\begin{equation}
\bra{VP_2}L_{VPP'}^{\rm F}\ket{P_1}=
\bra{VP_2}L_{V_{\bf 8}P_{\bf 8}P'_{\bf 8}}^{\rm F}\ket{P_1}=
G_{VP_2,P_1} \times (I_V I_V^z, I_{P_2} I_{P_2}^z|I_{P_1} I_{P_1}^z) , 
\label{eq:vpp-mat2}
\end{equation}
with
\begin{equation}
G_{P_2,VP_1} =
(-1)^{-(Y_{P_1}/2)+I_{P_1}-1}
\frac{1}{\sqrt{8}}
\sqrt{\frac{2I_V+1}{2I_{P_2}+1}}
g''
\left(
\begin{array}{cc|c}
8 & 8 & 8_{(2)} \\
I_{P_2} -Y_{P_2} & I_{P_1} Y_{P_1} & I_V -Y_V
\end{array}
\right),
\end{equation}
and $G_{VP_2,P_1} = G_{P_2,VP_1}$.
On the other hand, for the case that the operator $P$ ($P'$) contracts with 
the $P_1$ ($P_2$) meson in the ket (bra) state, we have
\begin{equation}
\bra{P_2}L_{VPP'}^{\rm F}\ket{VP_1}=
\bra{P_2}L_{V_{\bf 8}P_{\bf 8}P'_{\bf 8}}^{\rm F}\ket{VP_1}=
G'_{P_2,VP_1} \times (I_V I_V^z, I_{P_1} I_{P_1}^z|I_{P_2} I_{P_2}^z) , 
\label{eq:vpp-mat3}
\end{equation}
\begin{equation}
\bra{VP_2}L_{VPP'}^{\rm F}\ket{P_1}=
\bra{VP_2}L_{V_{\bf 8}P_{\bf 8}P'_{\bf 8}}^{\rm F}\ket{P_1}=
G'_{VP_2,P_1} \times (I_V I_V^z, I_{P_2} I_{P_2}^z|I_{P_1} I_{P_1}^z) , 
\label{eq:vpp-mat4}
\end{equation}
with $G'_{P_2,VP_1} = -G_{P_2,VP_1}$ and $G'_{VP_2,P_1} = -G_{VP_2,P_1}$.

Next consider the $V\to P_1 P_2$ and $P_1 P_2 \to V$ transitions.
For the case that the operator $P$ ($P'$) contracts with 
the $P_1$ ($P_2$) meson, we have
\begin{equation}
\bra{V}L_{VPP'}^{\rm F}\ket{P_1P_2}=
\bra{V}L_{V_{\bf 8}P_{\bf 8}P'_{\bf 8}}^{\rm F}\ket{P_1P_2}=
G_{V,P_1P_2} \times (I_{P_1} I_{P_1}^z, I_{P_2} I_{P_2}^z|I_{V} I_{V}^z) , 
\label{eq:vpp-mat5}
\end{equation}
\begin{equation}
\bra{P_1P_2}L_{VPP'}^{\rm F}\ket{V}=
\bra{P_1P_2}L_{V_{\bf 8}P_{\bf 8}P'_{\bf 8}}^{\rm F}\ket{V}=
G_{P_1P_2,V} \times (I_{P_1} I_{P_1}^z, I_{P_2} I_{P_2}^z|I_{V} I_{V}^z) , 
\label{eq:vpp-mat6}
\end{equation}
with
\begin{equation}
G_{V,P_1 P_2} =
-\frac{1}{\sqrt{8}}
g''
\left(
\begin{array}{cc|c}
8 & 8 & 8_{(2)} \\
I_{P_1}  Y_{P_1} & I_{P_2} Y_{P_2} & I_V  Y_V
\end{array}
\right),
\end{equation}
and $G_{P_1P_2,V} = -G_{V,P_1P_2}$.
For the case that the operator $P$ ($P'$) contracts with 
the $P_2$ ($P_1$) meson, however, we have
\begin{equation}
\bra{V}L_{VPP'}^{\rm F}\ket{P_1P_2}=
\bra{V}L_{V_{\bf 8}P_{\bf 8}P'_{\bf 8}}^{\rm F}\ket{P_1P_2}=
G'_{V,P_1P_2} \times (I_{P_1} I_{P_1}^z, I_{P_2} I_{P_2}^z|I_{V} I_{V}^z) , 
\label{eq:vpp-mat7}
\end{equation}
\begin{equation}
\bra{P_1P_2}L_{VPP'}^{\rm F}\ket{V}=
\bra{P_1P_2}L_{V_{\bf 8}P_{\bf 8}P'_{\bf 8}}^{\rm F}\ket{V}=
G'_{P_1P_2,V} \times (I_{P_1} I_{P_1}^z, I_{P_2} I_{P_2}^z|I_{V} I_{V}^z) , 
\label{eq:vpp-mat8}
\end{equation}
with $G'_{V,P_1P_2} = -G_{V,P_1P_2}$ and $G'_{P_1P_2,V} = -G_{P_1P_2,V}$.
The parameter $g_{vpp} = (-2)\times(\sqrt{6}/24)g''$,
which corresponds to $g_{\rho\pi\pi}$ in Ref.~\cite{knls13},
is varied freely and determined by the fits.

\subsection{$SPP'$ interaction}

The Lagrangian for the $SPP'$ interaction is expressed as
\begin{equation}
L_{SPP'} = L^{\rm L}_{SPP'} \times L^{\rm F}_{SPP'} .
\end{equation}
The Lorentz part is given by
\begin{equation}
L_{SPP'}^{\rm L} = - (\partial^\mu P) (\partial_\mu P') S .
\end{equation}
The flavor part of the $SPP'$ Lagrangian is derived from the following SU(3) singlet form,
\begin{equation}
L^{\rm F}_{S_{\bf 8}P_{\bf 8}P'_{\bf 8}} =
g'''[[P_{\bf 8} \otimes P'_{\bf 8}]^{({\bf 8}_1)} \otimes S_{\bf 8}]^{({\bf 1})}. 
\label{eq:spp-full-flavor}
\end{equation}
The necessary information for our calculation is the following matrix elements:
\begin{equation}
\bra{P_2}L_{SPP'}^{\rm F}\ket{SP_1}=
\bra{P_2}L_{S_{\bf 8}P_{\bf 8}P'_{\bf 8}}^{\rm F}\ket{SP_1}=
G_{P_2,SP_1} \times (I_S I_S^z, I_{P_1} I_{P_1}^z|I_{P_2} I_{P_2}^z) , 
\label{eq:spp-mat1}
\end{equation}
\begin{equation}
\bra{SP_2}L_{SPP'}^{\rm F}\ket{P_1}=
\bra{SP_2}L_{S_{\bf 8}P_{\bf 8}P'_{\bf 8}}^{\rm F}\ket{P_1}=
G_{SP_2,P_1} \times (I_S I_S^z, I_{P_2} I_{P_2}^z|I_{P_1} I_{P_1}^z) , 
\label{eq:spp-mat2}
\end{equation}
with
\begin{align}
G_{P_2,SP_1}& =
2\times (-1)^{-(Y_{P_1}/2)+I_{P_1}-1}
\frac{1}{\sqrt{8}}
\sqrt{\frac{2I_S+1}{2I_{P_2}+1}}
g'''
\left(
\begin{array}{cc|c}
8 & 8 & 8_{(1)} \\
I_{P_2} -Y_{P_2} & I_{P_1} Y_{P_1} & I_S -Y_S
\end{array}
\right),
\label{eq:spp-g}
\end{align}
and $ G_{SP_2,P_1} = G_{P_2,SP_1}$.
Note that the factor 2 appears in the right-hand side of Eq.~(\ref{eq:spp-g}).
For $P_1 \not= P_2$, it arises from the fact that
the full flavor-part-Lagrangian~(\ref{eq:spp-full-flavor}) contains two terms 
that can contract with a given $S$, $P_1$, and $P_2$.
For $P_1 = P_2$, however, only one term in Eq.~(\ref{eq:spp-full-flavor})
can contract with a given $S$, $P_1$, and $P_2$, but there are two ways of contractions with $P_1$ and $P_2$.
The parameter $g_{spp}\equiv (-2/3)\times (\sqrt{30}/40) g'''$, which corresponds to
$g_{\sigma\pi\pi}/(2m_\pi)$ in Ref.~\cite{knls13}, is varied freely and determined by the fits.

\subsection{Weinberg-Tomozawa interaction}
\label{app:wt-lag}

Finally, we present the so-called Weinberg-Tomozawa (WT) interaction
for the $PP'BB'$ four-point vertex, which is used for Model B.
The Lagrangian is expressed as
\begin{equation}
L_{\rm WT} = L_{\rm WT}^{\rm L} \times L_{\rm WT}^{\rm F} .
\end{equation}
The Lorentz part is given by
\begin{equation}
L_{\rm WT}^{\rm L} = iP'(\partial_\mu P)\bar B' \gamma^\mu B ,
\end{equation}
and the flavor part is derived from the following SU(3) singlet form,
\begin{equation}
L_{\rm WT;P'_{\bf 8} P_{\bf 8} B'_{\bf 8} B_{\bf 8}}^{\rm F} = 
g''''[[P'_{\bf 8} \otimes P_{\bf 8}]^{({\bf 8}_2)} 
\otimes
[B_{\bf 8}^{\prime\dag} \otimes B_{\bf 8}]^{({\bf 8}_2)}
]^{({\bf 1})}.
\label{eq:wt-f-lag}
\end{equation}
Here it is noted that the SU(3)$\times$SU(3) chiral symmetry
does not allow one to have the contribution from 
the combination of $B_{\bf 8}^{\prime\dag} \otimes B_{\bf 8} \to {\bf 8}_1$.
For the case that the operator 
$P'$ ($P$) contracts with $P_2$ ($P_1$) meson in the bra (ket) state,
we obtain the following matrix element,
\begin{eqnarray}
\bra{P_2B'}L_{\rm WT}^{\rm F}\ket{P_1B}
&=&
\bra{P_2 B'}L_{\rm WT;P'_{\bf 8} P_{\bf 8} B'_{\bf 8} B_{\bf 8}}^F \ket{P_1 B}
\nonumber\\
&=&
\sum_{T}
G^{{\rm WT}(T)}_{P_2B';P_1B} \times 
(I_{P_2} I_{P_2}^z, I_{B'} I_{B'}^z|T T^z)  
(I_{P_1} I_{P_1}^z, I_{B} I_{B}^z|T T^z)  ,
\label{eq:wt-mat1}
\end{eqnarray}
with
\begin{equation}
G^{{\rm WT}(T)}_{P_2B';P_1B} = \left(\frac{\sqrt{2}}{48} g'''' \right) \lambda^{T}_{P_2B';P_1B} ,
\end{equation}
and
\begin{align}
\lambda^{T}_{P_2B';P_1B} 
&=
(-12)(-1)^{-(Y_{B'}/2)+(Y_{P_1}/2)+I_{B'}+I_B - T}
\notag\\
&\times
\sum_{I} (2I+1)
\left(
\begin{array}{cc|c}
8              & 8                 & 8_{(2)} \\
I_{B'} -Y_{B'} & I_{B} Y_{B} & I Y
\end{array}
\right)
\left(
\begin{array}{cc|c}
8 & 8 & 8_{(2)} \\
I_{P_2} -Y_{P_2} & I_{P_1} Y_{P_1} & I -Y
\end{array}
\right)
\notag\\
&\times
W(I_{P_1} I_{P_2} I_{B} I_{B'}; I T) ,
\end{align}
where 
$T^z = I_{P_2}^z + I_{B'}^z = I_{P_1}^z + I_{B}^z $,
$Y = Y_B-Y_{B'} = Y_{P_2}-Y_{P_1} $, and
$W(abcd;ef)$ is the Racah coefficients.
On the other hand, for the case that the operator 
$P$ ($P'$) contracts with $P_2$ ($P_1$) meson in the bra (ket) state,
we have
\begin{eqnarray}
\bra{P_2B'}L_{WT}^{\rm F}\ket{P_1B}
&=&
\bra{P_2 B'}L_{\rm WT;P'_{\bf 8} P_{\bf 8} B'_{\bf 8} B_{\bf 8}}^F \ket{P_1 B}
\nonumber\\
&=&
\sum_{T}
G^{\prime WT(T)}_{P_2B';P_1B} \times 
(I_{P_2} I_{P_2}^z, I_{B'} I_{B'}^z|T T^z)  
(I_{P_1} I_{P_1}^z, I_{B} I_{B}^z|T T^z)  ,
\label{eq:wt-mat2}
\end{eqnarray}
with $G^{\prime WT(T)}_{P_2B';P_1B} = -G^{WT(T)}_{P_2B';P_1B}$.

Comparing with the leading order term of the SU(3)$\times$SU(3) chiral Lagrangian,
we find that the coupling constant $g''''$ in Eq.~(\ref{eq:wt-f-lag})
can be related to the low energy constant $f$, 
\begin{equation}
-\frac{1}{8f^2} = \frac{\sqrt{2}}{48} g'''' .
\end{equation}
The constant $f$ is known as the decay constant of 
the pseudoscalar Nambu-Goldstone bosons in the chiral limit, and in this work it is taken to be
$f = 92.4$ MeV.
Also, we multiply $-1/(8f^2)$ by a factor $\gamma_{\rm WT}$ and
vary the factor in the fits.

\section{Matrix elements of meson-baryon potentials}
\label{app:pot}

\begin{table}
\caption{\label{tab:vmbmb}
Exchanged particles considered in the potentials $v_{M'B',MB}$. 
}
\begin{ruledtabular}
\begin{tabular}{lclllllll}
             & & $\bar K N$                                & $\pi \Sigma$                    & $\pi \Lambda$        & $\eta \Lambda$            & $K \Xi$                                   & $\pi \Sigma^*$                  & $\bar K^* N$ \\ 
\hline
$\bar K N$   &$s$& $\Lambda$, $\Sigma$                       & $\Lambda$, $\Sigma$             & $\Sigma$             & $\Lambda$                 & $\Lambda$, $\Sigma$                       & $\Lambda$, $\Sigma$             & $\Lambda$, $\Sigma$  \\
             &$u$& -                                         & $N$, $\Delta$                   & $N$                  & $N$                       & $\Lambda$, $\Sigma$, $\Sigma^*$           & $N$, $\Delta$                   & -  \\
             &$t$& $\rho$, $\omega$, $\phi$, $\sigma$, $f_0$ & $K^*$, $\kappa$                 & $K^*$, $\kappa$      & $K^*$, $\kappa$           & -                                         & $K^*$                           & $\pi$ \\ 
\\
$\pi \Sigma$ &$s$&                                           & $\Lambda$, $\Sigma$             & $\Sigma$             & $\Lambda$                 & $\Lambda$, $\Sigma$                       & $\Lambda$, $\Sigma$             & $\Lambda$, $\Sigma$  \\
             &$u$&                                           & $\Lambda$, $\Sigma$, $\Sigma^*$ &$\Sigma$, $\Sigma^*$  & $\Sigma$, $\Sigma^*$      & $\Xi$, $\Xi^*$                            & $\Lambda$, $\Sigma$, $\Sigma^*$ & $N$, $\Delta$  \\
             &$t$&                                           & $\rho$, $\sigma$, $f_0$         & $\rho$               & -                         & $K^*$, $\kappa$                           & $\rho$                          & $K$ \\ 
\\
$\pi\Lambda$ &$s$&                                           &                                 & $\Sigma$             & -                         & $\Sigma$                                  & $\Sigma$                        & $\Sigma$  \\
             &$u$&                                           &                                 & $\Sigma$, $\Sigma^*$ & -                         & $\Xi$, $\Xi^*$                            & $\Sigma$, $\Sigma^*$            & $N$  \\
             &$t$&                                           &                                 & $\sigma$, $f_0$      & -                         & $K^*$, $\kappa$                           & $\rho$                          & $K$ \\ 
\\
$\eta\Lambda$&$s$&                                           &                                 &                      & $\Lambda$                 & $\Lambda$                                 & $\Lambda$                       & $\Lambda$  \\
             &$u$&                                           &                                 &                      & $\Lambda$                 & $\Xi$, $\Xi^*$                            & $\Sigma$                        & $N$  \\
             &$t$&                                           &                                 &                      & $\sigma$, $f_0$           & $\bar K^*$, $\bar \kappa$                 & -                               & $\bar K$ \\ 
\\
$K\Xi$       &$s$&                                           &                                 &                      &                           & $\Lambda$, $\Sigma$                       & $\Lambda$, $\Sigma$             & $\Lambda$, $\Sigma$  \\
             &$u$&                                           &                                 &                      &                           & $\Omega$                                  & $\Xi$, $\Xi^*$                  & $\Lambda$, $\Sigma$, $\Xi^*$ \\
             &$t$&                                           &                                 &                      &                           & $\rho$, $\omega$, $\phi$, $\sigma$, $f_0$ & $\bar K^*$                      & - \\ 
\\
$\pi \Sigma^*$& & &&&&&-&-\\
\\
$\bar K^* N$ &&&&&&&&-
\end{tabular}
\end{ruledtabular}
\end{table}

The plane-wave matrix elements for the meson-baryon exchange potentials $v_{M'B',MB}$
can be expressed as
\begin{eqnarray}
\bra{M'(k'),B'(p')}
v_{M'B',MB}
\ket{M(k),B(p)}
&=&
\frac{1}{(2\pi)^3}
\sqrt{\frac{m_{B'}}{E_{B'}(p')}}
\frac{1}{\sqrt{2E_{M'}(k')}}
\sqrt{\frac{m_{B}}{E_{B}(p)}}
\frac{1}{\sqrt{2E_{M}(k)}}
\nonumber\\
&&
\times
\sum_{T}
(I_{M'} I_{M'}^z, I_{B'} I_{B'}^z|TT^z)
(I_M I_M^z, I_B I_B^z|TT^z)
\nonumber\\
&&
\times 
V^{(T)} .
\label{eq:plane-pot}
\end{eqnarray}
The partial-wave decomposition of Eq.~(\ref{eq:plane-pot})
is explained in detail in Refs.~\cite{knls13,msl07} and is not presented here.
In the following, the explicit expressions of $V^{(T)}$
for $P + B \to P'+ B'$, $P + B \to P' + D$, and $P+ B \to V+ B'$ are presented,
while we omit those for $P' + D \to P + B$ and $V + B' \to P + B$ 
since those can be deduced from the corresponding inverse processes.

\subsection{$P(k) + B(p) \to P'(k') + B'(p')$}

\subsubsection{s-channel B exchange}

\begin{equation}
V^{1(T)}_{sB_{\rm ex}} =  
\sum_{B_{\rm ex}} 
C^{1(T)}_{sB_{\rm ex}}  
G_{P'B',B_{\rm ex}}G_{B_{\rm ex},PB}
\bar u_{B'}(p') \slas{k}' \gamma_5 S_{B_{\rm ex}}(p+k) \slas{k}\gamma_5 u_B (p),
\label{eq:pbpb-sb}
\end{equation}
\begin{equation}
C^{1(T)}_{sB_{\rm ex}} = \delta_{T I_{B_{\rm ex}}} ,
\end{equation}
where $u_B(p)$ is the Dirac spinor for the baryon $B$.
In evaluating the time component of the propagators, 
$S_{B}(p) = 1/(\not\! p - m_B)$ in the above as well as in the following, 
we follow the definite procedures defined by 
the unitary transformation method~\cite{sko,sl96}.
For more detail, see Appendix C of Ref.~\cite{msl07}.
Hereafter the particles exchanged are indicated with the subscript ``ex.''
The summation in Eq.~(\ref{eq:pbpb-sb}) runs over the 
spin-$\frac{1}{2}$ octet $B_{\rm ex}$ states listed in Table~\ref{tab:vmbmb}.

\subsubsection{u-channel B exchange}

\begin{equation}
V^{1(T)}_{uB_{\rm ex}} =  
\sum_{B_{\rm ex}} 
C^{1(T)}_{uB_{\rm ex}}  
G_{B',PB_{\rm ex}}G_{P'B_{\rm ex},B}
\bar u_{B'}(p')\slas{k} \gamma_5 S_{B_{\rm ex}}(p-k') \slas{k}'\gamma_5 u_B(p) , 
\end{equation}
\begin{equation}
C^{1(T)}_{uB_{\rm ex}} =
\sqrt{2I_B+1}\sqrt{2I_{B'}+1}
W(I_P I_{B'} I_{B} I_{P'}; I_{B_{\rm ex}} T) .
\end{equation}

\subsubsection{u-channel D exchange}

\begin{equation}
V^{1(T)}_{uD_{\rm ex}} =  
\sum_{D_{\rm ex}} 
C^{1(T)}_{uD_{\rm ex}}  
G_{B',PD_{\rm ex}}G_{P'D_{\rm ex},B}
\bar u_{B'}(p'){k}_\alpha S^{\alpha\beta}_{D_{\rm ex}}(p-k')k'_{\beta} u_B(p) , 
\end{equation}
\begin{equation}
C^{1(T)}_{uD_{\rm ex}}  =
\sqrt{2I_B+1}\sqrt{2I_{B'}+1}
W(I_P I_{B'} I_{B} I_{P'}; I_{D_{\rm ex}} T) ,
\end{equation}
where $S^{\alpha\beta}_{D_{\rm ex}}(p-k')$ is the propagator for the spin-$\frac{3}{2}$ Rarita-Schwinger field~\cite{knls13}.

\subsubsection{t-channel V exchange}
\label{app:pbpb-t-v}

\begin{eqnarray}
V^{1(T)}_{tV_{\rm ex}} &=& 
\sum_{V_{\rm ex}} 
C^{1(T)}_{tV_{\rm ex}}  
G_{P',V_{\rm ex}P}G_{V_{\rm ex}B',B}
\frac{-1}{q^2 - m_{V_{\rm ex}}^2}
\nonumber\\
&&
\times
\bar u_{B'}(p')
\left[
(\slas{k}+\slas{k}')+
\frac{\kappa_{VBB'}}{2(m_B + m_{B'})}
\{(\slas{k}+\slas{k}^\prime)\slas{q}
-\slas{q}(\slas{k}+\slas{k}^\prime)\}
\right] u_B(p) ,
\label{eq:1tv}
\end{eqnarray}
\begin{equation}
C^{1(T)}_{tV_{\rm ex}}  
 = (-1)^{I_B + I_P - T}
\sqrt{2I_{P'}+1}\sqrt{2I_{B}+1}
W(I_P I_{P'} I_{B} I_{B'}; I_{V_{\rm ex}} T) ,
\end{equation}
where the momentum transfer $q$ is defined by $q = k' - k$ or $q = p - p'$.

\subsubsection{t-channel S exchange}

\begin{equation}
V^{1(T)}_{tS_{\rm ex}} = 
\sum_{S_{\rm ex}} 
C^{1(T)}_{tS_{\rm ex}}  
G_{P',S_{\rm ex}P}G_{B'S_{\rm ex},B}
\frac{-k\cdot k'}{q^2-m_{S_{\rm ex}}^2} 
\bar u_{B'}(p') u_B(p)  ,
\end{equation}
\begin{equation}
C^{1(T)}_{tS_{\rm ex}}  
= (-1)^{I_B + I_P - T}
\sqrt{2I_{P'}+1}\sqrt{2I_{B}+1}
W(I_P I_{P'} I_{B} I_{B'}; I_{S_{\rm ex}} T) .
\end{equation}

\subsubsection{Modified t-channel V exchange used for Model B}
\label{app:pbpb-t-v-2}

It is known that in the $q \to 0$ limit the Lorentz structure of the 
vector-meson-exchange potentials~(\ref{eq:1tv}) reduces to the one
known as the WT interaction.
Recent studies suggest that this contact interaction plays an important role for understanding
the near-threshold phenomena in $S$-wave (see, e.g., Ref.~\cite{ucm-1}).
To make a clear connection with such studies,
in Model B we employ a modified vector-meson exchange potentials, 
instead of using the one described in Appendix~\ref{app:pbpb-t-v}.
The explicit form is,
\begin{equation}
V^{1(T)}_{tV_{\rm ex},{\rm mod}} = V^{(T)}_{\rm WT} + \bar V^{1(T)}_{tV_{\rm ex}}. 
\label{eq:1tvmod}
\end{equation}
Here, $V^{(T)}_{\rm WT}$ is the contribution from the WT term
described in Appendix~\ref{app:wt-lag},
\begin{equation}
V^{(T)}_{\rm WT} = C_{\rm WT}^{(T)}\left(-\frac{\gamma_{\rm WT}}{8f^2}\right)
\bar u_{B'}(p') (\slas{k}+\slas{k}') u_B(p) ,
\end{equation}
\begin{equation}
C_{\rm WT}^{(T)} = \lambda^T_{P'B';PB} .
\end{equation}
The second term in the right hand side of Eq.~(\ref{eq:1tvmod}), $\bar V^{1(T)}_{tV_{\rm ex}}$,
is same as $V^{1(T)}_{tV_{\rm ex}}$ but the following term is subtracted,
\begin{equation}
V^{1(T)}_{tV_{\rm ex}}|_{q\to 0} =
\sum_{V_{\rm ex}} 
C_{tV_{\rm ex}}^{1(T)}
G_{P',V_{\rm ex}P}G_{B'V_{\rm ex},B}
\left(\frac{1}{m_{V_{\rm ex}}^2}\right)
\bar u_{B'}(p') (\slas{k}+\slas{k}') u_B(p).
\end{equation}
This term corresponds to the WT term
in terms of the resonance saturation, 
and thus should be subtracted to avoid the double counting.
Here we note that for $V^{(T)}_{\rm WT}$ we attach the following combination of the form factors:
\begin{equation}
c^{\rm WT}_{P'B',PB}F(\vec q,\Lambda^{\rm WT}_{P'B',PB}) F(\vec q,\Lambda^{\rm WT}_{P'B',PB})
+
(1-c^{\rm WT}_{P'B',PB})F(\vec k',\Lambda^{\rm WT}_{P'B',PB}) F(\vec k,\Lambda^{\rm WT}_{P'B',PB}),
\end{equation}
where $\vec q = \vec k' - \vec k$, $F(\vec k,\Lambda)$ is defined in Eq.~(\ref{eq:ff}), 
and the coefficients $c^{\rm WT}_{P'B',PB}$ and cutoffs $\Lambda^{\rm WT}_{P'B',PB}$,
which satisfy $c^{\rm WT}_{PB,P'B'} =c^{\rm WT}_{P'B',PB}$ and $\Lambda^{\rm WT}_{PB,P'B'}=\Lambda^{\rm WT}_{P'B',PB}$,
respectively, are determined by the fits.

\subsection{$P(k) + B(p) \to P'(k') + D(p')$}

\subsubsection{s-channel B exchange}

\begin{equation}
V^{2(T)}_{sB_{\rm ex}} = 
\sum_{B_{\rm ex}} 
C^{2(T)}_{sB_{\rm ex}} 
G_{P'D,B_{\rm ex}}G_{B_{\rm ex},PB}
\bar U_{D}^\mu(p') k'_\mu S_{B_{\rm ex}}(p+k)\slas{k}\gamma_5  u_B(p), 
\end{equation}
\begin{equation}
C^{2(T)}_{sB_{\rm ex}} = \delta_{T I_{B_{\rm ex}}} ,
\end{equation}
where $U_{D}^\mu(p)$ is the Rarita-Schwinger vector-spinor for spin-$\frac{3}{2}$ baryon $D$ with 
the momentum $p$.

\subsubsection{u-channel B exchange}

\begin{equation}
V^{2(T)}_{uB_{\rm ex}} = 
\sum_{B_{\rm ex}} 
C^{2(T)}_{uB_{\rm ex}} 
G_{D,PB_{\rm ex}}G_{P'B_{\rm ex},B}
\bar U_{D}^\mu (p')k_\mu S_{B_{\rm ex}}(p-k')\slas{k}'\gamma_5 u_B(p),
\end{equation}
\begin{equation}
C^{2(T)}_{uB_{\rm ex}}
=
\sqrt{2I_B+1}\sqrt{2I_{D}+1}
W(I_P I_{D} I_{B} I_{P'}; I_{B_{\rm ex}} T) .
\end{equation}

\subsubsection{u-channel D exchange}

\begin{equation}
V^{2(T)}_{uD_{\rm ex}} = 
\sum_{D_{\rm ex}} 
C^{2(T)}_{uD_{\rm ex}} 
G_{D,PD_{\rm ex}}G_{P'D_{\rm ex},B}
(-1)
\bar U_{D\mu} (p')\slas{k}\gamma_5 
 S^{\mu\nu}_{D_{\rm ex}}(p- k') k'_\nu u_B(p),
\end{equation}
\begin{equation}
C^{2(T)}_{uD_{\rm ex}} =
\sqrt{2I_B+1}\sqrt{2I_{D}+1}
W(I_P I_{D} I_{B} I_{P'}; I_{D_{\rm ex}} T) .
\end{equation}

\subsubsection{t-channel V exchange}

\begin{equation}
V^{2(T)}_{tV_{\rm ex}} = 
\sum_{V_{\rm ex}} 
C^{2(T)}_{tV_{\rm ex}} 
G_{P',V_{\rm ex}P}G_{V_{\rm ex}D,B}
\frac{1}{q^2-m^2_{V_{\rm ex}}}
\bar U_{D}^\mu(p')
\left[
q_\mu (\slas{k}+\slas{k}^\prime)\gamma_5
-(k+k^\prime)_\mu \slas{q}\gamma_5
\right]
u_B(p) , 
\end{equation}
\begin{equation}
C^{2(T)}_{tV_{\rm ex}} 
= (-1)^{I_B + I_P - T}
\sqrt{2I_{P'}+1}\sqrt{2I_{B}+1}
W(I_P I_{P'} I_{B} I_{D}; I_{V_{\rm ex}} T) .
\end{equation}

\subsection{$P(k) + B(p) \to V(k') + B'(p')$}

\subsubsection{s-channel B exchange}

\begin{equation}
V^{3(T)}_{sB_{\rm ex}} = 
\sum_{B_{\rm ex}} 
C^{3(T)}_{sB_{\rm ex}} 
i G_{VB',B_{\rm ex}}G_{B_{\rm ex},PB}
\bar u_{B'}(p')
\Gamma_{VB_{\rm ex}B'}
 S_{B_{\rm ex}}(p+k) \slas{k}\gamma_5 u_B(p) , 
\end{equation}
\begin{equation}
C^{3(T)}_{sB_{\rm ex}} 
 = \delta_{T I_{B_{\rm ex}}} .
\end{equation}
Here we have introduced
\begin{equation}
\Gamma_{VB_{\rm ex} B'} =
\left[
\slas{\epsilon_{V}}^*+
\frac{\kappa_{VB_{\rm ex}B'}}{2(m_{B_{\rm ex}}+m_{B'})}
(\slas{\epsilon_{V}}^*\slas{k}^\prime
-\slas{k}^\prime\slas{\epsilon_{V}}^*)
\right]  ,
\end{equation}
and $\epsilon_V^\mu$ is the polarization vector of the vector meson $V$.

\subsubsection{u-channel B exchange}

\begin{equation}
V^{3(T)}_{uB_{\rm ex}} = 
\sum_{B_{\rm ex}} 
C^{3(T)}_{uB_{\rm ex}} 
i
G_{B',PB_{\rm ex}}G_{VB_{\rm ex},B}
\bar u_{B'}(p')
\slas{k}\gamma_5 S_{B_{\rm ex}}(p-k') 
\Gamma_{VBB_{\rm ex}}
u_{B}(p) ,
\end{equation}
\begin{equation}
C^{3(T)}_{uB_{\rm ex}} 
=
\sqrt{2I_B+1}\sqrt{2I_{B'}+1}
W(I_P I_{B'} I_{B} I_{V}; I_{B_{\rm ex}} T) .
\end{equation}

\subsubsection{t-channel P exchange}

\begin{equation}
V^{3(T)}_{tP_{\rm ex}} = 
\sum_{P_{\rm ex}} 
C^{3(T)}_{tP_{\rm ex}} 
i G_{V,P_{\rm ex}P}G_{P_{\rm ex}B',B}
\frac{1}{q^2-m_{P_{\rm ex}}^2}
\bar u_{B'}(p')
(q-k)\cdot\epsilon_{V}^* \slas{q}\gamma_5
u_B(p) ,
\end{equation}
\begin{equation}
C^{3(T)}_{tP_{\rm ex}} 
=
 (-1)^{I_B + I_P - T}
\sqrt{2I_{V}+1}\sqrt{2I_{B}+1}
W(I_P I_{V} I_{B} I_{B'}; I_{P_{\rm ex}} T) .
\end{equation}

\section{Self-energies in meson-baryon Green functions}
\label{app:self}

In this appendix, we give an expression of the self-energy $\Sigma_{MB}(k;W)$
appearing in the meson-baryon Green's function [Eq.~(\ref{eq:prop-unstab})] for 
the unstable channels $MB=\pi\Sigma^*, \bar K^* N$.
The self-energies are explicitly given by~\cite{kjlms09-1}
\begin{equation}
\Sigma_{\pi \Sigma^*}(k;W) = \frac{m_{\Sigma^*}}{E_{\Sigma^*}(k)}
\int_{C_3} q^2 dq \frac{ M_{\pi \Lambda}(q)}{[M^2_{\pi \Lambda}(q) + k^2]^{1/2}}
\frac{\left|f_{\pi\Lambda, \Sigma^*}(q)\right|^2}
{W-E_\pi(k) -[M^2_{\pi \Lambda}(q) + k^2]^{1/2} + i\epsilon},
\label{eq:self-pisstar}
\end{equation}
\begin{equation}
\Sigma_{\bar K^* N}(k;W) = \frac{m_{\bar K^*}}{E_{\bar K^*}(k)}
\int_{C_3} q^2 dq \frac{ M_{\pi\bar K}(q)}{[M^2_{\pi \bar K}(q) + k^2]^{1/2}}
\frac{\left|f_{\pi \bar K,\bar K^*}(q)\right|^2}
{W-E_N(k) -[M^2_{\pi \bar K}(q) + k^2]^{1/2} + i\epsilon},
\label{eq:self-kstarn}
\end{equation}
where $M_{MB}(q) = E_M (q) + E_B(q)$, and the momentum integral path $C_3$ is chosen
appropriately when one makes an analytic continuation of the scattering amplitudes.

The form factors $f_{\pi \Lambda,\Sigma^*}(q)$ and $f_{\pi \bar K, \bar K^*}(q)$ are for describing
the $\Sigma^*\to\pi \Lambda$ and $\bar K^* \to \pi \bar K$ decays 
in the $\Sigma^*$ and $\bar K^*$ rest frames,
respectively.
Those are parametrized as
\begin{equation}
f_{\pi \Lambda,\Sigma^*}(q) = -i\frac{\bar g_{\pi\Lambda\Sigma^*}}{(2\pi)^{3/2}}
  \sqrt{\frac{1}{2E_\pi(q)}}\sqrt{\frac{E_\Lambda(q)+m_\Lambda}{2E_\Lambda(q)}} 
\left( \frac{q}{m_\pi} \right)
  \left(\frac{\bar\Lambda_{\pi\Lambda\Sigma^*}^2}{\bar\Lambda_{\pi\Lambda\Sigma^*}^2 + q^2}\right)^2
\sqrt{\frac{4\pi}{3}}.
\end{equation}
\begin{equation}
f_{\pi\bar K,\bar K^\ast}(q) =
\frac{\bar g_{\pi\bar K\bar K^\ast} }{\sqrt{m_\pi}}
\left(\frac{q}{m_\pi}\right)
\left(\frac{\bar\Lambda_{\bar K^\ast \bar K \pi}^2}{\bar\Lambda_{\bar K^\ast \bar K\pi}^2 + q^2}\right)^{3/2}.
\end{equation}
The parameters associated with $f_{\pi \Lambda,\Sigma^*}(q)$ are determined such that
the pole mass of the decuplet $\Sigma^*$ baryon, $1381 -i20$ MeV~\cite{sstar-mass}, is reproduced.
The resulting value for the parameters are 
$m_{\Sigma^*} = 1435.2$ MeV, $\bar g_{\pi\Lambda\Sigma^*}=1.753$, and $\bar \Lambda_{\pi\Lambda\Sigma^*}=650$ MeV (fixed).
The parameters associated with $f_{\pi \bar K,\bar K^*}(q)$ are determined by fitting to 
the $\pi K$ scattering phase shift~\cite{pik} for the isospin $1/2$ and $P$ wave.
We then obtain
$m_{\bar K^*} = 930.4$ MeV, $\bar g_{\pi\bar K,\bar K^*}=-0.152$, and $\bar \Lambda_{\pi\bar K,\bar K^*}=341$ MeV.
With these parameters, we find the $\bar K^*$ pole mass becomes $899.3 -i29.7$ MeV.

Here it is noted that the decuplet $\Sigma^*$ baryon can decay also 
to $\pi\Sigma$ channel via the strong interaction,
although its decay ratio is known to be much smaller than 
the dominant $\pi\Lambda$ channel~\cite{pdg2012}.
As a first step, we only consider the $\Sigma^* \to \pi \Lambda$ decay in this work,
and the contribution of the $\Sigma^* \to \pi\Sigma$ process will be taken into account
in our future development.

\section{Model parameters}
\label{app:model-para}

In this appendix, we list the values of model parameters determined via our
analysis of the unpolarized and polarized observables of 
$K^- p \to \bar K N, \pi \Sigma, \pi \Lambda, \eta\Lambda, K \Xi$ up to $W= 2.1$ GeV.
The channel masses are presented in Table~\ref{tab:chn-mass}.
The parameters associated with the exchange potentials are listed
in Tables~\ref{tab:model-para-mass}-~\ref{tab:nres-wt},
while those associated with the bare $Y^*$ states are listed in 
Tables~\ref{tab:res-mass}-~\ref{tab:res-mb-modelb}.

\begin{table}
\caption{\label{tab:chn-mass}
Channel masses that are used for the meson-baryon Green's functions~(\ref{eq:prop-stab}) 
and~(\ref{eq:prop-unstab}), and for external particles in the exchange potentials $v_{M'B',MB}$.
}
\begin{ruledtabular}
\begin{tabular}{lr}
Masses & (MeV) \\
\hline
$m_N$ & 938.5 \\
$m_\Sigma$ & 1193.2 \\
$m_\Lambda$ & 1115.7 \\
$m_\Xi$ & 1318.3 \\
$m_\Sigma^*$ & 1435.2 \\
$m_{\pi}$ & 138.5 \\
$m_{\eta}$ & 547.9 \\
$m_{\bar K}$ & 493.7 \\
$m_{\bar K^*}$ & 930.4 
\end{tabular}
\end{ruledtabular}
\end{table}

\begin{table}
\caption{\label{tab:model-para-mass}
Masses used for the exchange particles in $v_{M'B',MB}$.
All masses are kept constant during the fits
except for the $\sigma$ ($m_\sigma$), $f_0$ ($m_{f_0}$), and $\kappa$ ($m_\kappa$) masses.
}
\begin{ruledtabular}
\begin{tabular}{lrlrr}
Masses &       & Masses & Model A & Model B\\
&(MeV) &&(MeV)&(MeV) \\
\hline
$m_N$  & 938.5  & $m_\sigma$ & 310.1& 310.0\\
$m_\Lambda$ & 1115.7& $m_{f_0}$ & 760.5& 950.0\\
$m_\Sigma$ & 1193.2 & $m_\kappa$ & 1150.0&1018.7\\
$m_{\Xi}$&    1318.3 &&& \\
$m_{\Delta}$&    1211.0&&&  \\
$m_{\Sigma^*}$&    1384.5&&&  \\
$m_{\Xi^*}$&    1533.4 &&& \\
$m_{\Omega}$&    1672.5&&&  \\
$m_K$ & 495.6 &&&\\
$m_\rho$ & 775.2&&& \\
$m_\omega$ & 782.7&&& \\
$m_\phi$ & 1019.5 &&&\\
$m_{K^*}$ & 893.8 &&&\\
\end{tabular}
\end{ruledtabular}
\end{table}

\begin{table}
\caption{\label{tab:nres-cc}
Fitted values of coupling constants associated with the exchange potentials $v_{M'B',MB}$.
}
\begin{ruledtabular}
\begin{tabular}{lrr}
Couplings & Model A& Model B\\
\hline
$g_v$ & $-$3.027&$-$2.969 \\
$g_s\times g_{spp}\times 2m_\pi$ & $-$17.123& $-$31.510\\
$\alpha_p^D$ & 0.627& 0.604\\
$\alpha_v^D$ & 0.200& 0.290\\
$\alpha_s^D$ & 0.802& 0.784\\
$g_{pbd}\times m_\pi$ & 1.240& 1.204\\
$g_{vbd}\times m_\rho$ & 14.910& 12.556\\
$g_{pdd}\times m_\pi$ & 0.298& 0.100\\
$g_{vpp}$ & 9.582& 10.258 \\
$\kappa_{\rho NN}$ & 1.181& 2.261\\
$\kappa_{\omega NN}$ & 0.502$\times\kappa_{\rho NN}$ & 2.718$\times\kappa_{\rho NN}$ \\
$\kappa_{\phi NN}$ &  3.000$\times\kappa_{\rho NN}$ & 2.986$\times\kappa_{\rho NN}$ \\
$\kappa_{\rho \Xi\Xi}/(2m_{\Xi})$ & 0.020$\times\kappa_{\rho NN}/(2m_N)$ & 2.934$\times\kappa_{\rho NN}/(2m_N)$ \\
$\kappa_{K^* N\Lambda }/(m_{N}+m_{\Lambda})$ & 0.650$\times\kappa_{\rho NN}/(2m_N)$ & 0.931$\times\kappa_{\rho NN}/(2m_N)$ \\
$\kappa_{K^* \Xi\Lambda }/(m_{\Xi}+m_{\Lambda})$ & 2.042$\times\kappa_{\rho NN}/(2m_N)$ & 0.994$\times\kappa_{\rho NN}/(2m_N)$ \\
$\kappa_{K^* \Xi\Sigma }/(m_{\Xi}+m_{\Sigma})$ & 2.936$\times\kappa_{\rho NN}/(2m_N)$ & 0.573$\times\kappa_{\rho NN}/(2m_N)$
\end{tabular}
\end{ruledtabular}
\end{table}

\begin{table}
\caption{\label{tab:nres-cu}
Fitted values of cutoff parameters associated with the exchange potentials $v_{M'B',MB}$.
}
\begin{ruledtabular}
\begin{tabular}{lrr}
Cutoffs & Model A & Model B \\
& (MeV) & (MeV)\\
\hline
$\Lambda_{\pi NN}~(\equiv\Lambda_{\pi\Xi\Xi})$             &  600 & 600 \\
$\Lambda_{\pi \Lambda\Sigma}$  &  1575 &1566 \\
$\Lambda_{\pi \Sigma\Sigma}$   &  1036 & 890\\
$\Lambda_{K N\Lambda}~(\equiv\Lambda_{K\Xi\Lambda},\Lambda_{K\Xi\Sigma})$         & 1242 & 1365\\
$\Lambda_{K N\Sigma}$          &  688 &  835\\
$\Lambda_{\eta NN}~(\equiv\Lambda_{\eta \Lambda\Lambda} ,\Lambda_{\eta \Sigma\Sigma} ,\Lambda_{\eta \Xi\Xi})$ &  509 & 840\\
$\Lambda_{\rho NN}~(\equiv\Lambda_{\rho \Xi\Xi})$          &  889& 951 \\
$\Lambda_{\rho \Lambda\Sigma}$    &  787 & 1051\\
$\Lambda_{\rho \Sigma\Sigma}$          & 611 & 654 \\
$\Lambda_{K^* N\Lambda}$         & 1675 & 1005\\
$\Lambda_{K^* N\Sigma}$          & 1129 & 1014\\
$\Lambda_{K^* \Xi\Lambda}$         & 1581 & 1254\\
$\Lambda_{K^* \Xi\Sigma}$          &  500&  1256\\
$\Lambda_{\omega NN}~(\equiv\Lambda_{\omega \Xi\Xi})$         & 1016 &  581\\
$\Lambda_{\phi NN}~(\equiv\Lambda_{\phi \Xi\Xi})$         &  873 &  537\\
$\Lambda_{\sigma NN}~(\equiv\Lambda_{\sigma \Lambda\Lambda},\Lambda_{\sigma \Sigma\Sigma},\Lambda_{\sigma \Xi\Xi})$ & 888 & 1488\\
$\Lambda_{f_0 NN}~(\equiv\Lambda_{f_0 \Lambda\Lambda},\Lambda_{f_0 \Sigma\Sigma},\Lambda_{f_0 \Xi\Xi})$ &  1229 & 1051 \\
$\Lambda_{\kappa N\Lambda}~(\equiv\Lambda_{\kappa \Xi\Lambda}, \Lambda_{\kappa \Xi\Sigma})$         & 1123 & 1102\\
$\Lambda_{\kappa N\Sigma}$         & 1024 & 1304\\
$\Lambda_{\pi N\Delta}~(\equiv\Lambda_{\pi \Xi\Xi^*})$          &  771 & 789\\
$\Lambda_{\pi \Sigma\Sigma^*}$          & 1724 & 926 \\
$\Lambda_{\pi \Lambda\Sigma^*}$          & 1589 & 1604\\
$\Lambda_{K \Sigma\Delta}~(\equiv\Lambda_{K \Xi\Sigma^*},\Lambda_{K \Lambda\Xi^*},\Lambda_{K \Sigma\Xi^*},\Lambda_{K \Xi\Omega})$          & 616 & 502 \\
$\Lambda_{K N\Sigma^*}$          & 751 & 1197\\
$\Lambda_{\eta \Sigma\Sigma^*}~(\equiv\Lambda_{\eta\Xi\Xi^*})$          &  1800& 677 \\
$\Lambda_{\rho N\Delta}~(\equiv\Lambda_{\rho \Sigma\Sigma^*}) $          & 1185 & 1348\\
$\Lambda_{\rho \Lambda\Sigma^*}$          & 1672 & 1753\\
$\Lambda_{K^* \Sigma\Delta}~(\equiv\Lambda_{K^* N\Sigma^*},\Lambda_{K^* \Xi\Sigma^*})$          & 500 & 722\\
$\Lambda_{\pi \Delta\Delta}~(\equiv\Lambda_{\pi \Sigma^*\Sigma^*})$          & 1674 &  526\\
$\Lambda_{K \Delta\Sigma^*}~(\equiv\Lambda_{K \Sigma^*\Xi^*})$           & 739 & 1003\\
$\Lambda_{\rho \pi\pi}$         & 1520  & 1763\\
$\Lambda_{\rho K K}$         &  845 & 505\\
$\Lambda_{K^* K\pi}$         &  1423 & 1205\\
$\Lambda_{K^* K \eta}$          & 667 & 1211\\
$\Lambda_{\omega K K}$         & 1011 & 501\\
$\Lambda_{\phi K K}$         & 860 & 503\\
$\Lambda_{\sigma \pi\pi}~(\equiv\Lambda_{\sigma \eta\eta})$         &  500 & 1425\\
$\Lambda_{\sigma K K}$         & 1579 &  1196\\
$\Lambda_{f_0\pi\pi}~(\equiv\Lambda_{f_0 \eta\eta}, \Lambda_{f_0 KK})$         & 1333  & 1557\\
$\Lambda_{\kappa K\pi}~(\equiv\Lambda_{\kappa K\eta})$         &  987 & 1062
\end{tabular}
\end{ruledtabular}
\end{table}

\begin{table}
\caption{\label{tab:nres-wt}
The parameters associated with the modified $t$-channel potentials in Appendix~\ref{app:pbpb-t-v-2}.
These are only relevant to Model B. 
}
\begin{ruledtabular}
\begin{tabular}{lrlr}
Cutoffs & (MeV) & Parameters &\\
\hline
$\Lambda^{\rm WT}_{\bar KN,\bar KN}~(\equiv\Lambda^{\rm WT}_{\pi\Sigma,\bar KN},\Lambda^{\rm WT}_{\pi\Lambda,\bar KN}, \Lambda^{\rm WT}_{\eta \Lambda,\bar KN})$ & 948 & 
$c^{\rm WT}_{\bar KN,\bar KN}$ & 0.760 \\
$\Lambda^{\rm WT}_{\pi\Sigma,\pi\Sigma}~(\equiv\Lambda^{\rm WT}_{K\Xi,\pi\Sigma})$ & 1014& 
$c^{\rm WT}_{\bar KN,\pi\Sigma}$ & 0.118 \\
$\Lambda^{\rm WT}_{K\Xi,\pi\Lambda}$ & 500 & 
$c^{\rm WT}_{\bar KN,\pi\Lambda}$ & 1.000\\
$\Lambda^{\rm WT}_{K\Xi,K\Xi}~(\equiv \Lambda^{\rm WT}_{\eta\Lambda,K\Xi})$ &  535 & 
$c^{\rm WT}_{\pi\Sigma,K\Xi}~(\equiv c^{\rm WT}_{\pi\Lambda,K\Xi},c^{\rm WT}_{K\Xi,K\Xi})$ &  0.710 \\
$\Lambda^{\rm WT}_{\eta\Lambda,\eta\Lambda}$ &1177 & 
$c^{\rm WT}_{\bar K N,\eta\Lambda}~(\equiv c^{\rm WT}_{K\Xi,\eta\Lambda})$ & 0.466 \\
&&$\gamma_{\rm WT}$ &0.800
\end{tabular}
\end{ruledtabular}
\end{table}

\begin{table}
\caption{\label{tab:res-mass}
Fitted values of bare mass $M_{Y^\ast}^0$ of the $Y^\ast$ states.
The numbers ($i=1,2$) in parentheses in the first column indicate
the $i$-th bare state in a given partial wave.
}
\begin{ruledtabular}
\begin{tabular}{lrr}
$l_{I2J}$  & \multicolumn{2}{c}{$M_{Y^\ast}^0$ (MeV)} \\ 
\cline{2-3}
& Model A & Model B\\
\hline
$S_{01}$ (1)& 1853 & 1857\\
$S_{01}$ (2)& 2155 & 2299\\
$P_{01}$ (1)& 1985 & 1909\\
$P_{01}$ (2)& 1990 & 1990\\
$P_{03}$    & 2392 & 2168\\
$D_{03}$ (1)& 1925 & 1835\\
$D_{03}$ (2)& 1970 & 1984\\
$D_{05}$ (1)& 2059 & 2125\\
$D_{05}$ (2)& 2394 & 2180\\
$F_{05}$    & 2289 & 2234\\
$F_{07}$    & 2135 & 2716\\
$S_{11}$ (1)& 2000 & 1928\\
$S_{11}$ (2)& 2508 & 2363\\
$P_{11}$ (1)& 1884 & 1800\\
$P_{11}$ (2)& 2046 & 1959\\
$P_{13}$ (1)& 1576 & 1619\\
$P_{13}$ (2)& 2471 & 2595\\
$D_{13}$ (1)& 1898 & 1810\\
$D_{13}$ (2)& 2006 & 1998\\
$D_{15}$    & 2285 & 2333\\
$F_{15}$    & 2345 & 2104\\
$F_{17}$    & 2214 & 2630
\end{tabular}
\end{ruledtabular}
\end{table}

\begin{table}
\caption{\label{tab:res-mb-modela}
Fitted values of cutoffs and coupling constants of the bare $Y^\ast \to MB$ vertices 
($MB = \bar K N, \pi \Sigma, \pi \Lambda, \eta \Lambda, K \Xi, \pi \Sigma^*, \bar K^* N$) for Model A. 
The corresponding $(LS)$ quantum numbers of each $MB$ state are shown in Table~\ref{tab:pw}.
The cutoff $\Lambda_{Y^*}$ is listed in the unit of MeV.
The numbers ($i=1,2$) in parentheses in the first column indicate
the $i$-th bare state in a given partial wave.
}
\begin{ruledtabular}
\begin{tabular}{lrrrrrrrrrrr}
$l_{I2J}$ &$\Lambda_{Y^*}$ & \multicolumn{9}{c}{$C_{MB(LS),Y^*}$}\\
\cline{3-12}
  & &$\bar K N$&$\pi \Sigma$&$\pi\Lambda$&$\eta\Lambda$&$K\Xi$& $(\pi\Sigma^*)_1$&$(\pi\Sigma^*)_2$&$(\bar K^* N)_1$&$(\bar K^* N)_2$&$(\bar K^* N)_3$\\
\hline
$S_{01}$ (1)&  977&   13.164&   12.298& -       &    4.508&   11.135& $-$0.451& -       &    0.759& $-$0.401& -       \\      
$S_{01}$ (2)& 1093&   13.500& $-$2.355& -       & $-$0.646& $-$6.706& $-$0.227& -       &    5.490&    0.785& -       \\      
$P_{01}$ (1)&  520&    0.009&    6.760& -       & $-$1.955&    0.868&   11.000& -       & $-$2.435& $-$1.869& -       \\      
$P_{01}$ (2)&  981&    4.590&    8.286& -       & $-$1.944&    2.674& $-$2.262& -       & $-$1.576& $-$5.278& -       \\      
$P_{03}$    &  988&    1.018& $-$0.985& -       &    0.319&    1.337&    3.213& $-$0.091& $-$2.653&    3.034& $-$0.068\\      
$D_{03}$ (1)& 1284&    0.480&    0.060& -       &    0.004&    0.196& $-$1.493& $-$0.111& $-$0.033&    4.004&    0.013\\      
$D_{03}$ (2)&  654&    0.785&    1.572& -       &    0.164&    1.790&    0.961& $-$3.335& $-$1.434&   10.996&    0.970\\      
$D_{05}$ (1)&  500&    0.892& $-$1.742& -       & $-$0.505&    2.791& $-$1.271&    0.034&    1.852& $-$1.631& $-$0.265\\      
$D_{05}$ (2)&  869&    0.366& $-$0.306& -       & $-$0.047& $-$0.705& $-$1.258&    0.009& $-$0.001&    0.071& $-$0.066\\      
$F_{05}$    & 1136&    0.095& $-$0.027& -       &    0.005& $-$0.013& $-$1.155& $-$0.102& $-$0.006&    0.982& $-$0.007\\      
$F_{07}$    &  654&    0.000&    0.705& -       & $-$0.088&    0.102& $-$0.619& $-$0.001&    0.382& $-$0.511&    0.031\\      
$S_{11}$ (1)&  500&    6.024&$-$12.706& $-$0.166& -       &    8.000& $-$1.013& -       &    9.861& $-$9.055& -       \\      
$S_{11}$ (2)& 1222&    8.681& $-$0.817&    9.155& -       & $-$4.961& $-$0.727& -       &    1.011&    0.689& -       \\      
$P_{11}$ (1)& 1801&    0.131&    0.176& $-$0.052& -       &    1.998& $-$0.593& -       &    0.120& $-$0.362& -       \\      
$P_{11}$ (2)&  681&    2.053& $-$5.357&    2.062& -       &    4.827& $-$5.404& -       & $-$7.797&    1.486& -       \\      
$P_{13}$ (1)&  704&    3.827& $-$1.604& $-$0.743& -       & $-$3.261&    8.293& $-$0.032&    2.706& $-$2.121&    0.196\\     
$P_{13}$ (2)&  715&    6.903&    1.211&    9.000& -       &    0.789& $-$0.719&    0.109& $-$4.589& $-$2.223& $-$0.036\\      
$D_{13}$ (1)&  501&    0.478&    3.182& $-$1.758& -       &    0.493&    0.255& $-$1.072& $-$5.100&    3.999& $-$4.483\\      
$D_{13}$ (2)&  904&    0.374& $-$0.112&    1.172& -       & $-$0.283& $-$9.752& $-$0.836& $-$0.297&    5.416& $-$0.328\\      
$D_{15}$    &  855&    0.718&    0.451&    0.168& -       &    0.587& $-$1.435&    0.011&    0.030&    1.156&    0.002\\      
$F_{15}$    & 1199& $-$0.013&    0.106&    0.031& -       & $-$0.008&    0.710& $-$0.023& $-$0.029& $-$0.913& $-$0.002\\      
$F_{17}$    &  745&    0.148& $-$0.060&    0.037& -       & $-$0.041&    0.060& $-$0.002&    0.189&    0.312&    0.001
\end{tabular}
\end{ruledtabular}
\end{table}

\begin{table}
\caption{\label{tab:res-mb-modelb}
Fitted values of cutoffs and coupling constants of the bare $Y^\ast \to MB$ vertices 
($MB = \bar K N, \pi \Sigma, \pi \Lambda, \eta \Lambda, K \Xi, \pi \Sigma^*, \bar K^* N$) for Model B. 
The corresponding $(LS)$ quantum numbers of each $MB$ state are shown in Table~\ref{tab:pw}.
The cutoff $\Lambda_{Y^*}$ is listed in the unit of MeV.
The numbers ($i=1,2$) in parentheses in the first column indicate
the $i$-th bare state in a given partial wave.
}
\begin{ruledtabular}
\begin{tabular}{lrrrrrrrrrrr}
$l_{I2J}$ &$\Lambda_{Y^*}$ & \multicolumn{9}{c}{$C_{MB(LS),Y^*}$}\\
\cline{3-12}
  & &$\bar K N$&$\pi \Sigma$&$\pi\Lambda$&$\eta\Lambda$&$K\Xi$& $(\pi\Sigma^*)_1$&$(\pi\Sigma^*)_2$&$(\bar K^* N)_1$&$(\bar K^* N)_2$&$(\bar K^* N)_3$\\
\hline
$S_{01}$ (1)&  694&    9.101&   14.775& -       &    3.959&   13.776&    1.348& -       & $-$5.270& $-$2.038& -       \\      
$S_{01}$ (2)& 1235&   10.426& $-$7.539& -       &    2.535& $-$8.016&    0.438& -       &    7.534&    0.326& -       \\      
$P_{01}$ (1)&  500&    0.001&    8.075& -       &    0.200&    7.431&   11.474& -       &    2.767&    7.824& -       \\      
$P_{01}$ (2)&  978&    7.131&    5.260& -       & $-$1.018&    5.280&    2.180& -       &    2.921& $-$3.130& -       \\      
$P_{03}$    &  674&    1.887&    1.027& -       &    8.882&    5.444& $-$2.644&    0.078& $-$4.752& $-$3.182& $-$0.105\\      
$D_{03}$ (1)&  981&    0.626&    0.242& -       &    0.186& $-$0.144& $-$1.767& $-$0.607&    0.566& $-$0.972& $-$0.238\\      
$D_{03}$ (2)&  614&    0.733&    2.135& -       &    0.061&    0.247& $-$2.016& $-$4.667& $-$0.321&   13.871& $-$1.194\\      
$D_{05}$ (1)&  739&    0.412& $-$1.285& -       & $-$0.035& $-$0.210&    0.186& $-$0.001&    1.388& $-$0.763& $-$0.135\\      
$D_{05}$ (2)&  792&    0.368&    0.160& -       &    0.030& $-$0.249& $-$1.156&    0.005&    0.677& $-$0.441&    0.031\\      
$F_{05}$    & 1258&    0.062& $-$0.033& -       &    0.022&    0.008&    1.558& $-$0.030& $-$0.023& $-$0.947& $-$0.008\\      
$F_{07}$    &  862&    0.046&    0.261& -       & $-$0.006&    0.333& $-$0.148& $-$0.001&    0.069& $-$0.076& $-$0.006\\      
$S_{11}$ (1)&  779&    7.992&    8.039&    8.973& -       & $-$0.607& $-$0.511& -       &    3.262&    0.992& -       \\      
$S_{11}$ (2)& 1709& $-$0.516& $-$7.985&   12.805& -       &    9.037& $-$0.233& -       &    7.645&    0.216& -       \\      
$P_{11}$ (1)&  887&    0.419&    0.305&    1.437& -       &    8.038& $-$1.480& -       &    0.718& $-$2.188& -       \\      
$P_{11}$ (2)&  500& $-$0.002& $-$3.503& $-$4.003& -       & $-$2.454& $-$7.046& -       & $-$2.321&    3.762& -       \\      
$P_{13}$ (1)&  795&    3.453& $-$1.148&    1.201& -       &    0.220& $-$7.558&    0.030& $-$1.343& $-$0.346&    0.049\\     
$P_{13}$ (2)&  736&    4.289& $-$9.734&    3.510& -       &    3.917&    1.026& $-$0.082&    3.696&    1.244& $-$0.006\\      
$D_{13}$ (1)&  502&    0.225&    2.899& $-$3.264& -       & $-$4.559& $-$3.235&    0.108& $-$2.364&$-$16.140& $-$0.192\\      
$D_{13}$ (2)&  674&    0.623&    0.270&    2.634& -       & $-$0.195& $-$2.014& $-$1.784&    2.095&$-$12.195&    0.359\\      
$D_{15}$    &  891&    0.708&    0.520&    0.458& -       &    0.208& $-$1.605&    0.005& $-$0.580& $-$0.322& $-$0.000\\      
$F_{15}$    &  673&    0.003&    0.631&    0.551& -       & $-$0.181& $-$0.107& $-$0.087& $-$0.119& $-$2.523&    0.009\\      
$F_{17}$    &  852&    0.136& $-$0.063& $-$0.073& -       & $-$0.046&    0.342& $-$0.004&    0.062&    0.085&    0.012
\end{tabular}
\end{ruledtabular}
\end{table}

\end{document}